\documentclass[aps,prl,twocolumn,superscriptaddress,longbibliography]{revtex4-2}
\usepackage{amssymb,amsthm,amsmath,amstext,amsbsy,amsopn}
\usepackage{physics}
\usepackage{graphicx}
\usepackage[dvipsnames]{xcolor}
\usepackage{colortbl}
\usepackage{bm}
\usepackage[breaklinks,colorlinks,linkcolor=blue,citecolor=blue,urlcolor=blue]{hyperref}

\begin{document}

\title{Efimovian Phonon Production for an Analog Coasting Universe in Bose-Einstein Condensates}
\author{Yunfei Xue}
\affiliation{MOE Key Laboratory for Nonequilibrium Synthesis and Modulation of Condensed Matter, Shaanxi Province Key Laboratory of Quantum
Information and Quantum Optoelectronic Devices, School of Physics, Xi'an Jiaotong University, Xi'an 710049, China}

\author{Jiabin Wang}
\affiliation{Institute of Theoretical Physics, State Key Laboratory of Quantum Optics and Quantum Optics Devices, Shanxi University, Taiyuan 030006, China}

\author{Li Chen}
\email{lchen@sxu.edu.cn}
\affiliation{Institute of Theoretical Physics, State Key Laboratory of Quantum Optics and Quantum Optics Devices, Shanxi University, Taiyuan 030006, China}

 \author{Chenwei Lv}
  \email{lcw205046@gmail.com}
\affiliation{Department of Physics, The Chinese University of Hong Kong, Shatin, New Territories, Hong Kong, China}
\affiliation{The State Key Laboratory of  Quantum Information Technologies and Materials, The Chinese University of Hong Kong, Shatin, New Territories, Hong Kong, China}
\affiliation{New Cornerstone Science Laboratory, The Chinese University of Hong Kong, Shatin, New Territories, Hong Kong, China}

\author{Ren Zhang}
\email{renzhang@xjtu.edu.cn}
\affiliation{MOE Key Laboratory for Nonequilibrium Synthesis and Modulation of Condensed Matter, Shaanxi Province Key Laboratory of Quantum
Information and Quantum Optoelectronic Devices, School of Physics, Xi'an Jiaotong University, Xi'an 710049, China}
\affiliation{Hefei National Laboratory, Hefei, 230088, China}

\date{\today}

\begin{abstract}
Efimov effects arise from scale invariance, a fundamental symmetry with universal implications. While spatial Efimov physics has been extensively studied, realizing its temporal counterpart remains challenging, as it requires a dynamical system that breaks time-translation symmetry yet preserves the essential time-scaling symmetry.
Analog cosmology offers a powerful platform to address this challenge, bridging the domains of Efimov physics and cosmology. Here, we predict a temporal Efimov effect in an analog 
linearly expanding universe realized with a quasi-two-dimensional Bose-Einstein condensate. The invariance of phonon mode equations under time rescaling leads to particle production with two distinct dynamics: power-law growth and log-periodic oscillations, with the latter being the hallmark signature of the Efimov effect. Furthermore, these dynamics map directly onto sub- and super-horizon cosmological modes. 
Our predictions can be directly verified through time-averaged measurements of the density-fluctuation spectrum $S_{k}(t)$ in current experiments.
\end{abstract}

\maketitle

{\it Introduction—}
The Efimov effect, first predicted in nuclear physics \cite{Efimov1970}, is a universal phenomenon observed across disciplines ranging from quantum gases to biological systems \cite{Kraemer2006, Braaten2007, Zaccanti2009, Gross2009, Stecher2009, Lompe2010, Maji2010, Tung2014, Pires2014, zheyu2015, Kunitski2015, Pal2015, Deng2016, Sun2017, Greene2017, Naidon2017, Johansen2017,pengfei2017, Deng2018, Wang2018, xiaoling2019, Hammer2020, ren2021, Chuang2025, Haiwen2025,pengfei2025}. 
Its hallmark log-periodic behavior has been observed in diverse systems, including the bound states in cold atomic gases \cite{Kraemer2006, Zaccanti2009, Tung2014, Greene2017, Naidon2017, Johansen2017, Chuang2025}, expanding Fermi gases \cite{Deng2016, Deng2018}, semimetals \cite{Wang2018, Haiwen2025}, and helical DNA \cite{Maji2010, Pal2015}. The root of the Efimov effect is scale invariance. This symmetry can manifest in either space or time, two equally fundamental domains. Yet, there is a stark imbalance:
While Efimov physics stemming from spatial scale invariance has been extensively investigated, its counterpart arising from {\it temporal} scale invariance has been addressed only in limited contexts, such as the expanding Fermi gases \cite{Deng2016, Deng2018}. This is because temporal scale invariance imposes stringent constraints on the dynamics. Consequently, a pressing challenge is to identify a tunable platform where such temporal symmetry can be engineered to reveal the temporal Efimov effect. 

To address this challenge, we bridge two seemingly distinct frontiers: Efimov physics and analog cosmology. While cosmic expansion inherently involves time-dependent metrics, engineering the precise temporal scaling 
is impractical in observational cosmology. Analog cosmology offers a powerful alternative for synthesizing curved spacetimes in controlled laboratory settings \cite{Unruh1981, Brumfiel2008, Chiao2018}. Ultracold atomic gases, a highly tunable system, have been demonstrated as a versatile platform for analog cosmology \cite{Garay2000, Garay2001, Fischer2003, Jain2007, Rousseaux2008, Carusotto2008, Macher2009, Lahav2010, Prain2010, Lahav2010, Jaskula2012, ChenLung2013, Steinhauer2014, Steinhauer2016, Eckel2018, Barcelo2018, Steinhauer2019, Gooding2020, Almheiri2021, Banik2022, TolosaSimeon2022, Viermann2022, Oberthaler2024, Gondret2025}. 
Using quasi-two-dimensional Bose–Einstein condensates (BECs), recent experiments have successfully simulated the (2+1)-dimensional Friedmann-Lema{\^i}tre-Robertson-Walker (FLRW) universe, described by the metric \cite{Viermann2022, Oberthaler2024, SM}
\begin{align}
ds^{2}=-dt^{2}+a^{2}(t)\left(dx^{2}+dy^{2}\right),
\end{align}
where the scale factor $a(t)$ is dynamically controlled by tuning atomic interactions via Feshbach resonances \cite{Chin2010}. 
This flexibility enables the realization of arbitrary scaling behaviors. A particularly intriguing case is the linear expansion with $a(t)\propto t$, known as the coasting universe \cite{Milne1934, Kolb1989, Padmanabhan1990, Abha2002, Melia2018, Wilkinson2025}. 
This model is of fundamental interest because its massless mode equation is invariant under time-scaling, hinting at the presence of Efimovian phenomena. However, for this symmetry to manifest as particle production, the dimensionality is a critical factor. Unlike the (1+1)-dimensional case, where the coincidence of minimal coupling and conformal symmetry prohibits particle production \cite{Fulling1974, Chitre1977, BunchT.1978, Parker2009}, a minimally coupled scalar field in (2+1)-dimension breaks this symmetry, allowing for particle production \cite{Viermann2022}. Nevertheless, the implication of temporal scale invariance on particle production within this (2+1)-dimensional setting remains unexplored.

In this Letter, we explicitly reveal the temporal Efimov effect in an analog coasting universe realized in quasi-two-dimensional BECs. We show that momentum-resolved particle production exhibits two distinct behaviors governed by the dimensionless parameter $kl$ (with $k$ being momentum and $l$ being a characteristic length scale defining the expansion rate): For $kl > 1/2$, the particle production is log-periodic in time, a phenomenon we term “Efimovian phonon production”; for $kl < 1/2$, it follows a power-law growth. These two behaviors are direct analogs of the sub- and super-horizon regimes in cosmology. Consistently, the momentum-integrated particle density in super-horizon regimes grows linearly with the expansion time. In contrast, in the sub‑horizon regime, it quickly saturates with a double‑logarithmic correction. 
The time-averaged density-fluctuation spectrum provides a direct probe of these phenomena.

{\it Efimovian phonon production-}
In a quasi-two-dimensional BEC, phonon excitations effectively propagate in an emergent curved spacetime, termed the phonon universe.
Specifically, the phonon field $\phi({\bm \rho},t)$ obeys the Klein-Gordon equation in a (2+1)-dimensional FLRW metric with the scale factor $a(t)=(m^{3}/(8\pi\omega_{z}\hbar^{3}n_{0}^{2}a_{s}^{2}(t)))^{1/4}$.
Here, $a_{s}(t)$ denotes the $s$-wave scattering length, $n_{0}$ is the background atomic number density, $m$ is the atomic mass, and $\omega_{z}$ is the trapping frequency along the $z$-axis \cite{Viermann2022, TolosaSimeon2022,Oberthaler2024}.
Upon quantization, the phonon field dynamics are governed by the equation of motion (EoM) for the mode function $u_{k}$: $\ddot u_{k}(t)+2[\dot{a}(t)/a(t)]\dot{u}_{k}(t)+[k^{2}/a^{2}(t)]u_{k}(t)=0$ \cite{TolosaSimeon2022}. The damping term, $2\dot{a}(t)/a(t)$, simulates the Hubble friction \cite{Eckel2018}, which breaks time-reversal symmetry and drives phonon production as the phonon universe expands. 
It has been demonstrated that the dynamics can be mapped onto a one-dimensional scattering problem, thereby establishing a relationship between particle production and the scattering amplitude \cite{Oberthaler2024}. This mapping is rooted in the underlying SU(1,1) symmetry.

\begin{figure}
    \includegraphics[width=0.43\textwidth]{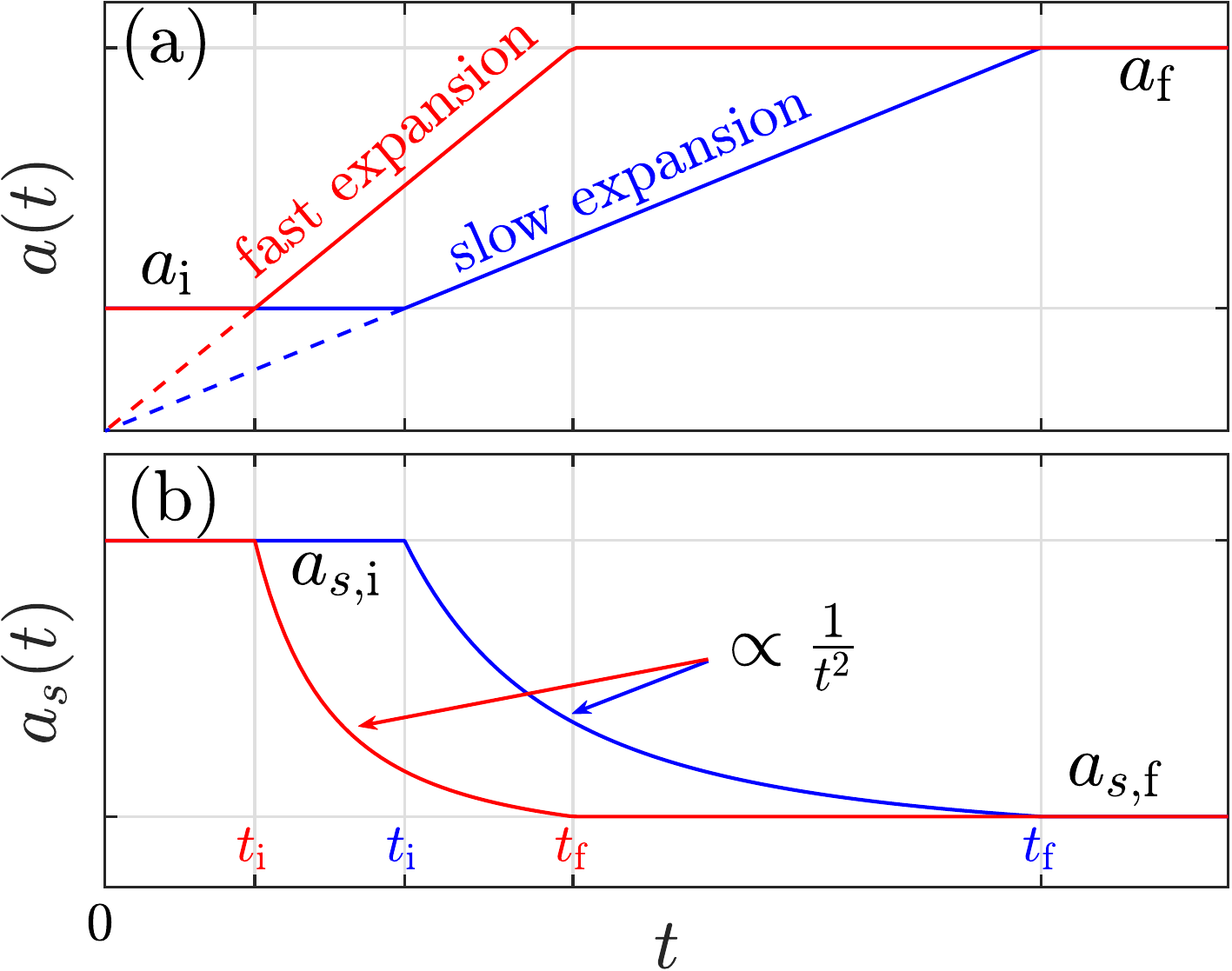}
    \caption{Schematic of the scale factor $a(t)$ and the $s$-wave scattering length $a_{s}(t)$. (a): The scale factor exhibits linear growth $a(t)= t/l$ during the interval $t_{\rm i}<t<t_{\rm f}$. (b) The required scattering length follows the form $a_{s}(t)\propto 1/t^{2}$ in this regime. Red and blue curves correspond to fast (small $l$) and slow (large $l$) expansion scenarios, respectively. Note that the initial time is defined as $t_{\rm i}\equiv la_{\rm i}$. Consequently, $t_{\rm i}$ for fast expansion (red) is smaller than that for slow expansion (blue). Similarly, for a fixed target scale factor $a_{\rm f}$, the final time $t_{\rm f}$ occurs earlier for fast expansion. However, the expansion ratio $t_{\rm f}/t_{\rm i}$ remains identical for both cases.}
    \label{figure1}
\end{figure}

The temporal Efimov effect requires a protocol that inherently possesses temporal scale invariance. The simplest choice is a linear expansion, corresponding to a coasting universe. We therefore consider the protocol schematized in Fig.~\ref{figure1}(a), characterized by a scale factor 
\begin{align}\label{scalingfactor}
    a(t)=t/l,\quad t_{\rm i}\leqslant t\leqslant t_{\rm f}.
\end{align}
Outside this interval, $a(t)$ remains constant, taking the values $a_{\rm i}$ and $a_{\rm f}$ when $t<t_{\rm i}$ and $t>t_{\rm f}$, respectively. Note that the zero intercept in Eq.~(\ref{scalingfactor}) is crucial for exact time-scaling invariance. Experimentally, this amounts to defining the time origin $t=0$ at the singularity. 
We set $t_{\rm i}\equiv la_{\rm i}$ and $t_{\rm f}\equiv la_{\rm f}$ to ensure continuity. This protocol can be implemented by tuning the $s$-wave scattering length according to $a_{s}(t)=\sqrt{m^{3}/(8\pi\omega_{z}\hbar^{3}n_{0}^{2})} l^2/t^{2}$(see Fig.~\ref{figure1}(b)). 
Substituting $a(t)$ into the EoM for the mode function yields, for $t_{\rm i}\leqslant t\leqslant t_{\rm f}$, 
\begin{align}
\label{modeequation}
\ddot{u}_{k}(t)+\frac{2}{t}\dot u_{k}(t)+\frac{(kl)^{2}}{t^{2}}u_{k}(t)=0.
\end{align}
This equation is invariant under the time scaling transformation $t\to\lambda t$ for arbitrary real $\lambda$. 
This invariance is underpinned by an SU(1,1) symmetry, in which
the time scaling is generated by $\hat K_2=-i(t\partial_t+3/2)/2$, while the dynamics are determined by the generators $\hat K_{0(1)}=\partial_t^2+(2/t)\partial_t+(kl)^2/t^2\mp t^2/16$. Crucially, this representation has a Bargmann index $(1\pm\sqrt{\cal B})/2$ (with ${\cal B}\equiv1/4-(kl)^2$), which signals the emergence of two distinct dynamical behaviors separated by the critical value $kl=1/2$~\cite{deAlfaro1976}.
Reflecting this bifurcation, the solutions of Eq.~(\ref{modeequation}) take different forms depending on $kl$.  For $kl<1/2$, the solution is a superposition of two power laws: $u_{k}(t)=t^{-1/2}(c_{1}t^{-\sqrt{\cal B}}+c_{2}t^{\sqrt{\cal B}})$; In contrast, for $kl>1/2$, the solution becomes: $u_{k}(t)=t^{-1/2}(c_{1}e^{-i\sqrt{|\cal B|}\ln t}+c_{2}e^{i\sqrt{|\cal B|}\ln t})$, a log-periodic function of $t$. At the critical point $kl=1/2$, $u_{k}(t)=t^{-1/2}(c_{1}+c_{2}\ln t)$. Here, $c_{1,2}$ are determined by the boundary condition at $t_{\rm i}$ and $t_{\rm f}$. Similar to the spatial Efimov effect, the boundary conditions break continuous scaling invariance down to a discrete symmetry. 

The temporal Efimov effect can be observed in current experiments by detecting the particle production following the expansion. The 
pre- and post-expansion mode functions, \( u_{k,\mathrm{i}}(t) \) and \( u_{k,\mathrm{f}}(t) \), defines two distinct sets of phonon operators, \( (\hat{b}_{\mathbf{k},\mathrm{i}},\hat{b}^{\dagger}_{\mathbf{k},\mathrm{i}}) \) and \( (\hat{b}_{\mathbf{k},\mathrm{f}},\hat{b}^{\dagger}_{\mathbf{k},\mathrm{f}}) \), which determine the pre- and post-expansion vacua, respectively. These sets are related via a Bogoliubov transformation, given by \( u_{k,\mathrm{f}}(t)=\alpha_{k} u_{k,\mathrm{i}}(t)+\beta_{k} u_{k,\mathrm{i}}^*(t) \) for the mode functions. Correspondingly, the operators transform as \( \hat{b}_{\mathbf{k},\mathrm{f}}=\alpha^*_{k}\hat{b}_{\mathbf{k},\mathrm{i}}-\beta^*_{k}\hat{b}^\dagger_{-\mathbf{k},\mathrm{i}} \), subject to the normalization condition \( |\alpha_k|^2-|\beta_k|^2=1 \).
Initially, the system is prepared in the ground state of the pre-expansion vacuum, satisfying $\hat{b}_{\bf k,{\rm i}}|\Omega\rangle=0$. After the expansion, the produced particle number $N_{k}\equiv\langle\Omega|\hat{b}^{\dagger}_{\bf k,{\rm f}}\hat{b}_{\bf k,{\rm f}}|\Omega\rangle=|\beta_k|^2$ is experimentally measurable. A straightforward calculation yields \cite{SM}
\begin{align}
\label{Nk}
   N_{k}=
    \begin{cases}
    \displaystyle
        \frac{\sin^2\left(\sqrt{|\cal B|}\ln (t_{\rm f}/t_{\rm i})\right)}{4|\cal B|},& kl>1/2\\
         \displaystyle
         \frac{\sinh^2\left(\sqrt{\cal B}\ln (t_{\rm f}/t_{\rm i})\right)}{4\cal B}.& kl<1/2
    \end{cases}
\end{align}
At the critical point $kl=1/2$, one has $N_{k}=(\ln t_{\rm f}/t_{\rm i})^{2}/4$, which is the common limit of the two branches.
Thus, particle production shows two distinct behaviors governed by the parameter $kl$: for $kl>1/2$, $N_{k}$ exhibits log-periodic oscillations with respect to $t_{\rm f}/t_{\rm i}$;  for $kl<1/2$, it follows a symmetric power-law growth.
From the perspective of one-dimensional scattering, $kl=1/2$ marks the boundary of the classical allowed and quantum tunneling regimes \cite{Oberthaler2024}.
\begin{figure}
    \includegraphics[width=0.43\textwidth]{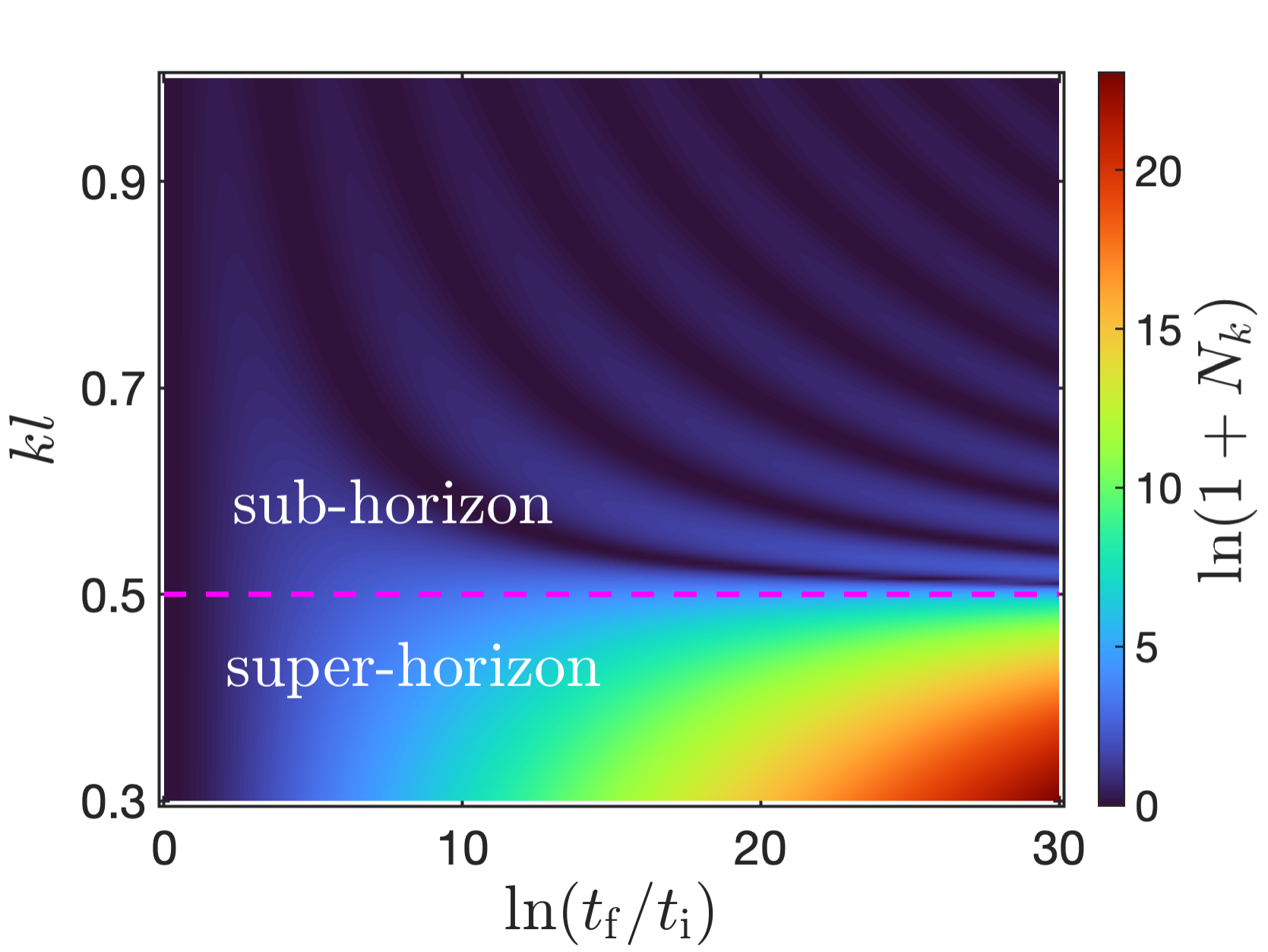}
    \caption{Post-expansion phonon production of the $(2+1)$ dimensional analog coasting universe for varying $kl$ and $\ln(t_{\rm f}/t_{\rm i})$. The color scale represents $\ln(N_{k}+1)$. According to the behaviors of $N_{k}$, two distinct regimes are identified: In the sub-horizon regime ($kl>1/2$), $N_{k}$ is a log-periodic function of $t_{\rm f}/t_{\rm i}$. In the super-horizon regime($kl<1/2$), $N_{k}$ exhibits symmetric power-law dependence on $t_{\rm f}/t_{\rm i}$. The purple-dashed line at $kl=1/2$ defines the boundary between these two regimes.
    \label{figure2}}    
\end{figure}
In Fig.~\ref{figure2}, we display the post-expansion particle number $N_{k}$ as functions of $kl$ and $\ln (t_{\rm f}/t_{\rm i})$. The log-periodic oscillations possess a period $T$ that is determined by the wavenumber. In the limit of $kl\gg1$, the period approximates $T\approx \pi/(kl)$, with an amplitude scaling as $\sim1/(2kl)^{2}$. Although the amplitude diminishes with increasing $kl$, the log-periodic regime is well within reach of current experiments. For instance, Ref.~\cite{Oberthaler2024} reports an expansion rate $\dot a^{2}(t)\approx 0.04\,\mu{\rm m}^{-2}$ (corresponding to $l\approx5\,\mu{\rm m}$) and observes phonon production at wavenumbers on the order of $k\sim 1\,\mu{\rm m}^{-1}$ (spanning a range from $0.4$ to $2.5\,\mu{\rm m}^{-1}$). These values correspond to $kl$ ranging approximately from $2$ to $12.5$, situated deeply within the Efimov regime. Extensions to higher spatial dimensions and the counterpart of Eq.~(\ref{Nk}) are provided in the End Matter.

We highlight a distinct advantage of observing the Efimov effect in our system. A long-standing dilemma in verifying Efimov physics lies in the difficulty of observing log-periodicity. For instance, in the three-body problem, at least three Efimov bound states are required to confirm the discrete scaling symmetry. However,  it is challenging to observe multiple Efimov bound states \cite{Kraemer2006}. Our scheme of Efimovian phonon production circumvents this challenge by manifesting multiple oscillations. Multi-oscillation signals can be accessed via two practical routes. (1) Varying expansion time: fixing the initial time $t_{\rm i}=la_{\rm i}$ (as illustrated in Fig.~\ref{figure1}) for a given initial scale factor $a_{\rm i}$ and expansion rate $\dot{a}(t)=1/l$, one can increase the expansion duration to observe more oscillations. (2) Varying expansion rate: When both $a_{\rm i}$ and $a_{\rm f}$ are given, the ratio $t_{\rm f}/t_{\rm i}$ is invariant for different expanding rates. While the observational window in logarithmic time is fixed, the oscillation period of $N_{k}$, given by $T=\pi/\sqrt{|\cal B|}$, depends on the expansion rate. A slower expansion (large $l$) or larger wavenumber $k$ reduces the period $T$, thereby allowing more oscillations within the same $\ln(t_{\rm f}/t_{\rm i})$ window. Equivalently, increasing $l$ lowers the Efimov threshold $k > 1/(2l)$, extending the Efimov regime to lower wavenumbers and facilitating observation with low‑$k$ phonons.

{\it Cosmology correspondence-} 
The phonon production, exhibiting either log-periodic or polynomial behaviors, maps directly onto the sub-horizon and super-horizon regimes in cosmology. In the context of cosmic expansion, sub-/super-horizon refers to modes whose comoving wavelength is much smaller/larger than the comoving Hubble radius $R_{H}=c/(a(t)H)$. These modes behave distinctly during expansion, satisfying the criteria $k^{-1}\ll R_{H}$ and $k^{-1}\gg R_{H}$ \cite{cos}. Here, $H=\dot a(t)/a(t)$ is the Hubble parameter and $c$ is the speed of light. 
In the analog coasting universe, the comoving Hubble radius $R_{H}=l$. The criteria for sub- and super-horizon are $kl\gg1$ and $kl\ll1$, which are consistent with our analytical results in Eq.~(\ref{Nk}). Specifically, in the quasi-two-dimensional BEC, we define the sub-horizon and super-horizon as $kl>1/2$ and $kl<1/2$, respectively, as depicted in Fig.~\ref{figure2}.

This correspondence can be understood in terms of the mode equation. By defining the comoving wavelength $\lambda\equiv 2\pi /k$, we rewrite the EoM for the mode function in Eq. (\ref{modeequation}) as 
$\ddot{u}_{k}(t)+{2\dot u_{k}(t)}/({a(t)R_{H}})+{4\pi^{2}u_{k}(t)}/({a^2(t)\lambda^{2}})=0$.
In the sub-horizon limit ($\lambda\ll R_{H}$), the Hubble‑friction term is negligible. The resulting  post-expansion phonon number is $N_{k}=\sin^2\left(kl\ln( t_{\rm f}/t_{\rm i})\right)/(2kl)^{2}$, matching the large $kl$ limit of Eq. (\ref{Nk}). Conversely, in the super-horizon limit ($\lambda\gg R_{H}$), the last term can be ignored. The solution simplifies to $u_{k}(t)=c_{1}+c_{2}/t$, leading to $N_{k}=\sinh^{2}(\ln(t_{\rm f}/t_{\rm i}))$, which is the limit of $kl\to0$ of Eq. (\ref{Nk}). Physically, the wavelength of sub‑horizon modes is so short that they are largely insensitive to the expansion. In contrast, super‑horizon modes are strongly affected by the expansion. Therefore, at late times, super-horizon modes dominate particle production.

\begin{figure}
    \includegraphics[width=0.43\textwidth]{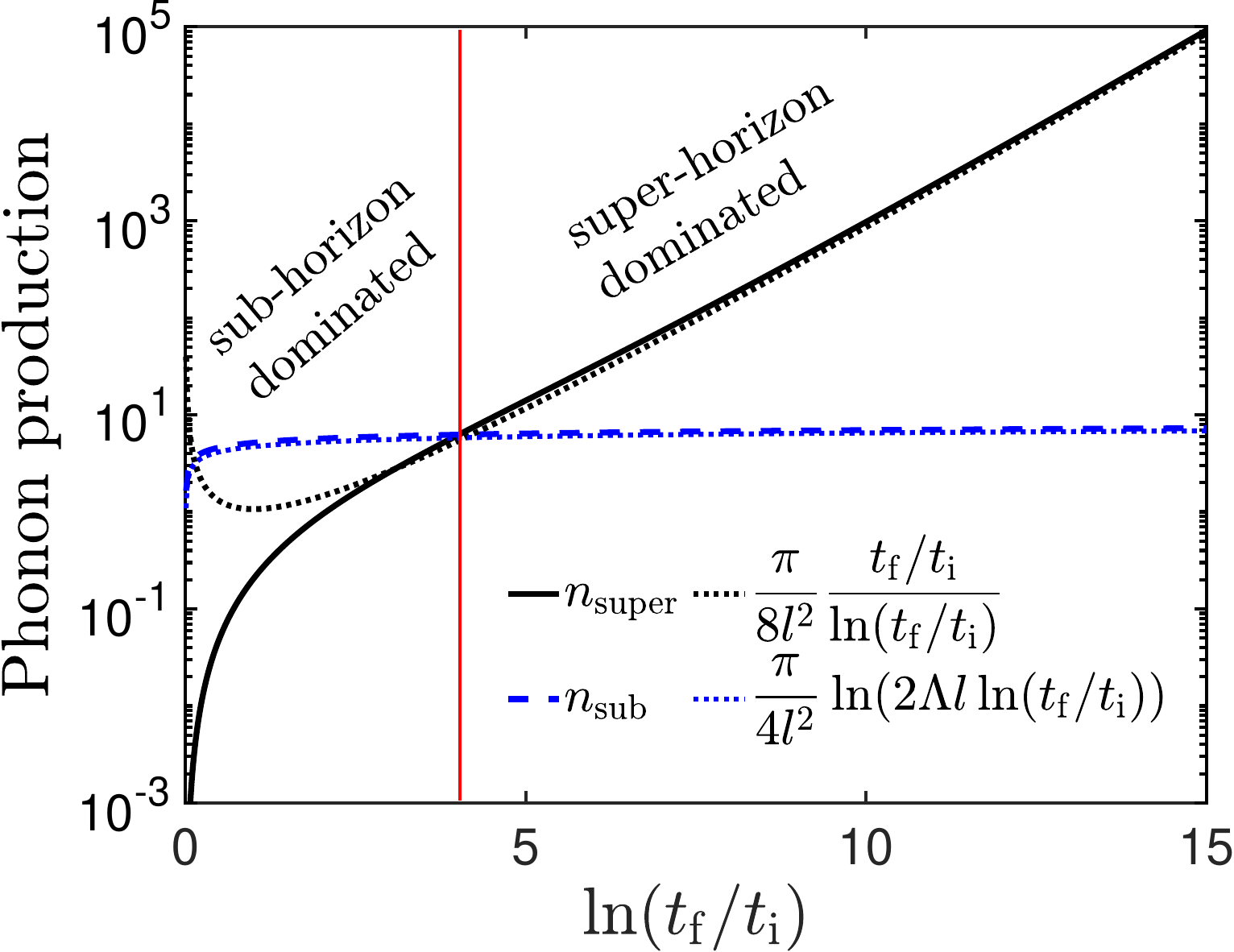}
    \caption{Phonon number densities in the sub- and super-horizon regimes as a function of $\ln( t_{\rm f}/t_{\rm i})$. The black solid and blue dashed curves represent the phonon number density in the super-horizon regime ($n_{\rm super}$) and the sub-horizon regime ($n_{\rm sub}$), respectively. The red vertical line separates the early-time, sub-horizon-dominated phase from the late-time, super-horizon-dominated phase. The black and blue dotted lines indicate the asymptotic behaviors of $n_{\rm super}$ and $n_{\rm sub}$, respectively. The ultraviolet cutoff $\Lambda=200/l$. We adopt $l$ as the unit of length.
    \label{figure3}}    
\end{figure}

We now compare the momentum-integrated phonon number density in the super- and sub-horizon modes as a function of $\ln (t_{\rm f}/t_{\rm i})$. The phonon numbers density are defined as $n_{\rm super}\equiv2\pi\int_{0}^{1/(2l)}kN_{k}dk$ and $n_{\rm sub}\equiv2\pi\int_{1/(2l)}^{\Lambda}kN_{k}dk$, where $\Lambda$ is the ultraviolet cutoff. The explicit expressions are given by
\begin{equation}
\begin{aligned}
\label{particledensity}
&n_{\rm sub}=\frac{\pi\left[\ln(s)-{\rm Ci}(s\ln(t_{\rm f}/t_{\rm i}))+\ln(\ln(t_{\rm f}/t_{\rm i}))+\gamma\right]}{4l^{2}},\\
&n_{\rm super}=\frac{\pi\left[{\rm Chi}(\ln(t_{\rm f}/t_{\rm i}))-\ln(\ln(t_{\rm f}/t_{\rm i}))-\gamma\right]}{4l^{2}},
\end{aligned}
\end{equation}
where $s\equiv\sqrt{(2\Lambda l)^{2}-1}$, and ${\rm Ci}(\cdot)$ and ${\rm Chi}(\cdot)$ denote the cosine and the hyperbolic cosine integral function, respectively, with $\gamma\approx 0.577216$ being Euler's constant. 
As illustrated in Fig.~\ref{figure3}, for a fixed $\Lambda$, the dynamics are divided into two regimes separated by the vertical red line. At a small $\ln (t_{\rm f}/t_{\rm i})$ regime, $n_{\rm super}$ is negligible compared to $n_{\rm sub}$, simply because the total number of sub-horizon modes is much greater than the super-horizon modes. However, $n_{\rm super}$ exhibits approximately exponential growth with respect to $\ln (t_{\rm f}/t_{\rm i})$, whereas $n_{\rm sub}$ rapidly saturates. We therefore define the regime $n_{\rm sub}>n_{\rm super}$ as the sub-horizon dominated regime.  At a critical expansion time $t_{\rm f}$ (set by $n_{\rm sub}=n_{\rm super}$), the dominance switches. For larger $t_{\rm f}$, $n_{\rm super}$ continues to increase, indicating substantial phonon production in the super-horizon regime. By contrast, $n_{\rm sub}$ remains approximately constant, reflecting that particle production is suppressed for modes in sub-horzion regime. Thus, the regime where $n_{\rm super}>n_{\rm sub}$ is defined as the super-horizon-dominated regime.
In the limit of large $\ln (t_{\rm f}/t_{\rm i})$, the asymptotic behavior of $n_{\rm super}$ is $n_{\rm super}\approx\pi(t_{\rm f}/t_{\rm i})/(8l^{2}\ln(t_{\rm f}/t_{\rm i}))$, a linear function in $t_{\rm f}/t_{\rm i}$ with an additional logarithmic correction, as indicated by the black dotted line in Fig.~\ref{figure3}. It can be reformulated into a form more aligned with cosmology conventions, $n_{\rm super}\propto \pi(a_{\rm f}/a_{\rm i})/(8l^{2}\ln(a_{\rm f}/a_{\rm i}))$. Conversely, in the combined limits of large $\ln (t_{\rm f}/t_{\rm i})$ and high ultraviolet cutoff $\Lambda l\gg1$, the sub-horizon contribution grows much more slowly, scaling as $n_{\rm sub}\approx \pi\ln(2\Lambda l\ln(t_{\rm f}/t_{\rm i}))/(4l^{2})$, as indicated by the blue dotted line in Fig.~\ref{figure3}. We note that these distinct asymptotic behaviors of particle density persist in higher spatial dimensions \cite{SM}. Explicit expressions are provided in the End Matter.

\begin{figure}[h]
    \includegraphics[width=0.43\textwidth]{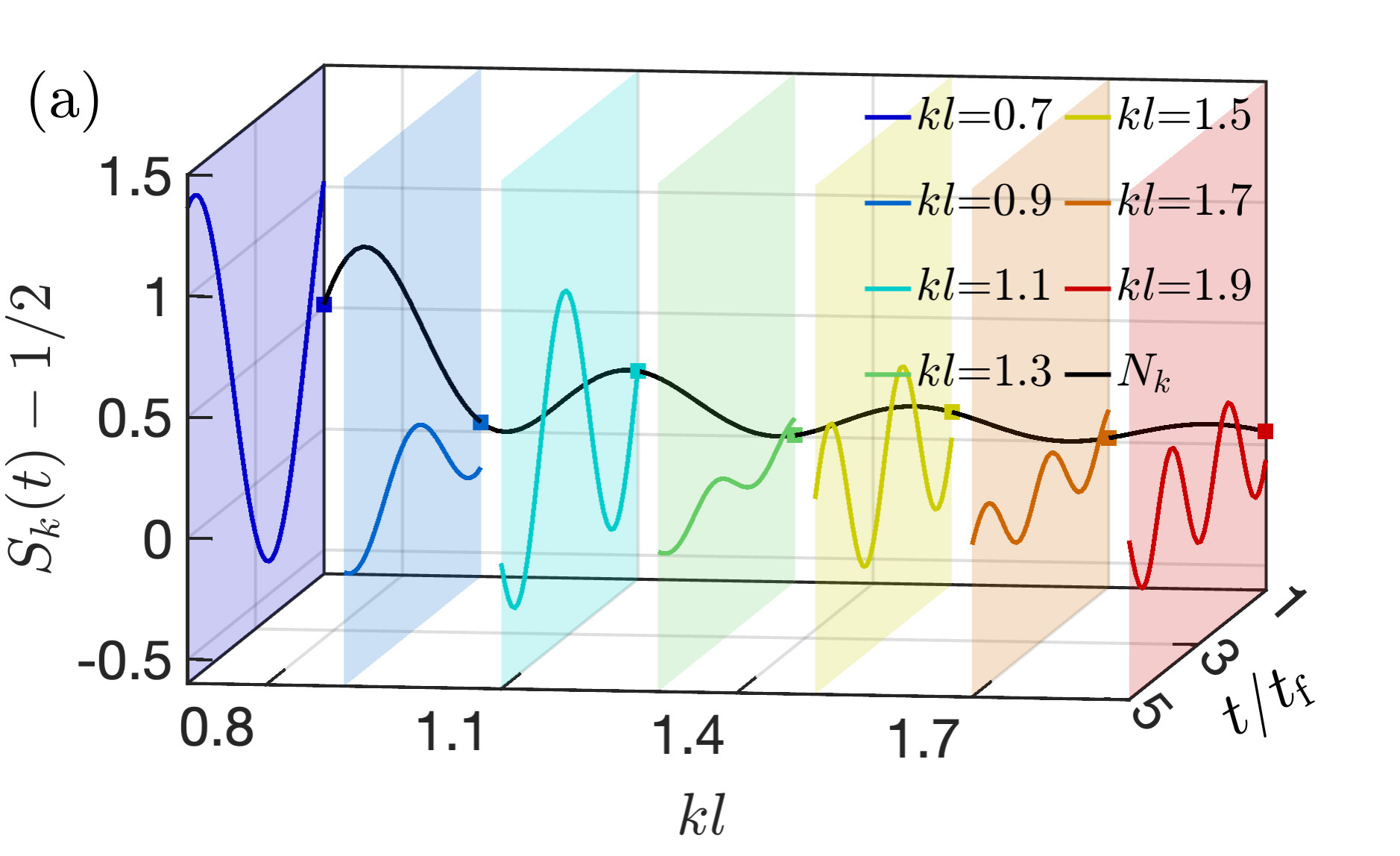}
    \includegraphics[width=0.43\textwidth]{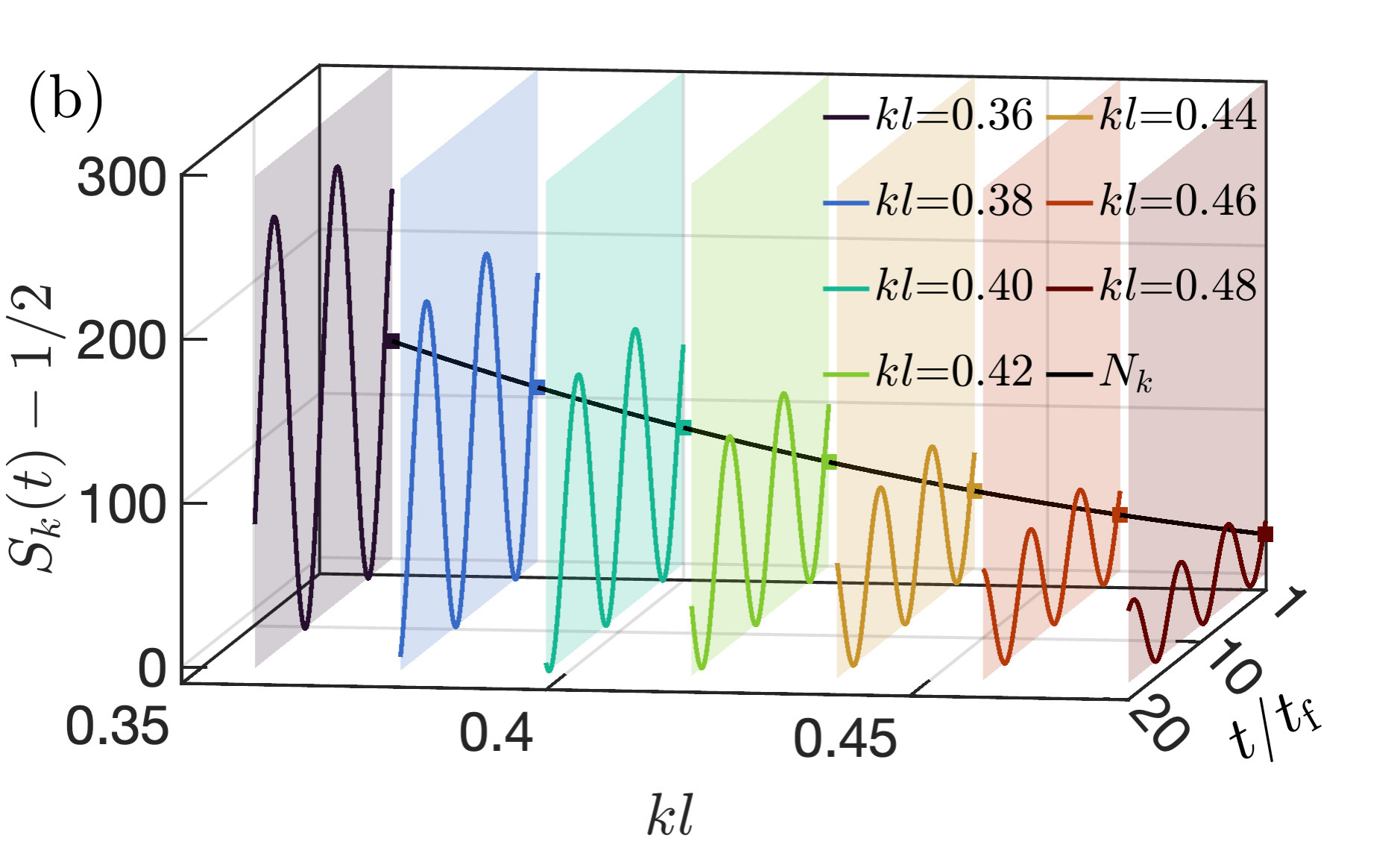}
    \caption{Density-fluctuation spectrum $S_{k}(t)$ and post-expansion phonon number $N_{k}$ in (a) sub-horizon regime and (b) super-horizon regime. The time-dependent spectra $S_{k}(t)$ are represented by colored curves, while the extracted $N_{k}$ values are shown as black curves with corresponding colored markers. In the sub-horizon regime, both $N_{k}$ and the amplitude of $S_{k}(t)$ are oscillatory as a function of $k$, while in the super-horizon regime, they exhibit monotonic growth. In our calculation, we set $\ln(t_{\rm f}/t_{\rm i})=8$  and zero temperature.
    \label{figure4}}    
\end{figure}

{\it Experimental Observable-} In current experiments, the momentum-resolved phonon number $N_{k}$ is not directly accessible; however, it can be inferred from the atomic density–density correlation function. In the hydrodynamic regime, the phonon field $\phi$ is related to the atomic density fluctuation via $\delta n\propto\dot\phi$. Consequently, the equal-time correlation function $\langle \delta n({\bm \rho},t)\delta n(0,t)\rangle$ encodes information about the phonons after the expansion. 
A straightforward calculation yields the correlation function $\langle\delta n({\bm \rho},t)\delta n(0,t)\rangle\propto\int k^{2}dkJ_{0}(k\rho)S_{k}(t)$, where $J_{0}(\cdot)$ is the 0th-order Bessel function of the first kind and $S_{k}(t)$ denotes the post-expansion density- fluctuation spectrum
$S_{k}(t)=1/2+N_{k}-{\cal A}_k\cos\left(2\omega^{\rm f}_{k}t+\theta_{\rm f}\right)$ \cite{TolosaSimeon2022}.
Here, ${\cal A}_k\equiv|\langle\Omega|\hat{b}^{\dagger}_{\bf k,{\rm f}}\hat{b}^{\dagger}_{\bf k,{\rm f}}|\Omega\rangle|=|{\alpha_k}\beta_k|$ represents the oscillation amplitude;  $\omega_{k}^{\rm f}=k/a_{\rm f}$ is the post-expansion eigenmode frequency, and $\theta_{\rm f}=\arg(\langle \hat{b}^{\dagger}_{\bf k,{\rm f}}\hat{b}^{\dagger}_{\bf k,{\rm f}}\rangle)$ is a time-independent phase.  The spectrum $S_{k}(t)$ is periodic in time, characterizing the well-known Sakharov oscillations \cite{ChenLung2020, ChenLung2021, Viermann2022}, which encodes the information regarding phonon production.

Time-averaging effectively eliminates the oscillatory cosine term in $S_{k}(t)$, allowing for the extraction of the phonon number via
\begin{align}
N_{k}=\langle S_{k}(t)\rangle_{t}-1/2,
\end{align}
where $\langle \cdot\rangle_{t}$ denotes a time average over multiple oscillation periods. Since multiple temporal cycles of $S_{k}(t)$ have already been resolved in recent experiments \cite{Oberthaler2024}, this extraction protocol is experimentally feasible. 
In Fig.~\ref{figure4}, we present representative temporal profiles of $S_{k}(t)$ generally (colored curves), alongside the time-averaged phonon number $N_{k}$ (black curves with matching colored markers).
As shown in Fig.~\ref{figure4}(a) for the sub-horizon regime, the Sakharov oscillation amplitude ${\cal A}_{k}$ varies non-monotonically with $kl$, a feature consistent with experimental observations~\cite{Oberthaler2024}. The resulting $N_{k}$ displays oscillations as a function of $kl$, in agreement with Eq.~\eqref{Nk}.
In contrast, the dynamics in the super-horizon regime, depicted in Fig.~\ref{figure4}(b), become monotonic. The amplitude ${\cal A}_{k}$ depends monotonically on $kl$, and $N_{k}$ grows exponentially with increasing $kl$. Notably, similar to $N_k$, ${\cal A}_k$ also manifests the Efimov effect, as detailed in the End Matter. 

{\it Summary-} 
We uncover the temporal Efimov effect in an analog coasting universe realized in a quasi‑two‑dimensional BEC. The post‑expansion phonon number $N_{k}$ displays two distinct behaviors: for $kl>1/2$, $N_{k}$ is log‑periodic in the expansion time, whereas for $kl<1/2$, it exhibits power-law growth. For this reason, we refer to this phenomenon as “Efimovian phonon production”. We further demonstrate that these two behaviors correspond to the sub- and super-horizon modes, respectively. Our predictions can be verified by measuring the density‑fluctuation spectrum $S_{k}(t)$ and extracting $N_{k}$ via time averaging, a procedure feasible with current experimental capabilities. Furthermore, the results naturally generalize to (n+1)‑dimensional spacetime. We expect these findings to stimulate further experimental and theoretical exploration of cosmological physics in laboratory settings.

\begin{acknowledgments}
\textit{Acknowledgments-} We are grateful to Shengli Ma, Yangqian Yan, and Zhigang Wu for stimulating discussions.
This work is supported by NSFC (Grants Nos. 12574299, 12174300, 12574296), the Innovation Program for Quantum Science and Technology (Grant No. 2021ZD0302001).
\end{acknowledgments}


\begin{thebibliography}{71}%
\makeatletter
\providecommand \@ifxundefined [1]{%
 \@ifx{#1\undefined}
}%
\providecommand \@ifnum [1]{%
 \ifnum #1\expandafter \@firstoftwo
 \else \expandafter \@secondoftwo
 \fi
}%
\providecommand \@ifx [1]{%
 \ifx #1\expandafter \@firstoftwo
 \else \expandafter \@secondoftwo
 \fi
}%
\providecommand \natexlab [1]{#1}%
\providecommand \enquote  [1]{``#1''}%
\providecommand \bibnamefont  [1]{#1}%
\providecommand \bibfnamefont [1]{#1}%
\providecommand \citenamefont [1]{#1}%
\providecommand \href@noop [0]{\@secondoftwo}%
\providecommand \href [0]{\begingroup \@sanitize@url \@href}%
\providecommand \@href[1]{\@@startlink{#1}\@@href}%
\providecommand \@@href[1]{\endgroup#1\@@endlink}%
\providecommand \@sanitize@url [0]{\catcode `\\12\catcode `\$12\catcode
  `\&12\catcode `\#12\catcode `\^12\catcode `\_12\catcode `\%12\relax}%
\providecommand \@@startlink[1]{}%
\providecommand \@@endlink[0]{}%
\providecommand \url  [0]{\begingroup\@sanitize@url \@url }%
\providecommand \@url [1]{\endgroup\@href {#1}{\urlprefix }}%
\providecommand \urlprefix  [0]{URL }%
\providecommand \Eprint [0]{\href }%
\providecommand \doibase [0]{https://doi.org/}%
\providecommand \selectlanguage [0]{\@gobble}%
\providecommand \bibinfo  [0]{\@secondoftwo}%
\providecommand \bibfield  [0]{\@secondoftwo}%
\providecommand \translation [1]{[#1]}%
\providecommand \BibitemOpen [0]{}%
\providecommand \bibitemStop [0]{}%
\providecommand \bibitemNoStop [0]{.\EOS\space}%
\providecommand \EOS [0]{\spacefactor3000\relax}%
\providecommand \BibitemShut  [1]{\csname bibitem#1\endcsname}%
\let\auto@bib@innerbib\@empty
\bibitem [{\citenamefont {Efimov}(1970)}]{Efimov1970}%
  \BibitemOpen
  \bibfield  {author} {\bibinfo {author} {\bibfnamefont {V.}~\bibnamefont
  {Efimov}},\ }\bibfield  {title} {\bibinfo {title} {Energy levels arising from
  resonant two-body forces in a three-body system},\ }\href
  {https://doi.org/https://doi.org/10.1016/0370-2693(70)90349-7} {\bibfield
  {journal} {\bibinfo  {journal} {Phys. Lett. B}\ }\textbf {\bibinfo {volume}
  {33}},\ \bibinfo {pages} {563} (\bibinfo {year} {1970})}\BibitemShut
  {NoStop}%
\bibitem [{\citenamefont {Kraemer}\ \emph {et~al.}(2006)\citenamefont
  {Kraemer}, \citenamefont {Mark}, \citenamefont {Waldburger}, \citenamefont
  {Danzl}, \citenamefont {Chin}, \citenamefont {Engeser}, \citenamefont
  {Lange}, \citenamefont {Pilch}, \citenamefont {Jaakkola}, \citenamefont
  {N{\"a}gerl},\ and\ \citenamefont {Grimm}}]{Kraemer2006}%
  \BibitemOpen
  \bibfield  {author} {\bibinfo {author} {\bibfnamefont {T.}~\bibnamefont
  {Kraemer}}, \bibinfo {author} {\bibfnamefont {M.}~\bibnamefont {Mark}},
  \bibinfo {author} {\bibfnamefont {P.}~\bibnamefont {Waldburger}}, \bibinfo
  {author} {\bibfnamefont {J.~G.}\ \bibnamefont {Danzl}}, \bibinfo {author}
  {\bibfnamefont {C.}~\bibnamefont {Chin}}, \bibinfo {author} {\bibfnamefont
  {B.}~\bibnamefont {Engeser}}, \bibinfo {author} {\bibfnamefont {A.~D.}\
  \bibnamefont {Lange}}, \bibinfo {author} {\bibfnamefont {K.}~\bibnamefont
  {Pilch}}, \bibinfo {author} {\bibfnamefont {A.}~\bibnamefont {Jaakkola}},
  \bibinfo {author} {\bibfnamefont {H.-C.}\ \bibnamefont {N{\"a}gerl}},\ and\
  \bibinfo {author} {\bibfnamefont {R.}~\bibnamefont {Grimm}},\ }\bibfield
  {title} {\bibinfo {title} {{Evidence for Efimov quantum states in an
  ultracold gas of caesium atoms}},\ }\href
  {https://doi.org/10.1038/nature04626} {\bibfield  {journal} {\bibinfo
  {journal} {Nature}\ }\textbf {\bibinfo {volume} {440}},\ \bibinfo {pages}
  {315} (\bibinfo {year} {2006})}\BibitemShut {NoStop}%
\bibitem [{\citenamefont {Braaten}\ and\ \citenamefont
  {Hammer}(2007)}]{Braaten2007}%
  \BibitemOpen
  \bibfield  {author} {\bibinfo {author} {\bibfnamefont {E.}~\bibnamefont
  {Braaten}}\ and\ \bibinfo {author} {\bibfnamefont {H.-W.}\ \bibnamefont
  {Hammer}},\ }\bibfield  {title} {\bibinfo {title} {Efimov physics in cold
  atoms},\ }\href {https://doi.org/10.1016/j.aop.2006.10.011} {\bibfield
  {journal} {\bibinfo  {journal} {Ann. Phys.}\ }\textbf {\bibinfo {volume}
  {322}},\ \bibinfo {pages} {120} (\bibinfo {year} {2007})}\BibitemShut
  {NoStop}%
\bibitem [{\citenamefont {Zaccanti}\ \emph {et~al.}(2009)\citenamefont
  {Zaccanti}, \citenamefont {Deissler}, \citenamefont {D'Errico}, \citenamefont
  {Fattori}, \citenamefont {Jona-Lasinio}, \citenamefont {M{\"u}ller},
  \citenamefont {Roati}, \citenamefont {Inguscio},\ and\ \citenamefont
  {Modugno}}]{Zaccanti2009}%
  \BibitemOpen
  \bibfield  {author} {\bibinfo {author} {\bibfnamefont {M.}~\bibnamefont
  {Zaccanti}}, \bibinfo {author} {\bibfnamefont {B.}~\bibnamefont {Deissler}},
  \bibinfo {author} {\bibfnamefont {C.}~\bibnamefont {D'Errico}}, \bibinfo
  {author} {\bibfnamefont {M.}~\bibnamefont {Fattori}}, \bibinfo {author}
  {\bibfnamefont {M.}~\bibnamefont {Jona-Lasinio}}, \bibinfo {author}
  {\bibfnamefont {S.}~\bibnamefont {M{\"u}ller}}, \bibinfo {author}
  {\bibfnamefont {G.}~\bibnamefont {Roati}}, \bibinfo {author} {\bibfnamefont
  {M.}~\bibnamefont {Inguscio}},\ and\ \bibinfo {author} {\bibfnamefont
  {G.}~\bibnamefont {Modugno}},\ }\bibfield  {title} {\bibinfo {title}
  {{Observation of an Efimov spectrum in an atomic system}},\ }\href
  {https://doi.org/10.1038/nphys1334} {\bibfield  {journal} {\bibinfo
  {journal} {Nat. Phys.}\ }\textbf {\bibinfo {volume} {5}},\ \bibinfo {pages}
  {586} (\bibinfo {year} {2009})}\BibitemShut {NoStop}%
\bibitem [{\citenamefont {Gross}\ \emph {et~al.}(2009)\citenamefont {Gross},
  \citenamefont {Shotan}, \citenamefont {Kokkelmans},\ and\ \citenamefont
  {Khaykovich}}]{Gross2009}%
  \BibitemOpen
  \bibfield  {author} {\bibinfo {author} {\bibfnamefont {N.}~\bibnamefont
  {Gross}}, \bibinfo {author} {\bibfnamefont {Z.}~\bibnamefont {Shotan}},
  \bibinfo {author} {\bibfnamefont {S.}~\bibnamefont {Kokkelmans}},\ and\
  \bibinfo {author} {\bibfnamefont {L.}~\bibnamefont {Khaykovich}},\ }\bibfield
   {title} {\bibinfo {title} {{Observation of Universality in Ultracold
  $^{7}\mathrm{Li}$ Three-Body Recombination}},\ }\href
  {https://doi.org/10.1103/PhysRevLett.103.163202} {\bibfield  {journal}
  {\bibinfo  {journal} {Phys. Rev. Lett.}\ }\textbf {\bibinfo {volume} {103}},\
  \bibinfo {pages} {163202} (\bibinfo {year} {2009})}\BibitemShut {NoStop}%
\bibitem [{\citenamefont {von Stecher}\ \emph {et~al.}(2009)\citenamefont {von
  Stecher}, \citenamefont {D'Incao},\ and\ \citenamefont
  {Greene}}]{Stecher2009}%
  \BibitemOpen
  \bibfield  {author} {\bibinfo {author} {\bibfnamefont {J.}~\bibnamefont {von
  Stecher}}, \bibinfo {author} {\bibfnamefont {J.~P.}\ \bibnamefont
  {D'Incao}},\ and\ \bibinfo {author} {\bibfnamefont {C.~H.}\ \bibnamefont
  {Greene}},\ }\bibfield  {title} {\bibinfo {title} {{Signatures of universal
  four-body phenomena and their relation to the Efimov effect}},\ }\href
  {https://doi.org/10.1038/nphys1253} {\bibfield  {journal} {\bibinfo
  {journal} {Nat. Phys.}\ }\textbf {\bibinfo {volume} {5}},\ \bibinfo {pages}
  {417} (\bibinfo {year} {2009})}\BibitemShut {NoStop}%
\bibitem [{\citenamefont {Lompe}\ \emph {et~al.}(2010)\citenamefont {Lompe},
  \citenamefont {Ottenstein}, \citenamefont {Serwane}, \citenamefont {Wenz},
  \citenamefont {Z{\"u}rn},\ and\ \citenamefont {Jochim}}]{Lompe2010}%
  \BibitemOpen
  \bibfield  {author} {\bibinfo {author} {\bibfnamefont {T.}~\bibnamefont
  {Lompe}}, \bibinfo {author} {\bibfnamefont {T.~B.}\ \bibnamefont
  {Ottenstein}}, \bibinfo {author} {\bibfnamefont {F.}~\bibnamefont {Serwane}},
  \bibinfo {author} {\bibfnamefont {A.~N.}\ \bibnamefont {Wenz}}, \bibinfo
  {author} {\bibfnamefont {G.}~\bibnamefont {Z{\"u}rn}},\ and\ \bibinfo
  {author} {\bibfnamefont {S.}~\bibnamefont {Jochim}},\ }\bibfield  {title}
  {\bibinfo {title} {{Radio-Frequency Association of Efimov Trimers}},\ }\href
  {https://doi.org/10.1126/science.1193148} {\bibfield  {journal} {\bibinfo
  {journal} {Science}\ }\textbf {\bibinfo {volume} {330}},\ \bibinfo {pages}
  {940} (\bibinfo {year} {2010})}\BibitemShut {NoStop}%
\bibitem [{\citenamefont {Maji}\ \emph {et~al.}(2010)\citenamefont {Maji},
  \citenamefont {Bhattacharjee}, \citenamefont {Seno},\ and\ \citenamefont
  {Trovato}}]{Maji2010}%
  \BibitemOpen
  \bibfield  {author} {\bibinfo {author} {\bibfnamefont {J.}~\bibnamefont
  {Maji}}, \bibinfo {author} {\bibfnamefont {S.~M.}\ \bibnamefont
  {Bhattacharjee}}, \bibinfo {author} {\bibfnamefont {F.}~\bibnamefont
  {Seno}},\ and\ \bibinfo {author} {\bibfnamefont {A.}~\bibnamefont
  {Trovato}},\ }\bibfield  {title} {\bibinfo {title} {{When a DNA triple helix
  melts: an analogue of the Efimov state}},\ }\href
  {https://doi.org/10.1088/1367-2630/12/8/083057} {\bibfield  {journal}
  {\bibinfo  {journal} {New J. Phys.}\ }\textbf {\bibinfo {volume} {12}},\
  \bibinfo {pages} {083057} (\bibinfo {year} {2010})}\BibitemShut {NoStop}%
\bibitem [{\citenamefont {Tung}\ \emph {et~al.}(2014)\citenamefont {Tung},
  \citenamefont {Jim\'enez-Garc\'{\i}a}, \citenamefont {Johansen},
  \citenamefont {Parker},\ and\ \citenamefont {Chin}}]{Tung2014}%
  \BibitemOpen
  \bibfield  {author} {\bibinfo {author} {\bibfnamefont {S.-K.}\ \bibnamefont
  {Tung}}, \bibinfo {author} {\bibfnamefont {K.}~\bibnamefont
  {Jim\'enez-Garc\'{\i}a}}, \bibinfo {author} {\bibfnamefont {J.}~\bibnamefont
  {Johansen}}, \bibinfo {author} {\bibfnamefont {C.~V.}\ \bibnamefont
  {Parker}},\ and\ \bibinfo {author} {\bibfnamefont {C.}~\bibnamefont {Chin}},\
  }\bibfield  {title} {\bibinfo {title} {{Geometric Scaling of Efimov States in
  a $^{6}\mathrm{Li}\text{\ensuremath{-}}^{133}\mathrm{Cs}$ Mixture}},\ }\href
  {https://doi.org/10.1103/PhysRevLett.113.240402} {\bibfield  {journal}
  {\bibinfo  {journal} {Phys. Rev. Lett.}\ }\textbf {\bibinfo {volume} {113}},\
  \bibinfo {pages} {240402} (\bibinfo {year} {2014})}\BibitemShut {NoStop}%
\bibitem [{\citenamefont {Pires}\ \emph {et~al.}(2014)\citenamefont {Pires},
  \citenamefont {Ulmanis}, \citenamefont {H{\"a}fner}, \citenamefont {Repp},
  \citenamefont {Arias}, \citenamefont {Kuhnle},\ and\ \citenamefont
  {Weidem{\"u}ller}}]{Pires2014}%
  \BibitemOpen
  \bibfield  {author} {\bibinfo {author} {\bibfnamefont {R.}~\bibnamefont
  {Pires}}, \bibinfo {author} {\bibfnamefont {J.}~\bibnamefont {Ulmanis}},
  \bibinfo {author} {\bibfnamefont {S.}~\bibnamefont {H{\"a}fner}}, \bibinfo
  {author} {\bibfnamefont {M.}~\bibnamefont {Repp}}, \bibinfo {author}
  {\bibfnamefont {A.}~\bibnamefont {Arias}}, \bibinfo {author} {\bibfnamefont
  {E.~D.}\ \bibnamefont {Kuhnle}},\ and\ \bibinfo {author} {\bibfnamefont
  {M.}~\bibnamefont {Weidem{\"u}ller}},\ }\bibfield  {title} {\bibinfo {title}
  {{Observation of Efimov Resonances in a Mixture with Extreme Mass
  Imbalance}},\ }\href {https://doi.org/10.1103/physrevlett.112.250404}
  {\bibfield  {journal} {\bibinfo  {journal} {Phys. Rev. Lett.}\ }\textbf
  {\bibinfo {volume} {112}},\ \bibinfo {pages} {250404} (\bibinfo {year}
  {2014})}\BibitemShut {NoStop}%
\bibitem [{\citenamefont {Shi}\ \emph {et~al.}(2015)\citenamefont {Shi},
  \citenamefont {Zhai},\ and\ \citenamefont {Cui}}]{zheyu2015}%
  \BibitemOpen
  \bibfield  {author} {\bibinfo {author} {\bibfnamefont {Z.-Y.}\ \bibnamefont
  {Shi}}, \bibinfo {author} {\bibfnamefont {H.}~\bibnamefont {Zhai}},\ and\
  \bibinfo {author} {\bibfnamefont {X.}~\bibnamefont {Cui}},\ }\bibfield
  {title} {\bibinfo {title} {Efimov physics and universal trimers in
  spin-orbit-coupled ultracold atomic mixtures},\ }\href
  {https://doi.org/10.1103/PhysRevA.91.023618} {\bibfield  {journal} {\bibinfo
  {journal} {Phys. Rev. A}\ }\textbf {\bibinfo {volume} {91}},\ \bibinfo
  {pages} {023618} (\bibinfo {year} {2015})}\BibitemShut {NoStop}%
\bibitem [{\citenamefont {Kunitski}\ \emph {et~al.}(2015)\citenamefont
  {Kunitski}, \citenamefont {Zeller}, \citenamefont {Voigtsberger},
  \citenamefont {Kalinin}, \citenamefont {Schmidt}, \citenamefont
  {Sch{\"o}ffler}, \citenamefont {Czasch}, \citenamefont {Sch{\"o}llkopf},
  \citenamefont {Grisenti}, \citenamefont {Jahnke}, \citenamefont {Blume},\
  and\ \citenamefont {D{\"o}rner}}]{Kunitski2015}%
  \BibitemOpen
  \bibfield  {author} {\bibinfo {author} {\bibfnamefont {M.}~\bibnamefont
  {Kunitski}}, \bibinfo {author} {\bibfnamefont {S.}~\bibnamefont {Zeller}},
  \bibinfo {author} {\bibfnamefont {J.}~\bibnamefont {Voigtsberger}}, \bibinfo
  {author} {\bibfnamefont {A.}~\bibnamefont {Kalinin}}, \bibinfo {author}
  {\bibfnamefont {L.~P.~H.}\ \bibnamefont {Schmidt}}, \bibinfo {author}
  {\bibfnamefont {M.}~\bibnamefont {Sch{\"o}ffler}}, \bibinfo {author}
  {\bibfnamefont {A.}~\bibnamefont {Czasch}}, \bibinfo {author} {\bibfnamefont
  {W.}~\bibnamefont {Sch{\"o}llkopf}}, \bibinfo {author} {\bibfnamefont
  {R.~E.}\ \bibnamefont {Grisenti}}, \bibinfo {author} {\bibfnamefont
  {T.}~\bibnamefont {Jahnke}}, \bibinfo {author} {\bibfnamefont
  {D.}~\bibnamefont {Blume}},\ and\ \bibinfo {author} {\bibfnamefont
  {R.}~\bibnamefont {D{\"o}rner}},\ }\bibfield  {title} {\bibinfo {title}
  {{Observation of the Efimov state of the helium trimer}},\ }\href
  {https://doi.org/10.1126/science.aaa5601} {\bibfield  {journal} {\bibinfo
  {journal} {Science}\ }\textbf {\bibinfo {volume} {348}},\ \bibinfo {pages}
  {551} (\bibinfo {year} {2015})}\BibitemShut {NoStop}%
\bibitem [{\citenamefont {Pal}\ \emph {et~al.}(2015)\citenamefont {Pal},
  \citenamefont {Sadhukhan},\ and\ \citenamefont {Bhattacharjee}}]{Pal2015}%
  \BibitemOpen
  \bibfield  {author} {\bibinfo {author} {\bibfnamefont {T.}~\bibnamefont
  {Pal}}, \bibinfo {author} {\bibfnamefont {P.}~\bibnamefont {Sadhukhan}},\
  and\ \bibinfo {author} {\bibfnamefont {S.~M.}\ \bibnamefont
  {Bhattacharjee}},\ }\bibfield  {title} {\bibinfo {title} {{Efimov-like phase
  of a three-stranded DNA and the renormalization-group limit cycle}},\ }\href
  {https://doi.org/10.1103/PhysRevE.91.042105} {\bibfield  {journal} {\bibinfo
  {journal} {Phys. Rev. E}\ }\textbf {\bibinfo {volume} {91}},\ \bibinfo
  {pages} {042105} (\bibinfo {year} {2015})}\BibitemShut {NoStop}%
\bibitem [{\citenamefont {Deng}\ \emph {et~al.}(2016)\citenamefont {Deng},
  \citenamefont {Shi}, \citenamefont {Diao}, \citenamefont {Yu}, \citenamefont
  {Zhai}, \citenamefont {Qi},\ and\ \citenamefont {Wu}}]{Deng2016}%
  \BibitemOpen
  \bibfield  {author} {\bibinfo {author} {\bibfnamefont {S.}~\bibnamefont
  {Deng}}, \bibinfo {author} {\bibfnamefont {Z.-Y.}\ \bibnamefont {Shi}},
  \bibinfo {author} {\bibfnamefont {P.}~\bibnamefont {Diao}}, \bibinfo {author}
  {\bibfnamefont {Q.}~\bibnamefont {Yu}}, \bibinfo {author} {\bibfnamefont
  {H.}~\bibnamefont {Zhai}}, \bibinfo {author} {\bibfnamefont {R.}~\bibnamefont
  {Qi}},\ and\ \bibinfo {author} {\bibfnamefont {H.}~\bibnamefont {Wu}},\
  }\bibfield  {title} {\bibinfo {title} {{Observation of the Efimovian
  expansion in scale-invariant Fermi gases}},\ }\href
  {https://doi.org/10.1126/science.aaf0666} {\bibfield  {journal} {\bibinfo
  {journal} {Science}\ }\textbf {\bibinfo {volume} {353}},\ \bibinfo {pages}
  {371} (\bibinfo {year} {2016})}\BibitemShut {NoStop}%
\bibitem [{\citenamefont {Sun}\ \emph {et~al.}(2017)\citenamefont {Sun},
  \citenamefont {Zhai},\ and\ \citenamefont {Cui}}]{Sun2017}%
  \BibitemOpen
  \bibfield  {author} {\bibinfo {author} {\bibfnamefont {M.}~\bibnamefont
  {Sun}}, \bibinfo {author} {\bibfnamefont {H.}~\bibnamefont {Zhai}},\ and\
  \bibinfo {author} {\bibfnamefont {X.}~\bibnamefont {Cui}},\ }\bibfield
  {title} {\bibinfo {title} {{Visualizing the Efimov Correlation in Bose
  Polarons}},\ }\href {https://doi.org/10.1103/physrevlett.119.013401}
  {\bibfield  {journal} {\bibinfo  {journal} {Phys. Rev. Lett.}\ }\textbf
  {\bibinfo {volume} {119}},\ \bibinfo {pages} {013401} (\bibinfo {year}
  {2017})}\BibitemShut {NoStop}%
\bibitem [{\citenamefont {Greene}\ \emph {et~al.}(2017)\citenamefont {Greene},
  \citenamefont {Giannakeas},\ and\ \citenamefont
  {P\'erez-R\'{\i}os}}]{Greene2017}%
  \BibitemOpen
  \bibfield  {author} {\bibinfo {author} {\bibfnamefont {C.~H.}\ \bibnamefont
  {Greene}}, \bibinfo {author} {\bibfnamefont {P.}~\bibnamefont {Giannakeas}},\
  and\ \bibinfo {author} {\bibfnamefont {J.}~\bibnamefont
  {P\'erez-R\'{\i}os}},\ }\bibfield  {title} {\bibinfo {title} {Universal
  few-body physics and cluster formation},\ }\href
  {https://doi.org/10.1103/RevModPhys.89.035006} {\bibfield  {journal}
  {\bibinfo  {journal} {Rev. Mod. Phys.}\ }\textbf {\bibinfo {volume} {89}},\
  \bibinfo {pages} {035006} (\bibinfo {year} {2017})}\BibitemShut {NoStop}%
\bibitem [{\citenamefont {Naidon}\ and\ \citenamefont
  {Endo}(2017)}]{Naidon2017}%
  \BibitemOpen
  \bibfield  {author} {\bibinfo {author} {\bibfnamefont {P.}~\bibnamefont
  {Naidon}}\ and\ \bibinfo {author} {\bibfnamefont {S.}~\bibnamefont {Endo}},\
  }\bibfield  {title} {\bibinfo {title} {Efimov physics: a review},\ }\href
  {https://doi.org/10.1088/1361-6633/aa50e8} {\bibfield  {journal} {\bibinfo
  {journal} {Rep. Prog. Phys.}\ }\textbf {\bibinfo {volume} {80}},\ \bibinfo
  {pages} {056001} (\bibinfo {year} {2017})}\BibitemShut {NoStop}%
\bibitem [{\citenamefont {Johansen}\ \emph {et~al.}(2017)\citenamefont
  {Johansen}, \citenamefont {DeSalvo}, \citenamefont {Patel},\ and\
  \citenamefont {Chin}}]{Johansen2017}%
  \BibitemOpen
  \bibfield  {author} {\bibinfo {author} {\bibfnamefont {J.}~\bibnamefont
  {Johansen}}, \bibinfo {author} {\bibfnamefont {B.~J.}\ \bibnamefont
  {DeSalvo}}, \bibinfo {author} {\bibfnamefont {K.}~\bibnamefont {Patel}},\
  and\ \bibinfo {author} {\bibfnamefont {C.}~\bibnamefont {Chin}},\ }\bibfield
  {title} {\bibinfo {title} {{Testing universality of Efimov physics across
  broad and narrow Feshbach resonances}},\ }\href
  {https://doi.org/10.1038/nphys4130} {\bibfield  {journal} {\bibinfo
  {journal} {Nat. Phys.}\ }\textbf {\bibinfo {volume} {13}},\ \bibinfo {pages}
  {731} (\bibinfo {year} {2017})}\BibitemShut {NoStop}%
\bibitem [{\citenamefont {Zhang}\ and\ \citenamefont {Yu}(2017)}]{pengfei2017}%
  \BibitemOpen
  \bibfield  {author} {\bibinfo {author} {\bibfnamefont {P.}~\bibnamefont
  {Zhang}}\ and\ \bibinfo {author} {\bibfnamefont {Z.}~\bibnamefont {Yu}},\
  }\bibfield  {title} {\bibinfo {title} {{Signature of the universal super
  Efimov effect: Three-body contact in two-dimensional Fermi gases}},\ }\href
  {https://doi.org/10.1103/PhysRevA.95.033611} {\bibfield  {journal} {\bibinfo
  {journal} {Phys. Rev. A}\ }\textbf {\bibinfo {volume} {95}},\ \bibinfo
  {pages} {033611} (\bibinfo {year} {2017})}\BibitemShut {NoStop}%
\bibitem [{\citenamefont {Deng}\ \emph {et~al.}(2018)\citenamefont {Deng},
  \citenamefont {Diao}, \citenamefont {Li}, \citenamefont {Yu}, \citenamefont
  {Yu},\ and\ \citenamefont {Wu}}]{Deng2018}%
  \BibitemOpen
  \bibfield  {author} {\bibinfo {author} {\bibfnamefont {S.}~\bibnamefont
  {Deng}}, \bibinfo {author} {\bibfnamefont {P.}~\bibnamefont {Diao}}, \bibinfo
  {author} {\bibfnamefont {F.}~\bibnamefont {Li}}, \bibinfo {author}
  {\bibfnamefont {Q.}~\bibnamefont {Yu}}, \bibinfo {author} {\bibfnamefont
  {S.}~\bibnamefont {Yu}},\ and\ \bibinfo {author} {\bibfnamefont
  {H.}~\bibnamefont {Wu}},\ }\bibfield  {title} {\bibinfo {title} {{Observation
  of Dynamical Super-Efimovian Expansion in a Unitary Fermi Gas}},\ }\href
  {https://doi.org/10.1103/PhysRevLett.120.125301} {\bibfield  {journal}
  {\bibinfo  {journal} {Phys. Rev. Lett.}\ }\textbf {\bibinfo {volume} {120}},\
  \bibinfo {pages} {125301} (\bibinfo {year} {2018})}\BibitemShut {NoStop}%
\bibitem [{\citenamefont {Wang}\ \emph {et~al.}(2018)\citenamefont {Wang},
  \citenamefont {Liu}, \citenamefont {Li}, \citenamefont {Liu}, \citenamefont
  {Wang}, \citenamefont {Liu}, \citenamefont {Dai}, \citenamefont {Wang},
  \citenamefont {Li}, \citenamefont {Yan}, \citenamefont {Mandrus},
  \citenamefont {Xie},\ and\ \citenamefont {Wang}}]{Wang2018}%
  \BibitemOpen
  \bibfield  {author} {\bibinfo {author} {\bibfnamefont {H.}~\bibnamefont
  {Wang}}, \bibinfo {author} {\bibfnamefont {H.}~\bibnamefont {Liu}}, \bibinfo
  {author} {\bibfnamefont {Y.}~\bibnamefont {Li}}, \bibinfo {author}
  {\bibfnamefont {Y.}~\bibnamefont {Liu}}, \bibinfo {author} {\bibfnamefont
  {J.}~\bibnamefont {Wang}}, \bibinfo {author} {\bibfnamefont {J.}~\bibnamefont
  {Liu}}, \bibinfo {author} {\bibfnamefont {J.-Y.}\ \bibnamefont {Dai}},
  \bibinfo {author} {\bibfnamefont {Y.}~\bibnamefont {Wang}}, \bibinfo {author}
  {\bibfnamefont {L.}~\bibnamefont {Li}}, \bibinfo {author} {\bibfnamefont
  {J.}~\bibnamefont {Yan}}, \bibinfo {author} {\bibfnamefont {D.}~\bibnamefont
  {Mandrus}}, \bibinfo {author} {\bibfnamefont {X.~C.}\ \bibnamefont {Xie}},\
  and\ \bibinfo {author} {\bibfnamefont {J.}~\bibnamefont {Wang}},\ }\bibfield
  {title} {\bibinfo {title} {Discovery of log-periodic oscillations in
  ultraquantum topological materials},\ }\href
  {https://doi.org/10.1126/sciadv.aau5096} {\bibfield  {journal} {\bibinfo
  {journal} {Sci. Adv.}\ }\textbf {\bibinfo {volume} {4}},\ \bibinfo {pages}
  {eaau5096} (\bibinfo {year} {2018})}\BibitemShut {NoStop}%
\bibitem [{\citenamefont {Sun}\ and\ \citenamefont {Cui}(2019)}]{xiaoling2019}%
  \BibitemOpen
  \bibfield  {author} {\bibinfo {author} {\bibfnamefont {M.}~\bibnamefont
  {Sun}}\ and\ \bibinfo {author} {\bibfnamefont {X.}~\bibnamefont {Cui}},\
  }\bibfield  {title} {\bibinfo {title} {{Efimov physics in the presence of a
  Fermi sea}},\ }\href {https://doi.org/10.1103/PhysRevA.99.060701} {\bibfield
  {journal} {\bibinfo  {journal} {Phys. Rev. A}\ }\textbf {\bibinfo {volume}
  {99}},\ \bibinfo {pages} {060701} (\bibinfo {year} {2019})}\BibitemShut
  {NoStop}%
\bibitem [{\citenamefont {Hammer}\ \emph {et~al.}(2020)\citenamefont {Hammer},
  \citenamefont {K{\"o}nig},\ and\ \citenamefont {van Kolck}}]{Hammer2020}%
  \BibitemOpen
  \bibfield  {author} {\bibinfo {author} {\bibfnamefont {H.-W.}\ \bibnamefont
  {Hammer}}, \bibinfo {author} {\bibfnamefont {S.}~\bibnamefont {K{\"o}nig}},\
  and\ \bibinfo {author} {\bibfnamefont {U.}~\bibnamefont {van Kolck}},\
  }\bibfield  {title} {\bibinfo {title} {{Nuclear effective field theory:
  Status and perspectives}},\ }\href
  {https://doi.org/10.1103/revmodphys.92.025004} {\bibfield  {journal}
  {\bibinfo  {journal} {Rev. Mod. Phys.}\ }\textbf {\bibinfo {volume} {92}},\
  \bibinfo {pages} {025004} (\bibinfo {year} {2020})}\BibitemShut {NoStop}%
\bibitem [{\citenamefont {Zhang}\ \emph {et~al.}(2021)\citenamefont {Zhang},
  \citenamefont {Lv}, \citenamefont {Yan},\ and\ \citenamefont
  {Zhou}}]{ren2021}%
  \BibitemOpen
  \bibfield  {author} {\bibinfo {author} {\bibfnamefont {R.}~\bibnamefont
  {Zhang}}, \bibinfo {author} {\bibfnamefont {C.}~\bibnamefont {Lv}}, \bibinfo
  {author} {\bibfnamefont {Y.}~\bibnamefont {Yan}},\ and\ \bibinfo {author}
  {\bibfnamefont {Q.}~\bibnamefont {Zhou}},\ }\bibfield  {title} {\bibinfo
  {title} {Efimov-like states and quantum funneling effects on synthetic
  hyperbolic surfaces},\ }\href
  {https://doi.org/https://doi.org/10.1016/j.scib.2021.06.017} {\bibfield
  {journal} {\bibinfo  {journal} {Sci. Bull.}\ }\textbf {\bibinfo {volume}
  {66}},\ \bibinfo {pages} {1967} (\bibinfo {year} {2021})}\BibitemShut
  {NoStop}%
\bibitem [{\citenamefont {Chuang}\ \emph {et~al.}(2025)\citenamefont {Chuang},
  \citenamefont {Bui}, \citenamefont {Christianen}, \citenamefont {Zhang},
  \citenamefont {Ni}, \citenamefont {Ahmed-Braun}, \citenamefont {Robens},\
  and\ \citenamefont {Zwierlein}}]{Chuang2025}%
  \BibitemOpen
  \bibfield  {author} {\bibinfo {author} {\bibfnamefont {A.~Y.}\ \bibnamefont
  {Chuang}}, \bibinfo {author} {\bibfnamefont {H.~Q.}\ \bibnamefont {Bui}},
  \bibinfo {author} {\bibfnamefont {A.}~\bibnamefont {Christianen}}, \bibinfo
  {author} {\bibfnamefont {Y.}~\bibnamefont {Zhang}}, \bibinfo {author}
  {\bibfnamefont {Y.}~\bibnamefont {Ni}}, \bibinfo {author} {\bibfnamefont
  {D.}~\bibnamefont {Ahmed-Braun}}, \bibinfo {author} {\bibfnamefont
  {C.}~\bibnamefont {Robens}},\ and\ \bibinfo {author} {\bibfnamefont
  {M.}~\bibnamefont {Zwierlein}},\ }\bibfield  {title} {\bibinfo {title}
  {{Observation of a Halo Trimer in an Ultracold Bose-Fermi Mixture}},\ }\href
  {https://doi.org/10.1103/flty-9d72} {\bibfield  {journal} {\bibinfo
  {journal} {Phys. Rev. X}\ }\textbf {\bibinfo {volume} {15}},\ \bibinfo
  {pages} {021098} (\bibinfo {year} {2025})}\BibitemShut {NoStop}%
\bibitem [{\citenamefont {Liu}\ \emph {et~al.}(2025)\citenamefont {Liu},
  \citenamefont {Jiang}, \citenamefont {Wang}, \citenamefont {Joynt},\ and\
  \citenamefont {Xie}}]{Haiwen2025}%
  \BibitemOpen
  \bibfield  {author} {\bibinfo {author} {\bibfnamefont {H.}~\bibnamefont
  {Liu}}, \bibinfo {author} {\bibfnamefont {H.}~\bibnamefont {Jiang}}, \bibinfo
  {author} {\bibfnamefont {Z.}~\bibnamefont {Wang}}, \bibinfo {author}
  {\bibfnamefont {R.}~\bibnamefont {Joynt}},\ and\ \bibinfo {author}
  {\bibfnamefont {X.~C.}\ \bibnamefont {Xie}},\ }\bibfield  {title} {\bibinfo
  {title} {Discrete scale invariance in topological semimetals},\ }\href
  {https://doi.org/10.1103/ks9f-grjd} {\bibfield  {journal} {\bibinfo
  {journal} {Phys. Rev. B}\ }\textbf {\bibinfo {volume} {112}},\ \bibinfo
  {pages} {035115} (\bibinfo {year} {2025})}\BibitemShut {NoStop}%
\bibitem [{\citenamefont {Sun}\ \emph {et~al.}()\citenamefont {Sun},
  \citenamefont {Feng},\ and\ \citenamefont {Zhang}}]{pengfei2025}%
  \BibitemOpen
  \bibfield  {author} {\bibinfo {author} {\bibfnamefont {N.}~\bibnamefont
  {Sun}}, \bibinfo {author} {\bibfnamefont {L.}~\bibnamefont {Feng}},\ and\
  \bibinfo {author} {\bibfnamefont {P.}~\bibnamefont {Zhang}},\ }\href@noop {}
  {\bibinfo {title} {{Efimov Effect in Long-range Quantum Spin Chains}}},\
  \Eprint {https://arxiv.org/abs/2502.20759} {arXiv:2502.20759} \BibitemShut
  {NoStop}%
\bibitem [{\citenamefont {Unruh}(1981)}]{Unruh1981}%
  \BibitemOpen
  \bibfield  {author} {\bibinfo {author} {\bibfnamefont {W.~G.}\ \bibnamefont
  {Unruh}},\ }\bibfield  {title} {\bibinfo {title} {Experimental black-hole
  evaporation?},\ }\href {https://doi.org/10.1103/PhysRevLett.46.1351}
  {\bibfield  {journal} {\bibinfo  {journal} {Phys. Rev. Lett.}\ }\textbf
  {\bibinfo {volume} {46}},\ \bibinfo {pages} {1351} (\bibinfo {year}
  {1981})}\BibitemShut {NoStop}%
\bibitem [{\citenamefont {Brumfiel}(2008)}]{Brumfiel2008}%
  \BibitemOpen
  \bibfield  {author} {\bibinfo {author} {\bibfnamefont {G.}~\bibnamefont
  {Brumfiel}},\ }\bibfield  {title} {\bibinfo {title} {{Experimental cosmology:
  Cosmos in a bottle}},\ }\href {https://doi.org/10.1038/451236a} {\bibfield
  {journal} {\bibinfo  {journal} {Nature}\ }\textbf {\bibinfo {volume} {451}},\
  \bibinfo {pages} {236} (\bibinfo {year} {2008})}\BibitemShut {NoStop}%
\bibitem [{\citenamefont {Chiao}(2018)}]{Chiao2018}%
  \BibitemOpen
  \bibfield  {author} {\bibinfo {author} {\bibfnamefont {M.}~\bibnamefont
  {Chiao}},\ }\bibfield  {title} {\bibinfo {title} {Tabletop cosmology},\
  }\href {https://doi.org/10.1038/s41550-018-0491-3} {\bibfield  {journal}
  {\bibinfo  {journal} {Nat. Astron.}\ }\textbf {\bibinfo {volume} {2}},\
  \bibinfo {pages} {442} (\bibinfo {year} {2018})}\BibitemShut {NoStop}%
\bibitem [{\citenamefont {Garay}\ \emph {et~al.}(2000)\citenamefont {Garay},
  \citenamefont {Anglin}, \citenamefont {Cirac},\ and\ \citenamefont
  {Zoller}}]{Garay2000}%
  \BibitemOpen
  \bibfield  {author} {\bibinfo {author} {\bibfnamefont {L.~J.}\ \bibnamefont
  {Garay}}, \bibinfo {author} {\bibfnamefont {J.~R.}\ \bibnamefont {Anglin}},
  \bibinfo {author} {\bibfnamefont {J.~I.}\ \bibnamefont {Cirac}},\ and\
  \bibinfo {author} {\bibfnamefont {P.}~\bibnamefont {Zoller}},\ }\bibfield
  {title} {\bibinfo {title} {{Sonic Analog of Gravitational Black Holes in
  Bose-Einstein Condensates}},\ }\href
  {https://doi.org/10.1103/PhysRevLett.85.4643} {\bibfield  {journal} {\bibinfo
   {journal} {Phys. Rev. Lett.}\ }\textbf {\bibinfo {volume} {85}},\ \bibinfo
  {pages} {4643} (\bibinfo {year} {2000})}\BibitemShut {NoStop}%
\bibitem [{\citenamefont {Garay}\ \emph {et~al.}(2001)\citenamefont {Garay},
  \citenamefont {Anglin}, \citenamefont {Cirac},\ and\ \citenamefont
  {Zoller}}]{Garay2001}%
  \BibitemOpen
  \bibfield  {author} {\bibinfo {author} {\bibfnamefont {L.~J.}\ \bibnamefont
  {Garay}}, \bibinfo {author} {\bibfnamefont {J.~R.}\ \bibnamefont {Anglin}},
  \bibinfo {author} {\bibfnamefont {J.~I.}\ \bibnamefont {Cirac}},\ and\
  \bibinfo {author} {\bibfnamefont {P.}~\bibnamefont {Zoller}},\ }\bibfield
  {title} {\bibinfo {title} {{Sonic black holes in dilute Bose-Einstein
  condensates}},\ }\href {https://doi.org/10.1103/PhysRevA.63.023611}
  {\bibfield  {journal} {\bibinfo  {journal} {Phys. Rev. A}\ }\textbf {\bibinfo
  {volume} {63}},\ \bibinfo {pages} {023611} (\bibinfo {year}
  {2001})}\BibitemShut {NoStop}%
\bibitem [{\citenamefont {Fedichev}\ and\ \citenamefont
  {Fischer}(2003)}]{Fischer2003}%
  \BibitemOpen
  \bibfield  {author} {\bibinfo {author} {\bibfnamefont {P.~O.}\ \bibnamefont
  {Fedichev}}\ and\ \bibinfo {author} {\bibfnamefont {U.~R.}\ \bibnamefont
  {Fischer}},\ }\bibfield  {title} {\bibinfo {title} {{Gibbons-Hawking Effect
  in the Sonic de Sitter Space-Time of an Expanding Bose-Einstein-Condensed
  Gas}},\ }\href {https://doi.org/10.1103/PhysRevLett.91.240407} {\bibfield
  {journal} {\bibinfo  {journal} {Phys. Rev. Lett.}\ }\textbf {\bibinfo
  {volume} {91}},\ \bibinfo {pages} {240407} (\bibinfo {year}
  {2003})}\BibitemShut {NoStop}%
\bibitem [{\citenamefont {Jain}\ \emph {et~al.}(2007)\citenamefont {Jain},
  \citenamefont {Weinfurtner}, \citenamefont {Visser},\ and\ \citenamefont
  {Gardiner}}]{Jain2007}%
  \BibitemOpen
  \bibfield  {author} {\bibinfo {author} {\bibfnamefont {P.}~\bibnamefont
  {Jain}}, \bibinfo {author} {\bibfnamefont {S.}~\bibnamefont {Weinfurtner}},
  \bibinfo {author} {\bibfnamefont {M.}~\bibnamefont {Visser}},\ and\ \bibinfo
  {author} {\bibfnamefont {C.~W.}\ \bibnamefont {Gardiner}},\ }\bibfield
  {title} {\bibinfo {title} {{Analog model of a Friedmann-Robertson-Walker
  universe in Bose-Einstein condensates: Application of the classical field
  method}},\ }\href {https://doi.org/10.1103/PhysRevA.76.033616} {\bibfield
  {journal} {\bibinfo  {journal} {Phys. Rev. A}\ }\textbf {\bibinfo {volume}
  {76}},\ \bibinfo {pages} {033616} (\bibinfo {year} {2007})}\BibitemShut
  {NoStop}%
\bibitem [{\citenamefont {Rousseaux}\ \emph {et~al.}(2008)\citenamefont
  {Rousseaux}, \citenamefont {Mathis}, \citenamefont {Ma{\"i}ssa},
  \citenamefont {Philbin},\ and\ \citenamefont {Leonhardt}}]{Rousseaux2008}%
  \BibitemOpen
  \bibfield  {author} {\bibinfo {author} {\bibfnamefont {G.}~\bibnamefont
  {Rousseaux}}, \bibinfo {author} {\bibfnamefont {C.}~\bibnamefont {Mathis}},
  \bibinfo {author} {\bibfnamefont {P.}~\bibnamefont {Ma{\"i}ssa}}, \bibinfo
  {author} {\bibfnamefont {T.~G.}\ \bibnamefont {Philbin}},\ and\ \bibinfo
  {author} {\bibfnamefont {U.}~\bibnamefont {Leonhardt}},\ }\bibfield  {title}
  {\bibinfo {title} {Observation of negative-frequency waves in a water tank: a
  classical analogue to the {H}awking effect?},\ }\href
  {https://doi.org/10.1088/1367-2630/10/5/053015} {\bibfield  {journal}
  {\bibinfo  {journal} {New J. Phys.}\ }\textbf {\bibinfo {volume} {10}},\
  \bibinfo {pages} {053015} (\bibinfo {year} {2008})}\BibitemShut {NoStop}%
\bibitem [{\citenamefont {Carusotto}\ \emph {et~al.}(2008)\citenamefont
  {Carusotto}, \citenamefont {Fagnocchi}, \citenamefont {Recati}, \citenamefont
  {Balbinot},\ and\ \citenamefont {Fabbri}}]{Carusotto2008}%
  \BibitemOpen
  \bibfield  {author} {\bibinfo {author} {\bibfnamefont {I.}~\bibnamefont
  {Carusotto}}, \bibinfo {author} {\bibfnamefont {S.}~\bibnamefont
  {Fagnocchi}}, \bibinfo {author} {\bibfnamefont {A.}~\bibnamefont {Recati}},
  \bibinfo {author} {\bibfnamefont {R.}~\bibnamefont {Balbinot}},\ and\
  \bibinfo {author} {\bibfnamefont {A.}~\bibnamefont {Fabbri}},\ }\bibfield
  {title} {\bibinfo {title} {{Numerical observation of Hawking radiation from
  acoustic black holes in atomic Bose-Einstein condensates}},\ }\href
  {https://doi.org/10.1088/1367-2630/10/10/103001} {\bibfield  {journal}
  {\bibinfo  {journal} {New J. Phys.}\ }\textbf {\bibinfo {volume} {10}},\
  \bibinfo {pages} {103001} (\bibinfo {year} {2008})}\BibitemShut {NoStop}%
\bibitem [{\citenamefont {Macher}\ and\ \citenamefont
  {Parentani}(2009)}]{Macher2009}%
  \BibitemOpen
  \bibfield  {author} {\bibinfo {author} {\bibfnamefont {J.}~\bibnamefont
  {Macher}}\ and\ \bibinfo {author} {\bibfnamefont {R.}~\bibnamefont
  {Parentani}},\ }\bibfield  {title} {\bibinfo {title} {{Black-hole radiation
  in Bose-Einstein condensates}},\ }\href
  {https://doi.org/10.1103/PhysRevA.80.043601} {\bibfield  {journal} {\bibinfo
  {journal} {Phys. Rev. A}\ }\textbf {\bibinfo {volume} {80}},\ \bibinfo
  {pages} {043601} (\bibinfo {year} {2009})}\BibitemShut {NoStop}%
\bibitem [{\citenamefont {Lahav}\ \emph {et~al.}(2010)\citenamefont {Lahav},
  \citenamefont {Itah}, \citenamefont {Blumkin}, \citenamefont {Gordon},
  \citenamefont {Rinott}, \citenamefont {Zayats},\ and\ \citenamefont
  {Steinhauer}}]{Lahav2010}%
  \BibitemOpen
  \bibfield  {author} {\bibinfo {author} {\bibfnamefont {O.}~\bibnamefont
  {Lahav}}, \bibinfo {author} {\bibfnamefont {A.}~\bibnamefont {Itah}},
  \bibinfo {author} {\bibfnamefont {A.}~\bibnamefont {Blumkin}}, \bibinfo
  {author} {\bibfnamefont {C.}~\bibnamefont {Gordon}}, \bibinfo {author}
  {\bibfnamefont {S.}~\bibnamefont {Rinott}}, \bibinfo {author} {\bibfnamefont
  {A.}~\bibnamefont {Zayats}},\ and\ \bibinfo {author} {\bibfnamefont
  {J.}~\bibnamefont {Steinhauer}},\ }\bibfield  {title} {\bibinfo {title}
  {{Realization of a Sonic Black Hole Analog in a Bose-Einstein Condensate}},\
  }\href {https://doi.org/10.1103/physrevlett.105.240401} {\bibfield  {journal}
  {\bibinfo  {journal} {Phys. Rev. Lett.}\ }\textbf {\bibinfo {volume} {105}},\
  \bibinfo {pages} {240401} (\bibinfo {year} {2010})}\BibitemShut {NoStop}%
\bibitem [{\citenamefont {Prain}\ \emph {et~al.}(2010)\citenamefont {Prain},
  \citenamefont {Fagnocchi},\ and\ \citenamefont {Liberati}}]{Prain2010}%
  \BibitemOpen
  \bibfield  {author} {\bibinfo {author} {\bibfnamefont {A.}~\bibnamefont
  {Prain}}, \bibinfo {author} {\bibfnamefont {S.}~\bibnamefont {Fagnocchi}},\
  and\ \bibinfo {author} {\bibfnamefont {S.}~\bibnamefont {Liberati}},\
  }\bibfield  {title} {\bibinfo {title} {{Analogue cosmological particle
  creation: Quantum correlations in expanding Bose-Einstein condensates}},\
  }\href {https://doi.org/10.1103/physrevd.82.105018} {\bibfield  {journal}
  {\bibinfo  {journal} {Phys. Rev. D}\ }\textbf {\bibinfo {volume} {82}},\
  \bibinfo {pages} {105018} (\bibinfo {year} {2010})}\BibitemShut {NoStop}%
\bibitem [{\citenamefont {Jaskula}\ \emph {et~al.}(2012)\citenamefont
  {Jaskula}, \citenamefont {Partridge}, \citenamefont {Bonneau}, \citenamefont
  {Lopes}, \citenamefont {Ruaudel}, \citenamefont {Boiron},\ and\ \citenamefont
  {Westbrook}}]{Jaskula2012}%
  \BibitemOpen
  \bibfield  {author} {\bibinfo {author} {\bibfnamefont {J.-C.}\ \bibnamefont
  {Jaskula}}, \bibinfo {author} {\bibfnamefont {G.~B.}\ \bibnamefont
  {Partridge}}, \bibinfo {author} {\bibfnamefont {M.}~\bibnamefont {Bonneau}},
  \bibinfo {author} {\bibfnamefont {R.}~\bibnamefont {Lopes}}, \bibinfo
  {author} {\bibfnamefont {J.}~\bibnamefont {Ruaudel}}, \bibinfo {author}
  {\bibfnamefont {D.}~\bibnamefont {Boiron}},\ and\ \bibinfo {author}
  {\bibfnamefont {C.~I.}\ \bibnamefont {Westbrook}},\ }\bibfield  {title}
  {\bibinfo {title} {{Acoustic Analog to the Dynamical Casimir Effect in a
  Bose-Einstein Condensate}},\ }\href
  {https://doi.org/10.1103/PhysRevLett.109.220401} {\bibfield  {journal}
  {\bibinfo  {journal} {Phys. Rev. Lett.}\ }\textbf {\bibinfo {volume} {109}},\
  \bibinfo {pages} {220401} (\bibinfo {year} {2012})}\BibitemShut {NoStop}%
\bibitem [{\citenamefont {Hung}\ \emph {et~al.}(2013)\citenamefont {Hung},
  \citenamefont {Gurarie},\ and\ \citenamefont {Chin}}]{ChenLung2013}%
  \BibitemOpen
  \bibfield  {author} {\bibinfo {author} {\bibfnamefont {C.-L.}\ \bibnamefont
  {Hung}}, \bibinfo {author} {\bibfnamefont {V.}~\bibnamefont {Gurarie}},\ and\
  \bibinfo {author} {\bibfnamefont {C.}~\bibnamefont {Chin}},\ }\bibfield
  {title} {\bibinfo {title} {{From Cosmology to Cold Atoms: Observation of
  Sakharov Oscillations in a Quenched Atomic Superfluid}},\ }\href
  {https://doi.org/10.1126/science.1237557} {\bibfield  {journal} {\bibinfo
  {journal} {Science}\ }\textbf {\bibinfo {volume} {341}},\ \bibinfo {pages}
  {1213} (\bibinfo {year} {2013})}\BibitemShut {NoStop}%
\bibitem [{\citenamefont {Steinhauer}(2014)}]{Steinhauer2014}%
  \BibitemOpen
  \bibfield  {author} {\bibinfo {author} {\bibfnamefont {J.}~\bibnamefont
  {Steinhauer}},\ }\bibfield  {title} {\bibinfo {title} {{Observation of
  self-amplifying Hawking radiation in an analogue black-hole laser}},\ }\href
  {https://doi.org/10.1038/nphys3104} {\bibfield  {journal} {\bibinfo
  {journal} {Nat. Phys.}\ }\textbf {\bibinfo {volume} {10}},\ \bibinfo {pages}
  {864} (\bibinfo {year} {2014})}\BibitemShut {NoStop}%
\bibitem [{\citenamefont {Steinhauer}(2016)}]{Steinhauer2016}%
  \BibitemOpen
  \bibfield  {author} {\bibinfo {author} {\bibfnamefont {J.}~\bibnamefont
  {Steinhauer}},\ }\bibfield  {title} {\bibinfo {title} {{Observation of
  quantum Hawking radiation and its entanglement in an analogue black hole}},\
  }\href {https://doi.org/10.1038/nphys3863} {\bibfield  {journal} {\bibinfo
  {journal} {Nat. Phys.}\ }\textbf {\bibinfo {volume} {12}},\ \bibinfo {pages}
  {959} (\bibinfo {year} {2016})}\BibitemShut {NoStop}%
\bibitem [{\citenamefont {Eckel}\ \emph {et~al.}(2018)\citenamefont {Eckel},
  \citenamefont {Kumar}, \citenamefont {Jacobson}, \citenamefont {Spielman},\
  and\ \citenamefont {Campbell}}]{Eckel2018}%
  \BibitemOpen
  \bibfield  {author} {\bibinfo {author} {\bibfnamefont {S.}~\bibnamefont
  {Eckel}}, \bibinfo {author} {\bibfnamefont {A.}~\bibnamefont {Kumar}},
  \bibinfo {author} {\bibfnamefont {T.}~\bibnamefont {Jacobson}}, \bibinfo
  {author} {\bibfnamefont {I.~B.}\ \bibnamefont {Spielman}},\ and\ \bibinfo
  {author} {\bibfnamefont {G.~K.}\ \bibnamefont {Campbell}},\ }\bibfield
  {title} {\bibinfo {title} {{A Rapidly Expanding Bose-Einstein Condensate: An
  Expanding Universe in the Lab}},\ }\href
  {https://doi.org/10.1103/PhysRevX.8.021021} {\bibfield  {journal} {\bibinfo
  {journal} {Phys. Rev. X}\ }\textbf {\bibinfo {volume} {8}},\ \bibinfo {pages}
  {021021} (\bibinfo {year} {2018})}\BibitemShut {NoStop}%
\bibitem [{\citenamefont {Barcel{\'o}}(2018)}]{Barcelo2018}%
  \BibitemOpen
  \bibfield  {author} {\bibinfo {author} {\bibfnamefont {C.}~\bibnamefont
  {Barcel{\'o}}},\ }\bibfield  {title} {\bibinfo {title} {Analogue black-hole
  horizons},\ }\href {https://doi.org/10.1038/s41567-018-0367-6} {\bibfield
  {journal} {\bibinfo  {journal} {Nat. Phys.}\ }\textbf {\bibinfo {volume}
  {15}},\ \bibinfo {pages} {210} (\bibinfo {year} {2018})}\BibitemShut
  {NoStop}%
\bibitem [{\citenamefont {Mu{\~n}oz~de Nova}\ \emph {et~al.}(2019)\citenamefont
  {Mu{\~n}oz~de Nova}, \citenamefont {Golubkov}, \citenamefont {Kolobov},\ and\
  \citenamefont {Steinhauer}}]{Steinhauer2019}%
  \BibitemOpen
  \bibfield  {author} {\bibinfo {author} {\bibfnamefont {J.~R.}\ \bibnamefont
  {Mu{\~n}oz~de Nova}}, \bibinfo {author} {\bibfnamefont {K.}~\bibnamefont
  {Golubkov}}, \bibinfo {author} {\bibfnamefont {V.~I.}\ \bibnamefont
  {Kolobov}},\ and\ \bibinfo {author} {\bibfnamefont {J.}~\bibnamefont
  {Steinhauer}},\ }\bibfield  {title} {\bibinfo {title} {{Observation of
  thermal Hawking radiation and its temperature in an analogue black hole}},\
  }\href {https://doi.org/10.1038/s41586-019-1241-0} {\bibfield  {journal}
  {\bibinfo  {journal} {Nature}\ }\textbf {\bibinfo {volume} {569}},\ \bibinfo
  {pages} {688} (\bibinfo {year} {2019})}\BibitemShut {NoStop}%
\bibitem [{\citenamefont {Gooding}\ \emph {et~al.}(2020)\citenamefont
  {Gooding}, \citenamefont {Biermann}, \citenamefont {Erne}, \citenamefont
  {Louko}, \citenamefont {Unruh}, \citenamefont {Schmiedmayer},\ and\
  \citenamefont {Weinfurtner}}]{Gooding2020}%
  \BibitemOpen
  \bibfield  {author} {\bibinfo {author} {\bibfnamefont {C.}~\bibnamefont
  {Gooding}}, \bibinfo {author} {\bibfnamefont {S.}~\bibnamefont {Biermann}},
  \bibinfo {author} {\bibfnamefont {S.}~\bibnamefont {Erne}}, \bibinfo {author}
  {\bibfnamefont {J.}~\bibnamefont {Louko}}, \bibinfo {author} {\bibfnamefont
  {W.~G.}\ \bibnamefont {Unruh}}, \bibinfo {author} {\bibfnamefont
  {J.}~\bibnamefont {Schmiedmayer}},\ and\ \bibinfo {author} {\bibfnamefont
  {S.}~\bibnamefont {Weinfurtner}},\ }\bibfield  {title} {\bibinfo {title}
  {{Interferometric Unruh Detectors for Bose-Einstein Condensates}},\ }\href
  {https://doi.org/10.1103/PhysRevLett.125.213603} {\bibfield  {journal}
  {\bibinfo  {journal} {Phys. Rev. Lett.}\ }\textbf {\bibinfo {volume} {125}},\
  \bibinfo {pages} {213603} (\bibinfo {year} {2020})}\BibitemShut {NoStop}%
\bibitem [{\citenamefont {Almheiri}\ \emph {et~al.}(2021)\citenamefont
  {Almheiri}, \citenamefont {Hartman}, \citenamefont {Maldacena}, \citenamefont
  {Shaghoulian},\ and\ \citenamefont {Tajdini}}]{Almheiri2021}%
  \BibitemOpen
  \bibfield  {author} {\bibinfo {author} {\bibfnamefont {A.}~\bibnamefont
  {Almheiri}}, \bibinfo {author} {\bibfnamefont {T.}~\bibnamefont {Hartman}},
  \bibinfo {author} {\bibfnamefont {J.}~\bibnamefont {Maldacena}}, \bibinfo
  {author} {\bibfnamefont {E.}~\bibnamefont {Shaghoulian}},\ and\ \bibinfo
  {author} {\bibfnamefont {A.}~\bibnamefont {Tajdini}},\ }\bibfield  {title}
  {\bibinfo {title} {{The entropy of Hawking radiation}},\ }\href
  {https://doi.org/10.1103/revmodphys.93.035002} {\bibfield  {journal}
  {\bibinfo  {journal} {Rev. Mod. Phys.}\ }\textbf {\bibinfo {volume} {93}},\
  \bibinfo {pages} {035002} (\bibinfo {year} {2021})}\BibitemShut {NoStop}%
\bibitem [{\citenamefont {Banik}\ \emph {et~al.}(2022)\citenamefont {Banik},
  \citenamefont {Galan}, \citenamefont {Sosa-Martinez}, \citenamefont
  {Anderson}, \citenamefont {Eckel}, \citenamefont {Spielman},\ and\
  \citenamefont {Campbell}}]{Banik2022}%
  \BibitemOpen
  \bibfield  {author} {\bibinfo {author} {\bibfnamefont {S.}~\bibnamefont
  {Banik}}, \bibinfo {author} {\bibfnamefont {M.~G.}\ \bibnamefont {Galan}},
  \bibinfo {author} {\bibfnamefont {H.}~\bibnamefont {Sosa-Martinez}}, \bibinfo
  {author} {\bibfnamefont {M.~J.}\ \bibnamefont {Anderson}}, \bibinfo {author}
  {\bibfnamefont {S.}~\bibnamefont {Eckel}}, \bibinfo {author} {\bibfnamefont
  {I.~B.}\ \bibnamefont {Spielman}},\ and\ \bibinfo {author} {\bibfnamefont
  {G.~K.}\ \bibnamefont {Campbell}},\ }\bibfield  {title} {\bibinfo {title}
  {{Accurate Determination of Hubble Attenuation and Amplification in Expanding
  and Contracting Cold-Atom Universes}},\ }\href
  {https://doi.org/10.1103/PhysRevLett.128.090401} {\bibfield  {journal}
  {\bibinfo  {journal} {Phys. Rev. Lett.}\ }\textbf {\bibinfo {volume} {128}},\
  \bibinfo {pages} {090401} (\bibinfo {year} {2022})}\BibitemShut {NoStop}%
\bibitem [{\citenamefont {Tolosa-Sime{\'o}n}\ \emph {et~al.}(2022)\citenamefont
  {Tolosa-Sime{\'o}n}, \citenamefont {Parra-L{\'o}pez}, \citenamefont
  {S{\'a}nchez-Kuntz}, \citenamefont {Haas}, \citenamefont {Viermann},
  \citenamefont {Sparn}, \citenamefont {Liebster}, \citenamefont {Hans},
  \citenamefont {Kath}, \citenamefont {Strobel}, \citenamefont {Oberthaler},\
  and\ \citenamefont {Floerchinger}}]{TolosaSimeon2022}%
  \BibitemOpen
  \bibfield  {author} {\bibinfo {author} {\bibfnamefont {M.}~\bibnamefont
  {Tolosa-Sime{\'o}n}}, \bibinfo {author} {\bibfnamefont {{\'A}.}~\bibnamefont
  {Parra-L{\'o}pez}}, \bibinfo {author} {\bibfnamefont {N.}~\bibnamefont
  {S{\'a}nchez-Kuntz}}, \bibinfo {author} {\bibfnamefont {T.}~\bibnamefont
  {Haas}}, \bibinfo {author} {\bibfnamefont {C.}~\bibnamefont {Viermann}},
  \bibinfo {author} {\bibfnamefont {M.}~\bibnamefont {Sparn}}, \bibinfo
  {author} {\bibfnamefont {N.}~\bibnamefont {Liebster}}, \bibinfo {author}
  {\bibfnamefont {M.}~\bibnamefont {Hans}}, \bibinfo {author} {\bibfnamefont
  {E.}~\bibnamefont {Kath}}, \bibinfo {author} {\bibfnamefont {H.}~\bibnamefont
  {Strobel}}, \bibinfo {author} {\bibfnamefont {M.~K.}\ \bibnamefont
  {Oberthaler}},\ and\ \bibinfo {author} {\bibfnamefont {S.}~\bibnamefont
  {Floerchinger}},\ }\bibfield  {title} {\bibinfo {title} {{Curved and
  expanding spacetime geometries in Bose-Einstein condensates}},\ }\href
  {https://doi.org/10.1103/physreva.106.033313} {\bibfield  {journal} {\bibinfo
   {journal} {Phys. Rev. A}\ }\textbf {\bibinfo {volume} {106}},\ \bibinfo
  {pages} {033313} (\bibinfo {year} {2022})}\BibitemShut {NoStop}%
\bibitem [{\citenamefont {Viermann}\ \emph {et~al.}(2022)\citenamefont
  {Viermann}, \citenamefont {Sparn}, \citenamefont {Liebster}, \citenamefont
  {Hans}, \citenamefont {Kath}, \citenamefont {Parra-L{\'o}pez}, \citenamefont
  {Tolosa-Sime{\'o}n}, \citenamefont {S{\'a}nchez-Kuntz}, \citenamefont {Haas},
  \citenamefont {Strobel}, \citenamefont {Floerchinger},\ and\ \citenamefont
  {Oberthaler}}]{Viermann2022}%
  \BibitemOpen
  \bibfield  {author} {\bibinfo {author} {\bibfnamefont {C.}~\bibnamefont
  {Viermann}}, \bibinfo {author} {\bibfnamefont {M.}~\bibnamefont {Sparn}},
  \bibinfo {author} {\bibfnamefont {N.}~\bibnamefont {Liebster}}, \bibinfo
  {author} {\bibfnamefont {M.}~\bibnamefont {Hans}}, \bibinfo {author}
  {\bibfnamefont {E.}~\bibnamefont {Kath}}, \bibinfo {author} {\bibfnamefont
  {{\'A}.}~\bibnamefont {Parra-L{\'o}pez}}, \bibinfo {author} {\bibfnamefont
  {M.}~\bibnamefont {Tolosa-Sime{\'o}n}}, \bibinfo {author} {\bibfnamefont
  {N.}~\bibnamefont {S{\'a}nchez-Kuntz}}, \bibinfo {author} {\bibfnamefont
  {T.}~\bibnamefont {Haas}}, \bibinfo {author} {\bibfnamefont {H.}~\bibnamefont
  {Strobel}}, \bibinfo {author} {\bibfnamefont {S.}~\bibnamefont
  {Floerchinger}},\ and\ \bibinfo {author} {\bibfnamefont {M.~K.}\ \bibnamefont
  {Oberthaler}},\ }\bibfield  {title} {\bibinfo {title} {Quantum field
  simulator for dynamics in curved spacetime},\ }\href
  {https://doi.org/10.1038/s41586-022-05313-9} {\bibfield  {journal} {\bibinfo
  {journal} {Nature}\ }\textbf {\bibinfo {volume} {611}},\ \bibinfo {pages}
  {260} (\bibinfo {year} {2022})}\BibitemShut {NoStop}%
\bibitem [{\citenamefont {Sparn}\ \emph {et~al.}(2024)\citenamefont {Sparn},
  \citenamefont {Kath}, \citenamefont {Liebster}, \citenamefont {Duchene},
  \citenamefont {Schmidt}, \citenamefont {Tolosa-Sime{\'o}n}, \citenamefont
  {Parra-L{\'o}pez}, \citenamefont {Floerchinger}, \citenamefont {Strobel},\
  and\ \citenamefont {Oberthaler}}]{Oberthaler2024}%
  \BibitemOpen
  \bibfield  {author} {\bibinfo {author} {\bibfnamefont {M.}~\bibnamefont
  {Sparn}}, \bibinfo {author} {\bibfnamefont {E.}~\bibnamefont {Kath}},
  \bibinfo {author} {\bibfnamefont {N.}~\bibnamefont {Liebster}}, \bibinfo
  {author} {\bibfnamefont {J.}~\bibnamefont {Duchene}}, \bibinfo {author}
  {\bibfnamefont {C.~F.}\ \bibnamefont {Schmidt}}, \bibinfo {author}
  {\bibfnamefont {M.}~\bibnamefont {Tolosa-Sime{\'o}n}}, \bibinfo {author}
  {\bibfnamefont {{\'A}.}~\bibnamefont {Parra-L{\'o}pez}}, \bibinfo {author}
  {\bibfnamefont {S.}~\bibnamefont {Floerchinger}}, \bibinfo {author}
  {\bibfnamefont {H.}~\bibnamefont {Strobel}},\ and\ \bibinfo {author}
  {\bibfnamefont {M.~K.}\ \bibnamefont {Oberthaler}},\ }\bibfield  {title}
  {\bibinfo {title} {{Experimental Particle Production in Time-Dependent
  Spacetimes: A One-Dimensional Scattering Problem}},\ }\href
  {https://doi.org/10.1103/PhysRevLett.133.260201} {\bibfield  {journal}
  {\bibinfo  {journal} {Phys. Rev. Lett.}\ }\textbf {\bibinfo {volume} {133}},\
  \bibinfo {pages} {260201} (\bibinfo {year} {2024})}\BibitemShut {NoStop}%
\bibitem [{\citenamefont {Gondret}\ \emph {et~al.}(2025)\citenamefont
  {Gondret}, \citenamefont {Lamirault}, \citenamefont {Dias}, \citenamefont
  {Camier}, \citenamefont {Micheli}, \citenamefont {Leprince}, \citenamefont
  {Marolleau}, \citenamefont {Rullier}, \citenamefont {Robertson},
  \citenamefont {Boiron},\ and\ \citenamefont {Westbrook}}]{Gondret2025}%
  \BibitemOpen
  \bibfield  {author} {\bibinfo {author} {\bibfnamefont {V.}~\bibnamefont
  {Gondret}}, \bibinfo {author} {\bibfnamefont {C.}~\bibnamefont {Lamirault}},
  \bibinfo {author} {\bibfnamefont {R.}~\bibnamefont {Dias}}, \bibinfo {author}
  {\bibfnamefont {L.}~\bibnamefont {Camier}}, \bibinfo {author} {\bibfnamefont
  {A.}~\bibnamefont {Micheli}}, \bibinfo {author} {\bibfnamefont
  {C.}~\bibnamefont {Leprince}}, \bibinfo {author} {\bibfnamefont
  {Q.}~\bibnamefont {Marolleau}}, \bibinfo {author} {\bibfnamefont {J.-R.}\
  \bibnamefont {Rullier}}, \bibinfo {author} {\bibfnamefont {S.}~\bibnamefont
  {Robertson}}, \bibinfo {author} {\bibfnamefont {D.}~\bibnamefont {Boiron}},\
  and\ \bibinfo {author} {\bibfnamefont {C.~I.}\ \bibnamefont {Westbrook}},\
  }\bibfield  {title} {\bibinfo {title} {{Observation of Entanglement in a Cold
  Atom Analog of Cosmological Preheating}},\ }\href
  {https://doi.org/10.1103/h7ws-g9z2} {\bibfield  {journal} {\bibinfo
  {journal} {Phys. Rev. Lett.}\ }\textbf {\bibinfo {volume} {135}},\ \bibinfo
  {pages} {240603} (\bibinfo {year} {2025})}\BibitemShut {NoStop}%
\bibitem [{SM()}]{SM}%
  \BibitemOpen
  \href@noop {} {}\bibinfo {note} {{See the Supplemental Material for detailed
  derivations on: (i) the metric of an analogue FLRW universe in
  quasi-$2+1$-dimensional BEC; (ii) the momentum-resolved particle production
  $N_k$; and (iii) the momentum-integrated particle density after expansion,
  together with their asymptotic expressions.}}\BibitemShut {Stop}%
\bibitem [{\citenamefont {Chin}\ \emph {et~al.}(2010)\citenamefont {Chin},
  \citenamefont {Grimm}, \citenamefont {Julienne},\ and\ \citenamefont
  {Tiesinga}}]{Chin2010}%
  \BibitemOpen
  \bibfield  {author} {\bibinfo {author} {\bibfnamefont {C.}~\bibnamefont
  {Chin}}, \bibinfo {author} {\bibfnamefont {R.}~\bibnamefont {Grimm}},
  \bibinfo {author} {\bibfnamefont {P.}~\bibnamefont {Julienne}},\ and\
  \bibinfo {author} {\bibfnamefont {E.}~\bibnamefont {Tiesinga}},\ }\bibfield
  {title} {\bibinfo {title} {Feshbach resonances in ultracold gases},\ }\href
  {https://doi.org/10.1103/revmodphys.82.1225} {\bibfield  {journal} {\bibinfo
  {journal} {Rev. Mod. Phys.}\ }\textbf {\bibinfo {volume} {82}},\ \bibinfo
  {pages} {1225} (\bibinfo {year} {2010})}\BibitemShut {NoStop}%
\bibitem [{\citenamefont {Milne}(1934)}]{Milne1934}%
  \BibitemOpen
  \bibfield  {author} {\bibinfo {author} {\bibfnamefont {E.~A.}\ \bibnamefont
  {Milne}},\ }\bibfield  {title} {\bibinfo {title} {{A Newtonian expanding
  universe}},\ }\href {https://doi.org/10.1093/qmath/os-5.1.64} {\bibfield
  {journal} {\bibinfo  {journal} {Q. J. Math.}\ }\textbf {\bibinfo {volume}
  {Oxford 5}},\ \bibinfo {pages} {64} (\bibinfo {year} {1934})}\BibitemShut
  {NoStop}%
\bibitem [{\citenamefont {Kolb}(1989)}]{Kolb1989}%
  \BibitemOpen
  \bibfield  {author} {\bibinfo {author} {\bibfnamefont {E.~W.}\ \bibnamefont
  {Kolb}},\ }\bibfield  {title} {\bibinfo {title} {A coasting cosmology},\
  }\href {https://doi.org/10.1086/167825} {\bibfield  {journal} {\bibinfo
  {journal} {Astrophys. J.}\ }\textbf {\bibinfo {volume} {344}},\ \bibinfo
  {pages} {543} (\bibinfo {year} {1989})}\BibitemShut {NoStop}%
\bibitem [{\citenamefont {Padmanabhan}(1990)}]{Padmanabhan1990}%
  \BibitemOpen
  \bibfield  {author} {\bibinfo {author} {\bibfnamefont {T.}~\bibnamefont
  {Padmanabhan}},\ }\bibfield  {title} {\bibinfo {title} {Physical
  interpretation of quantum field theory in noninertial coordinate systems},\
  }\href {https://doi.org/10.1103/physrevlett.64.2471} {\bibfield  {journal}
  {\bibinfo  {journal} {Phys. Rev. Lett.}\ }\textbf {\bibinfo {volume} {64}},\
  \bibinfo {pages} {2471} (\bibinfo {year} {1990})}\BibitemShut {NoStop}%
\bibitem [{\citenamefont {Dev}\ \emph {et~al.}(2002)\citenamefont {Dev},
  \citenamefont {Safonova}, \citenamefont {Jain},\ and\ \citenamefont
  {Lohiya}}]{Abha2002}%
  \BibitemOpen
  \bibfield  {author} {\bibinfo {author} {\bibfnamefont {A.}~\bibnamefont
  {Dev}}, \bibinfo {author} {\bibfnamefont {M.}~\bibnamefont {Safonova}},
  \bibinfo {author} {\bibfnamefont {D.}~\bibnamefont {Jain}},\ and\ \bibinfo
  {author} {\bibfnamefont {D.}~\bibnamefont {Lohiya}},\ }\bibfield  {title}
  {\bibinfo {title} {Cosmological tests for a linear coasting cosmology},\
  }\href {https://doi.org/https://doi.org/10.1016/S0370-2693(02)02814-9}
  {\bibfield  {journal} {\bibinfo  {journal} {Phys. Lett. B}\ }\textbf
  {\bibinfo {volume} {548}},\ \bibinfo {pages} {12} (\bibinfo {year}
  {2002})}\BibitemShut {NoStop}%
\bibitem [{\citenamefont {Melia}(2018)}]{Melia2018}%
  \BibitemOpen
  \bibfield  {author} {\bibinfo {author} {\bibfnamefont {F.}~\bibnamefont
  {Melia}},\ }\bibfield  {title} {\bibinfo {title} {A comparison of the ${R}_h
  = ct$ and {$\Lambda$CDM} cosmologies using the cosmic distance duality
  relation},\ }\href {https://doi.org/10.1093/mnras/sty2596} {\bibfield
  {journal} {\bibinfo  {journal} {Mon. Not. R. Astron Soc.}\ }\textbf {\bibinfo
  {volume} {481}},\ \bibinfo {pages} {4855} (\bibinfo {year}
  {2018})}\BibitemShut {NoStop}%
\bibitem [{\citenamefont {Wilkinson}\ and\ \citenamefont
  {Louko}(2025)}]{Wilkinson2025}%
  \BibitemOpen
  \bibfield  {author} {\bibinfo {author} {\bibfnamefont {A.~S.}\ \bibnamefont
  {Wilkinson}}\ and\ \bibinfo {author} {\bibfnamefont {J.}~\bibnamefont
  {Louko}},\ }\bibfield  {title} {\bibinfo {title} {{Local quantum detection of
  the cosmological expansion: Unruh-DeWitt detectors in spatially compact Milne
  cosmology}},\ }\href {https://doi.org/10.1103/physrevd.111.025008} {\bibfield
   {journal} {\bibinfo  {journal} {Phys. Rev. D}\ }\textbf {\bibinfo {volume}
  {111}},\ \bibinfo {pages} {025008} (\bibinfo {year} {2025})}\BibitemShut
  {NoStop}%
\bibitem [{\citenamefont {Fulling}\ \emph {et~al.}(1974)\citenamefont
  {Fulling}, \citenamefont {Parker},\ and\ \citenamefont {Hu}}]{Fulling1974}%
  \BibitemOpen
  \bibfield  {author} {\bibinfo {author} {\bibfnamefont {S.~A.}\ \bibnamefont
  {Fulling}}, \bibinfo {author} {\bibfnamefont {L.}~\bibnamefont {Parker}},\
  and\ \bibinfo {author} {\bibfnamefont {B.~L.}\ \bibnamefont {Hu}},\
  }\bibfield  {title} {\bibinfo {title} {Conformal energy-momentum tensor in
  curved spacetime: {A}diabatic regularization and renormalization},\ }\href
  {https://doi.org/10.1103/physrevd.10.3905} {\bibfield  {journal} {\bibinfo
  {journal} {Phys. Rev. D}\ }\textbf {\bibinfo {volume} {10}},\ \bibinfo
  {pages} {3905} (\bibinfo {year} {1974})}\BibitemShut {NoStop}%
\bibitem [{\citenamefont {Chitre}\ and\ \citenamefont
  {Hartle}(1977)}]{Chitre1977}%
  \BibitemOpen
  \bibfield  {author} {\bibinfo {author} {\bibfnamefont {D.~M.}\ \bibnamefont
  {Chitre}}\ and\ \bibinfo {author} {\bibfnamefont {J.~B.}\ \bibnamefont
  {Hartle}},\ }\bibfield  {title} {\bibinfo {title} {Path-integral quantization
  and cosmological particle production: {A}n example},\ }\href
  {https://doi.org/10.1103/physrevd.16.251} {\bibfield  {journal} {\bibinfo
  {journal} {Phys. Rev. D}\ }\textbf {\bibinfo {volume} {16}},\ \bibinfo
  {pages} {251} (\bibinfo {year} {1977})}\BibitemShut {NoStop}%
\bibitem [{\citenamefont {S.}\ and\ \citenamefont {W.}(1978)}]{BunchT.1978}%
  \BibitemOpen
  \bibfield  {author} {\bibinfo {author} {\bibfnamefont {B.~T.}\ \bibnamefont
  {S.}}\ and\ \bibinfo {author} {\bibfnamefont {D.~P.~C.}\ \bibnamefont {W.}},\
  }\bibfield  {title} {\bibinfo {title} {Quantum field theory in de {S}itter
  space: renormalization by point-splitting},\ }\href
  {https://doi.org/10.1098/rspa.1978.0060} {\bibfield  {journal} {\bibinfo
  {journal} {Proc. R. Soc. A}\ }\textbf {\bibinfo {volume} {360}},\ \bibinfo
  {pages} {117} (\bibinfo {year} {1978})}\BibitemShut {NoStop}%
\bibitem [{\citenamefont {Parker}\ and\ \citenamefont
  {Toms}(2009)}]{Parker2009}%
  \BibitemOpen
  \bibfield  {author} {\bibinfo {author} {\bibfnamefont {L.}~\bibnamefont
  {Parker}}\ and\ \bibinfo {author} {\bibfnamefont {D.}~\bibnamefont {Toms}},\
  }\href@noop {} {\emph {\bibinfo {title} {Quantum Field Theory in Curved
  Spacetime: Quantized Fields and Gravity}}},\ Cambridge Monographs on
  Mathematical Physics\ (\bibinfo  {publisher} {Cambridge University Press},\
  \bibinfo {year} {2009})\BibitemShut {NoStop}%
\bibitem [{\citenamefont {de~Alfaro}\ \emph {et~al.}(1976)\citenamefont
  {de~Alfaro}, \citenamefont {Fubini},\ and\ \citenamefont
  {Furlan}}]{deAlfaro1976}%
  \BibitemOpen
  \bibfield  {author} {\bibinfo {author} {\bibfnamefont {V.}~\bibnamefont
  {de~Alfaro}}, \bibinfo {author} {\bibfnamefont {S.}~\bibnamefont {Fubini}},\
  and\ \bibinfo {author} {\bibfnamefont {G.}~\bibnamefont {Furlan}},\
  }\bibfield  {title} {\bibinfo {title} {Conformal invariance in quantum
  mechanics},\ }\href {https://doi.org/10.1007/bf02785666} {\bibfield
  {journal} {\bibinfo  {journal} {Il Nuovo Cimento A}\ }\textbf {\bibinfo
  {volume} {34}},\ \bibinfo {pages} {569–612} (\bibinfo {year}
  {1976})}\BibitemShut {NoStop}%
\bibitem [{\citenamefont {Bucher}(2015)}]{cos}%
  \BibitemOpen
  \bibfield  {author} {\bibinfo {author} {\bibfnamefont {M.}~\bibnamefont
  {Bucher}},\ }\bibfield  {title} {\bibinfo {title} {Physics of the cosmic
  microwave background anisotropy},\ }\href
  {https://doi.org/10.1142/s0218271815300049} {\bibfield  {journal} {\bibinfo
  {journal} {Int. J. Mod. Phys. D}\ }\textbf {\bibinfo {volume} {24}},\
  \bibinfo {pages} {1530004} (\bibinfo {year} {2015})}\BibitemShut {NoStop}%
\bibitem [{\citenamefont {Chen}\ and\ \citenamefont
  {Hung}(2020)}]{ChenLung2020}%
  \BibitemOpen
  \bibfield  {author} {\bibinfo {author} {\bibfnamefont {C.-A.}\ \bibnamefont
  {Chen}}\ and\ \bibinfo {author} {\bibfnamefont {C.-L.}\ \bibnamefont
  {Hung}},\ }\bibfield  {title} {\bibinfo {title} {Observation of universal
  quench dynamics and townes soliton formation from modulational instability in
  two-dimensional bose gases},\ }\href
  {https://doi.org/10.1103/PhysRevLett.125.250401} {\bibfield  {journal}
  {\bibinfo  {journal} {Phys. Rev. Lett.}\ }\textbf {\bibinfo {volume} {125}},\
  \bibinfo {pages} {250401} (\bibinfo {year} {2020})}\BibitemShut {NoStop}%
\bibitem [{\citenamefont {Chen}\ \emph {et~al.}(2021)\citenamefont {Chen},
  \citenamefont {Khlebnikov},\ and\ \citenamefont {Hung}}]{ChenLung2021}%
  \BibitemOpen
  \bibfield  {author} {\bibinfo {author} {\bibfnamefont {C.-A.}\ \bibnamefont
  {Chen}}, \bibinfo {author} {\bibfnamefont {S.}~\bibnamefont {Khlebnikov}},\
  and\ \bibinfo {author} {\bibfnamefont {C.-L.}\ \bibnamefont {Hung}},\
  }\bibfield  {title} {\bibinfo {title} {Observation of quasiparticle pair
  production and quantum entanglement in atomic quantum gases quenched to an
  attractive interaction},\ }\href
  {https://doi.org/10.1103/PhysRevLett.127.060404} {\bibfield  {journal}
  {\bibinfo  {journal} {Phys. Rev. Lett.}\ }\textbf {\bibinfo {volume} {127}},\
  \bibinfo {pages} {060404} (\bibinfo {year} {2021})}\BibitemShut {NoStop}%
\bibitem [{\citenamefont {Davies}\ and\ \citenamefont
  {Fulling}(1977)}]{DaviesP.C.1977}%
  \BibitemOpen
  \bibfield  {author} {\bibinfo {author} {\bibfnamefont {P.~C.~W.}\
  \bibnamefont {Davies}}\ and\ \bibinfo {author} {\bibfnamefont {S.~A.}\
  \bibnamefont {Fulling}},\ }\bibfield  {title} {\bibinfo {title} {Quantum
  vacuum energy in two dimensional space-times},\ }\href
  {https://doi.org/10.1098/rspa.1977.0056} {\bibfield  {journal} {\bibinfo
  {journal} {Proc. R. Soc. A}\ }\textbf {\bibinfo {volume} {354}},\ \bibinfo
  {pages} {59} (\bibinfo {year} {1977})}\BibitemShut {NoStop}%
\bibitem [{\citenamefont {Bunch}\ \emph {et~al.}(1978)\citenamefont {Bunch},
  \citenamefont {Christensen},\ and\ \citenamefont {Fulling}}]{Bunch1978}%
  \BibitemOpen
  \bibfield  {author} {\bibinfo {author} {\bibfnamefont {T.~S.}\ \bibnamefont
  {Bunch}}, \bibinfo {author} {\bibfnamefont {S.~M.}\ \bibnamefont
  {Christensen}},\ and\ \bibinfo {author} {\bibfnamefont {S.~A.}\ \bibnamefont
  {Fulling}},\ }\bibfield  {title} {\bibinfo {title} {{Massive quantum field
  theory in two-dimensional Robertson-Walker space-time}},\ }\href
  {https://doi.org/10.1103/physrevd.18.4435} {\bibfield  {journal} {\bibinfo
  {journal} {Phys. Rev. D}\ }\textbf {\bibinfo {volume} {18}},\ \bibinfo
  {pages} {4435} (\bibinfo {year} {1978})}\BibitemShut {NoStop}%
\end{thebibliography}
%

\appendix
\renewcommand{\theequation}{A\arabic{equation}}
\setcounter{equation}{0} 

\onecolumngrid 
\section{\large End Matter}
\twocolumngrid

\subsection{Momentum-resolved particle production for \texorpdfstring{$(n+1)$}{(n+1)} dimensions}

The universality of the Efimov effect extends across spacetime dimensions in a linearly expanding FLRW universe. We consider a generic $(n+1)$-dimensional FLRW spacetime with the metric
\begin{equation}
    ds^2=-dt^2+a^2(t)\sum_{i=1}^{n}d{x_i}^2,
\end{equation}
where the scale factor $a(t)$ follows the linear expansion described in the main text (see Eq. (\ref{scalingfactor})), and $n$ is an arbitrary positive number. The mode function $u_{k}(t)$ of a scalar field $\hat{\phi}({\bf x},t)$ satisfies the equation
\begin{equation}
    \ddot{u}_{k}(t)+\frac{n}{t}\dot{u}_{k}(t)+\frac{(kl)^2}{t^2}u_{k}(t)=0,
\end{equation}
which is invariant under the time rescaling $t\to\lambda t$, hinting at the emergence of Efimov physics.

A similar calculation yields the momentum‑resolved particle production
\begin{align}
    \label{Nkn}
    N_k=
    \begin{cases}
        \displaystyle\frac{\sin^2\left(\sqrt{(kl)^2-\mathcal{C}^2}\ln(t_{\rm f}/t_{\rm i})\right)}{(kl)^2/\mathcal{C}^2-1}, & kl>\mathcal{C} \\
        \displaystyle\frac{\sinh^2\left(\sqrt{\mathcal{C}^2-(kl)^2}\ln(t_{\rm f}/t_{\rm i})\right)}{1-(kl)^2/\mathcal{C}^2}, & kl<\mathcal{C}
    \end{cases}
\end{align}
where $\mathcal{C}=(n-1)/2$. Obviously, $N_{k}$ exhibits distinct behaviors across the horizon boundary at $kl=\mathcal{C}$. For super-horizon modes ($kl>{\cal C}$), $N_{k}$ is a log-periodic function of $\ln(t_{\rm f}/t_{\rm i})$, manifesting the hallmark of the Efimov effect. Conversely, for sub-horizon modes ($kl<{\cal C}$), $N_{k}$ grows exponentially with $\ln(t_{\rm f}/t_{\rm i})$. These distinct dynamical behaviors are consistent with the SU(1,1) symmetry analysis \cite{SM}.
Specific instances align with established results:
\begin{itemize}
    \item For $n=1$, $N_{k}$ vanishes, confirming the absence of particle production in a $(1+1)$-dimensional FLRW universe. This result aligns with the conformal symmetry analysis established in the literature \cite{Fulling1974, Chitre1977, DaviesP.C.1977, Bunch1978, BunchT.1978}.
    \item For $n=2$, we have $\mathcal{C}=1/2$, and Eq. (\ref{Nkn}) reduces to Eq. (\ref{Nk}) in the main text.
\end{itemize}
These distinct regimes of particle production persist for all higher dimensions with $n \geq 2$.

\subsection{Momentum-integrated particle density of sub- and super-horizon modes}

In the $(2+1)$-dimensional case, the long-time asymptotic behavior of the momentum-integrated particle density is illustrated in Fig.~\ref{figure3}. The corresponding analytical expressions are provided in Eq. (\ref{particledensity}), utilizing the cosine integral function ${\rm Ci}(\cdot)$ and the hyperbolic cosine integral function ${\rm Chi}(\cdot)$, defined as:
\begin{align}
    \mathrm{Ci}(x)&\equiv\gamma+\ln(x)+\int_0^x\frac{\cos(t)-1}{t}dt, \\
    \mathrm{Chi}(x)&\equiv\gamma+\ln(x)+\int_0^x\frac{\cosh(t)-1}{t}dt,
\end{align}
where $\gamma\approx 0.577216$ is the Euler constant. In the large-$x$ limit, their asymptotic expansions are
\begin{align}
    \mathrm{Ci}(x)=&\frac{\sin(x)}{x}+\left[\frac{\sin(x)}{x}+\cos(x)\right]{\cal O}(x^{-2}), \label{asympci}\\
    \mathrm{Chi}(x)=&\frac{\sinh(x)}{x}+\left[\frac{\sinh(x)}{x}+\cosh(x)\right]{\cal O}(x^{-2})-\frac{i\pi}{2}. \label{asympchi}
\end{align}
Consequently, for large positive $x$, $\mathrm{Chi}(x)\approx e^{x}/(2x)$, while $\mathrm{Ci}(x)$ exhibits oscillatory decay scaling as $\sim\sin x/x$. These mathematical properties underlie the contrasting asymptotics between super- and sub-horizon particle densities discussed in the main text.

Combining Eqs. (\ref{asympci}), (\ref{asympchi}) with Eq. (\ref{particledensity}), we derive the asymptotic behaviors of $n_{\rm super}$ and $n_{\rm sub}$ in the limits of large $\ln(t_{\rm f}/t_{\rm i})$ and large $\Lambda l$:
\begin{align}
    n_{\rm super}\Big|_{\ln(t_{\rm f}/t_{\rm i})\to\infty} &\approx\frac{\pi }{8l^2}\frac{ t_{\rm f}/t_{\rm i}}{\ln(t_{\rm f}/t_{\rm i})}, \\
    n_{\rm sub}\Big|_{\ln(t_{\rm f}/t_{\rm i}),\Lambda l\to\infty} &\approx\frac{\pi}{4l^{2}}\ln\left(2\Lambda l\ln(t_{\rm f}/t_{\rm i})\right).
\end{align}
Figure \ref{figure3} compares these asymptotic approximations with the exact numerical results, demonstrating excellent agreement.

Generalizing to $(n+1)$-dimensional spacetime, the momentum-integrated particle densities are defined as
\begin{align}
\label{n1density}
    n_{\rm super}=&\int_{0}^{\mathcal{C}/l}k^{n-1}S_{n}N_{k}dk, \\
    \label{n2density}
    n_{\rm sub}=&\int_{\mathcal{C}/l}^{\Lambda }k^{n-1}S_{n}N_{k}dk,
\end{align}
where $S_{n}=n\pi^{n/2}/(n/2)!$ denotes the surface area of the unit $n$-sphere. The exact analytical expression for $n_{\rm super}$ is given by
\begin{equation}
    n_{\rm super}=\frac{S_{n}\mathcal{C}^n}{l^n}\frac{(\mathcal{C}r)^2}{n}\;_2F_3\left(1,1;\frac{3}{2},2,\frac{n}{2}+1;(\mathcal{C}r)^2\right),
\end{equation}
where $_2F_3$ is the generalized hypergeometric function and $r\equiv\ln (t_{\rm f}/t_{\rm i})$.

Although a closed-form expression for $n_{\rm sub}$ is unavailable, we can derive an approximate one. By introducing the variable $t = \sqrt{(kl)^2 - \mathcal{C}^2}$, the integral is approximated as
\begin{equation}
\begin{aligned}
    n_{\rm sub}&\approx\frac{S_{n}\mathcal{C}^n}{l^n}\left[
    \int_{0}^{\mathcal{C}}\frac{\sin^2(tr)}{t}dt+ \int_{\mathcal{C}}^{s}\frac{\sin^2(tr)}{t}\left(\frac{t}{\mathcal{C}}\right)^{n-2}dt\right],
\end{aligned}
\end{equation}
where $s=\sqrt{(\Lambda l)^2-\mathcal{C}^2}$ and we assume $n>2$. Both integrals can be evaluated analytically. While the second term converges rapidly, the first term corresponds to near-horizon modes and yields a double-logarithmic contribution to the density.
We analytically derive and numerically verify the following asymptotic behaviors \cite{SM} in the limits of large $\ln(t_{\rm f}/t_{\rm i})$ and large $\Lambda l$:
\begin{align}
    n_{\rm super}\Big|_{\ln(t_{\rm f}/t_{\rm i})\to\infty}&\approx\frac{{\left(\pi\mathcal{C}\right)}^{n/2}}{4l^n}\frac{(t_{\rm f}/t_{\rm i})^{n-1}}{\ln^{n/2}(t_{\rm f}/t_{\rm i})}, \\
    n_{\rm sub}\Big|_{\ln(t_{\rm f}/t_{\rm i}),\Lambda l\to\infty}&\approx \frac{S_{n}\mathcal{C}^n}{2l^n}\left[\ln\!\left[\ln(t_{\rm f}/t_{\rm i})\right]
    +{\cal W}\right],
\end{align}
with ${\cal W}\equiv\left(\Lambda l/{\mathcal{C}}\right)^{n-2}/(n-2)$. Consequently, we observe that the $\ln(t_{\rm f}/t_{\rm i})$ dependence of $n_{\rm sub}$ is universal, whereas $n_{\rm super}$ exhibits a faster growth rate in higher dimensions.
\subsection{Efimov effect in Sakharov oscillation}
The amplitude of Sakharov oscillations exhibits distinct behaviors in the super- and sub-horizon regimes, with the latter manifesting characteristic Efimovian features. Recalling that the produced phonon number is given by $N_k=|\beta_k|^2$ and the oscillation amplitude by ${\cal A}_k=|\alpha_k\beta_k|$ (where $|\alpha_k|^2-|\beta_k|^2=1$), we can relate the amplitude directly to the phonon number via 
\begin{align}
{\cal A}_k=\sqrt{N_k(N_k+1)}.
\end{align}
\begin{figure}[h]
    \includegraphics[width=0.43\textwidth]{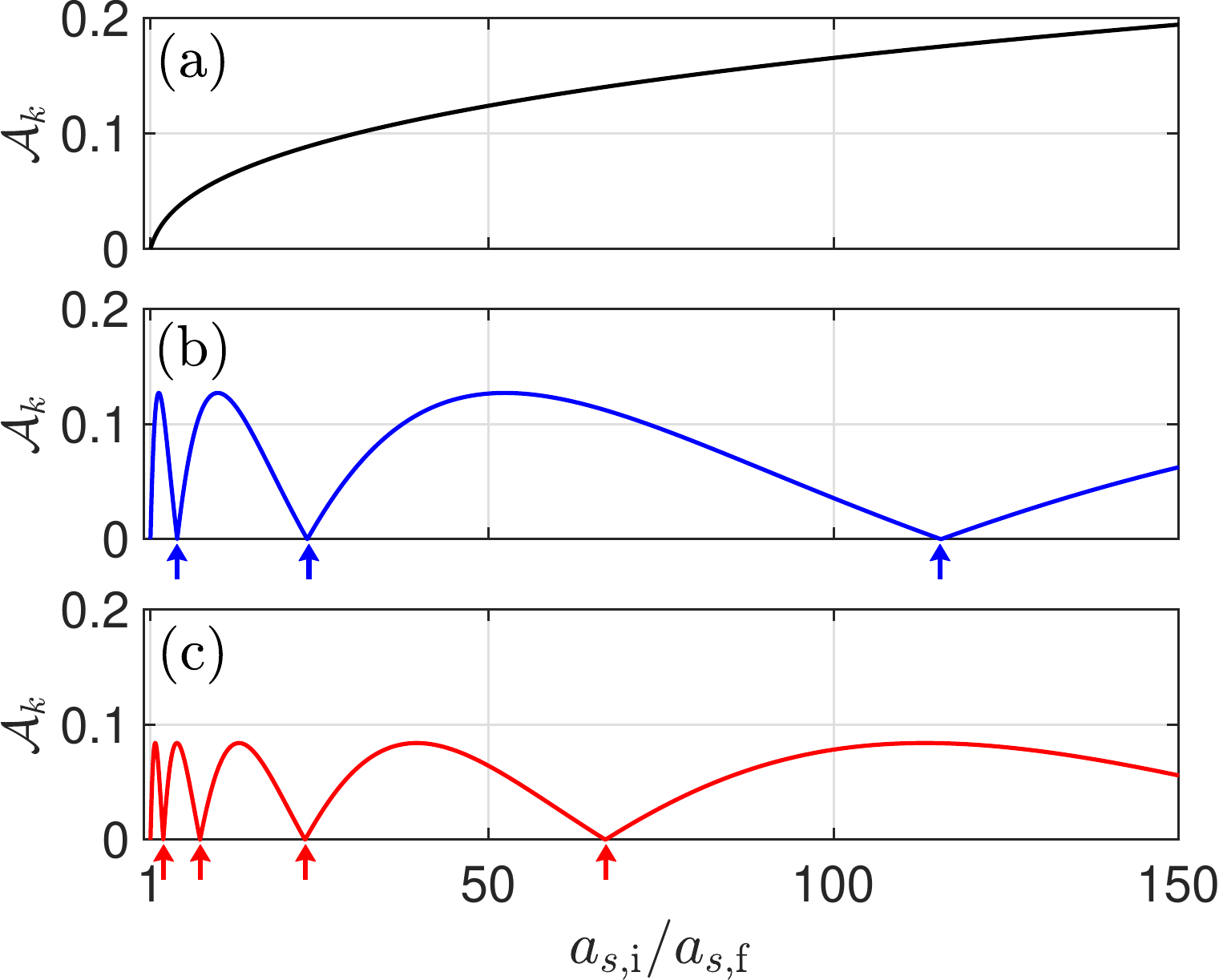}
    \caption{Amplitude of Sakharov oscillations as a function of the scattering length ratio $a_{s,{\rm i}}/a_{s,{\rm f}}$ in the (a) super-horizon and (b, c) sub-horizon regimes. The arrows indicate the specific values of $a_{s,{\rm i}}/a_{s,{\rm f}}$ where the amplitude ${\cal A}_k$ vanishes. These zero-crossing points form a geometric sequence with a common ratio of $\exp(2\pi/\sqrt{(kl)^2-1/4})$. The parameters are set to $kl=0.4$ for (a), $kl=4$ for (b), and $kl=6$ for (c).
    \label{figureem}}    
\end{figure}
In Fig. \ref{figureem}, we plot ${\cal A}_k$ as a function of the scattering length ratio $a_{s,{\rm i}}/a_{s,{\rm f}}$. In the super-horizon regime [Fig. \ref{figureem}(a)], ${\cal A}_k$ grows monotonically with an increasing ratio of $a_{s,{\rm i}}/a_{s,{\rm f}}$. Conversely, ${\cal A}_k$ is oscillatory in the sub-horizon regime [Fig. \ref{figureem}(b, c)], and the values of the ratio where ${\cal A}_k$ vanishes form a geometric sequence, as indicated by the arrows. Using Eq.~\eqref{Nk} and the relation $a_{s,{\rm i}}/a_{s,{\rm f}}=(t_{\rm f}/t_{\rm i})^2$, one can analytically show that the common ratio of this sequence is $\exp(2\pi/\sqrt{(kl)^2-1/4})$. This result is numerically verified for $kl=4$ and $kl=6$, as shown in Figs. \ref{figureem}(b) and (c), respectively.

\end{document}


\title{Supplementary Material on ``Efimovian Phonon Production for an Analog Coasting Universe in Bose-Einstein Condensates''}

\author{Yunfei Xue}
\affiliation{MOE Key Laboratory for Nonequilibrium Synthesis and Modulation of Condensed Matter, Shaanxi Province Key Laboratory of Quantum
Information and Quantum Optoelectronic Devices, School of Physics, Xi'an Jiaotong University, Xi'an 710049, China}

\author{Jiabin Wang}
\affiliation{Institute of Theoretical Physics, State Key Laboratory of Quantum Optics and Quantum Optics Devices, Shanxi University, Taiyuan 030006, China}

\author{Li Chen}
\email{lchen@sxu.edu.cn}
\affiliation{Institute of Theoretical Physics, State Key Laboratory of Quantum Optics and Quantum Optics Devices, Shanxi University, Taiyuan 030006, China}

 \author{Chenwei Lv}
  \email{lcw205046@gmail.com}
\affiliation{Department of Physics, The Chinese University of Hong Kong, Shatin, New Territories, Hong Kong, China}
\affiliation{The State Key Laboratory of  Quantum Information Technologies and Materials, The Chinese University of Hong Kong, Shatin, New Territories, Hong Kong, China}
\affiliation{New Cornerstone Science Laboratory, The Chinese University of Hong Kong, Shatin, New Territories, Hong Kong, China}

\author{Ren Zhang}
\email{renzhang@xjtu.edu.cn}
\affiliation{MOE Key Laboratory for Nonequilibrium Synthesis and Modulation of Condensed Matter, Shaanxi Province Key Laboratory of Quantum
Information and Quantum Optoelectronic Devices, School of Physics, Xi'an Jiaotong University, Xi'an 710049, China}
\affiliation{Hefei National Laboratory, Hefei, 230088, China}

\maketitle
\setcounter{equation}{0}
\setcounter{figure}{0}
\setcounter{table}{0}
\onecolumngrid

In this Supplementary Material, we present detailed derivations for: (I) the metric of an analogue Friedmann–Lemaître–Robertson–Walker (FLRW) universe in quasi-two-dimensional Bose-Einstein condensates (BEC); (II) the momentum-resolved particle production in \((n+1)\)-dimensional curved FLRW spacetime; and (III) the momentum-integrated particle density in flat space, including their asymptotic expressions. Throughout this material, we adopt the metric signature \((-, +, \cdots, +)\).

\subsection{I. Phonon universe of quasi-two-dimensional BEC}
The dynamics of a quasi-two-dimensional Bose-Einstein condensate (BEC) are governed by the Gross-Pitaevskii (GP) equation:
\begin{equation}
    i\hbar\partial_{t}\Psi({\bm \rho},t)=-\frac{\hbar^{2}}{2m}\nabla_{\bm \rho}^2\Psi({\bm \rho},t)+V\Psi({\bm \rho},t)+U(t)|\Psi({\bm \rho},t)|^{2}\Psi({\bm \rho},t),
\end{equation}
where \(\Psi({\bm \rho},t)\) denotes the wave function of the BEC (with \({\bm \rho}\equiv(x,y)\)), \(m\) represents the atomic mass, \(V\) is the trapping potential, \(\nabla_{\bm \rho}^2\) is the two-dimensional spatial Laplacian, and \(U(t)\) is the time-dependent pairwise interaction strength. Experimentally, \(U(t)\) is a tunable parameter in quasi-two-dimensional BECs, controlled by adjusting the \(s\)-wave scattering length via a Feshbach resonance. The corresponding Lagrangian density is defined as
\begin{equation}\label{LBEC}
    \mathcal{L}_{\rm BEC}=i\hbar\Psi^*({\bm \rho},t)\partial_{t}\Psi({\bm \rho},t)-\frac{\hbar^{2}}{2m}\nabla_{\bm \rho}\Psi^*({\bm \rho},t)\nabla_{\bm \rho}\Psi({\bm \rho},t)-V|\Psi({\bm \rho},t)|^2-\frac{U(t)}{2}(\Psi({\bm \rho},t)\Psi^*({\bm \rho},t))^{2}. 
\end{equation}
To investigate the hydrodynamic excitations, we adopt the Madelung representation of the bosonic field $\Psi({\bm \rho},t)=\sqrt{n({\bm \rho},t)}e^{i\theta({\bm \rho},t)}$, where $n({\bm \rho},t)$ and $\theta({\bm \rho},t)$ correspond to the density and phase of BEC, respectively. For brevity, we omit the arguments of $n({\bm \rho},t)$ and $\theta({\bm \rho},t)$ whenever this causes no ambiguity. Substituting this ansatz into the GP equation yields
\begin{equation}
    i\hbar\frac{\partial_{t}n}{2n}-\hbar\partial_{t}\theta=-\frac{\hbar^{2}}{2m}\left[\frac{\nabla_{\bm \rho}^2 \sqrt{n}}{\sqrt{n}}
    -|\nabla_{\bm \rho}\theta|^2+i\frac{\nabla_{\bm \rho}\cdot \left(n\nabla_{\bm \rho}\theta\right)}{n}\right]+V+U(t)n.
\end{equation}
For a uniform Bose gas in the long-wavelength limit, the quantum pressure term (proportional to the second spatial derivative of the density)  is negligible. Separating the real and imaginary parts leads to  the hydrodynamic equations: 
\begin{equation}\label{hydro}
\begin{aligned}
    &-\hbar\partial_{t}\theta=\frac{\hbar^{2}(\nabla_{\bm \rho}\theta)^2}{2m}+V+U(t)n,\\
    &\partial_{t}n+\frac{\hbar}{m}\nabla_{\bm \rho}\cdot \left(n\nabla_{\bm \rho}\theta\right)=0.
\end{aligned}
\end{equation}
To investigate low-energy excitation, we derive the dynamical equations for the density and phase fluctuations.
Specifically, we decompose the fields into a steady background ($n_0$, $\theta_0$) and small fluctuations ($\delta n$,$\delta\theta$), i.e., $n=n_0 + \delta n$ and $\theta=\theta_0+\delta\theta$. Assuming a stationary background superfluid with velocity $\hbar\nabla_{\bm \rho}\theta_0/m=0$, the background fields $n_0$ and $\theta_0$ satisfy
\begin{equation}\label{mean}
    -\hbar\partial_{t}\theta_0=V+U(t)n_0; \quad \partial_{t}n_0=0,  
\end{equation}
implying that the background density $n_0$ is time-independent.
 
Substituting this decomposition into Eq.~(\ref{hydro}) and retaining terms up to linear order in fluctuations, we obtain the equation of the fluctuation fields
\begin{align}
    \delta n=&-\frac{\hbar\partial_{t}\delta\theta}{U(t)}; \label{q1}\\
    \partial_{t}\delta n=&-\frac{\hbar}{m}\nabla_{\bm \rho}\cdot\left(n_0\nabla_{\bm \rho}\delta\theta\right). \label{q2}
\end{align}
Here, second-order terms are ignored.
Eliminating $\delta n$ by combining Eq.~(\ref{q1}) and Eq.~(\ref{q2}) leads to the wave equation for the phase fluctuation: 
\begin{equation}\label{kg}
    \partial_{t}\left(\frac{m}{U(t)n_0}\partial_t\delta\theta\right)=\nabla_{\bm \rho}^2\delta\theta. 
\end{equation}
Correspondingly, the Lagrangian density in Eq.~(\ref{LBEC}) simplifies to:
\begin{equation}\label{Ld}
      \mathcal{L}_\delta=\frac{\hbar^2}{2U(t)}(\partial_t\delta\theta)^2-\frac{\hbar^2 n_0}{2m}(\nabla_{\bm \rho}\delta\theta)^2.
\end{equation}
Applying the Euler-Lagrange equation to this density also yields the wave equation governing the phase fluctuation.

Equation~(\ref{kg}) is formally equivalent to the equation of motion for a massless scalar field in a $(2+1)$-dimensional FLRW spacetime with the metric
\begin{equation}
    ds^2 =g_{\mu\nu}dx^\mu dx^\nu= -dt^2 + a^2(t)\left(dx^2+dy^2\right), 
\end{equation}
where \( a(t) \) serves as the scale factor.
Under minimal coupling, the Klein-Gordon equation for a massless scalar field \(\phi\) in an FLRW metric is derived from the Euler-Lagrange equation, \(\frac{1}{\sqrt{|g|}}\partial_\mu\left(\sqrt{|g|}g^{\mu\nu}\partial_\nu\phi\right)=0\). For the (2+1)-dimensional spacetime simulated in the experiment, this yields:
\begin{equation}\label{kg2}
    \partial_{t}\left(a^2(t)\partial_t\phi\right) - \nabla_{\bm \rho}^2\phi=0.
\end{equation}
By comparing Eq.~(\ref{kg}) with Eq.~(\ref{kg2}), we identify the effective scale factor of the phonon universe as $a^2(t)=m/(U(t)n_0)$. Since the interaction strength $U(t)$ can be experimentally modulated, the scale factor $a(t)$ is highly tunable, enabling the simulation of various cosmological expansion histories. In this work, we specifically simulate a linear expansion profile, \( a(t) \propto t \), corresponding to the coasting universe, which is governed by time-scaling invariant dynamics.


\subsection{II. Momentum-resolved particle production in $(n+1)$ dimensional curved FLRW spacetime}
In this section, we generalize the analysis to an \((n+1)\)-dimensional FLRW spacetime with spatial curvature. Using comoving spherical coordinates $x^\mu=(t,r,\theta_{1},\cdots,\theta_{n-1})$, the metric is defined as 
\begin{equation}
\begin{aligned}
    ds^2 = g_{\mu\nu}dx^\mu dx^\nu = -dt^2 + a^2(t)f_{\alpha\beta}dr^\alpha dr^\beta &, \\
    f_{\alpha\beta}dr^\alpha dr^\beta = \frac{dr^2}{1-\kappa r^2}+r^2 \sum^{n-1}_{i=1}\prod^{n-1}_{j=i+1}\sin^2\theta_j d\theta_i^2 &, 
\end{aligned}
\end{equation}
where \( a(t) \) denotes the scale factor, \(\kappa \) is the spatial curvature, and $f_{\alpha\beta}$ is the metric of \(n\)-dimensional curved space of spherical coordinates $r^\alpha=(r,\theta_{1},\cdots,\theta_{n-1})$. 
The geometry is spherical for positive curvature ($\kappa > 0$), hyperbolic for negative curvature ($\kappa < 0$), and Euclidean (flat) for $\kappa=0$.


Based on this spacetime background, we now proceed to detail: (1) the derivation of the Klein–Gordon equation Eq.~(\ref{nkg}); (2) the mode expansion Eq.~(\ref{modeex}) including orthonormal and commutation relations; (3) the calculation of particle production Eq.~(\ref{Nk}) under a linear scale factor. 

\subsubsection{\rm 1. Klein–Gordon equation}
The Klein–Gordon equation follows from the principle of stationary action. For a massless scalar field \( \phi(x^\mu) \) in curved spacetime, the Lagrangian density with minimal coupling is given by
\begin{equation}\label{L}
    \mathcal{L}=-\frac{\hbar}{2}\sqrt{|g|}g^{\mu\nu}\partial_\mu\phi\partial_\nu\phi.
\end{equation}
The action $S$ is defined as the integral over the $(n+1)$-dimensional manifold $\mathcal{M}$:
\begin{equation}\label{action}
S=\int_{\mathcal{M}}\mathcal{L}d^{n+1}x=\int_{\mathcal{M}}\frac{\mathcal{L}}{\sqrt{|g|}}\epsilon,
\end{equation}
where $\int_{\mathcal{M}}$ represents the integral over the spacetime and $\epsilon=\sqrt{|g|}d^{n+1}x$ is the invariant volume element. 
The Lagrangian density $\mathcal{L}$ is not invariant under a coordinate transformation. Nevertheless, $\mathcal{L}/\sqrt{|g|}$ is a scalar, and thus the action $S$ is independent of the coordinate choice. Minimizing the action ($\delta S=0$) yields the covariant Klein–Gordon equation 
\begin{equation}
    \frac{1}{\sqrt{|g|}}\partial_\mu\left(\sqrt{|g|}g^{\mu\nu}\partial_\nu\phi\right)=0. 
\end{equation}
To write the Klein–Gordon equation in the curved FLRW spacetime, we first explicitly write the matrix form of the metric $g_{\mu\nu}$ and the inverse metric $g^{\mu\nu}$ as 
\begin{align}
    [g_{\mu\nu}]=
    \begin{pmatrix}
        -1&0 \quad \cdots \quad 0\\
        \begin{matrix}
            0\\
            \vdots\\
            0
        \end{matrix}&a^2(t) f_{\alpha\beta}
    \end{pmatrix}; \quad 
    [g^{\mu\nu}]=
    \begin{pmatrix}
        -1&0 \quad \cdots \quad 0\\
        \begin{matrix}
            0\\
            \vdots\\
            0
        \end{matrix}&a^{-2}(t) f^{\alpha\beta}
    \end{pmatrix}, 
\end{align}
where $f_{\alpha\beta}$, an element of $n\times n$ matrix, denotes the co-variant spatial metric, and $f^{\alpha\beta}$ represents the contra-variant metric. The determinant of the metric $g_{\mu\nu}$ is given by
\begin{equation}
    g = \det[g_{\mu\nu}]= -\det[a^2(t)f_{\alpha\beta}]= -a^{2n}(t)f, 
\end{equation}
where $f=\det[f_{\alpha\beta}]$ is the determinant of the spatial metric $f_{\alpha\beta}$, which is independent of time $t$. The Lagrangian density in this spacetime background takes the form 
\begin{equation}
    \mathcal{L}=\frac{\hbar}{2}\sqrt{f}\left[a^{n}(t)(\partial_t\phi)^2-a^{n-2}(t)f^{\alpha\beta}\partial_\alpha\phi\partial_\beta\phi\right]. 
\end{equation}
Upon substituting the explicit metric components into the action, the Klein–Gordon equation in the curved FLRW spacetime is expressed as 
\begin{equation}
     \frac{1}{a^n(t)\sqrt{f}} \left(\partial_t(a^n(t)\sqrt{f}\partial_t\phi)-\partial_\alpha(a^{n-2}(t)\sqrt{f}f^{\alpha\beta}\partial_\beta\phi)\right)=0
\end{equation}
Simplifying this expression yields the evolution equation:
\begin{equation}\label{nkg}
\partial_t^2\phi + \frac{n\dot{a}(t)}{a(t)}\partial_t\phi - \frac{\nabla^2\phi}{a^2(t)}=0, 
\end{equation}
where $\nabla^2\phi={\sqrt{|f|}}^{-1} \partial_\alpha(\sqrt{|f|}f^{\alpha\beta}\partial_\beta\phi)$ denotes the Laplace-Beltrami operator associated with the spatial metric $f_{\alpha\beta}$. This equation can be solved using the method of separation
of variables. Consequently, the general solution for the field $\phi$ is decomposed as 
\begin{equation}
    \phi(t,\mathbf{x}) = \int dk \sum_{m_1}\cdots\sum_{m_{n-1}}
    \left[ u_{k}(t)\mathcal{H}_{\mathbf{k}}(\mathbf{x}){b}_\mathbf{k}
    + u_{k}^*(t)\mathcal{H}^*_{\mathbf{k}}(\mathbf{x}){b}^*_\mathbf{k} \right].
\end{equation}
Here, we define the coordinates $\mathbf{x}=(r,\theta_1,\cdots,\theta_{n-1})$ and the mode indices $\mathbf{k}=(k,m_1,\cdots,m_{n-1})$. The terms ${b}_\mathbf{k}$ represent the superposition coefficients (or operators in the quantized theory), \( u_{k}(t) \) are the temporal mode functions, and $\mathcal{H}_{\mathbf{k}}(\mathbf{x})$ are the spatial eigenfunctions of the Laplacian $\nabla^2$. 
The temporal mode functions \( u_{k}(t) \) are governed by the evolution equation
\begin{equation}\label{eom}
    \ddot{u}_{k}(t) + \frac{n\dot{a}(t)}{a(t)}\dot{u}_{k}(t) + \frac{k^2}{a^2(t)}u_{k}(t) = 0,
\end{equation}
while the spatial eigenfunctions $\mathcal{H}_{\mathbf{k}}(\mathbf{x})$ satisfy the Helmholtz equation:
\begin{equation}\label{Hk}
    \nabla^2 \mathcal{H}_{\mathbf{k}}(\mathbf{x})=-k^2\mathcal{H}_{\mathbf{k}}(\mathbf{x}).
\end{equation}
Subsequently, we perform the canonical quantization of the field $\phi$ by imposing orthogonality conditions on the temporal mode functions \( u_{k}(t) \) and the spatial eigenfunctions $\mathcal{H}_{\mathbf{k}}(\mathbf{x})$. The explicit forms of $\mathcal{H}_{\mathbf{k}}(\mathbf{x})$ and the admissible range of wavenumbers $k$ will be determined in the derivation that follows. 

\subsubsection{\rm 2. Mode expansion of the field operator}
The quantized field operator \( \hat{\phi} \) admits the following expansion in terms of the creation and annihilation operators \( \{\hat{b}^\dagger_\mathbf{k}, \hat{b}_\mathbf{k}\} \):
\begin{equation}\label{modeex}
    \hat{\phi}(t,\mathbf{x})
    = \int dk \sum_{m_1}\cdots\sum_{m_{n-1}}
    \left[ u_{k}(t)\mathcal{H}_{\mathbf{k}}(\mathbf{x})\hat{b}_\mathbf{k}
    + u_{k}^*(t)\mathcal{H}^*_{\mathbf{k}}(\mathbf{x})\hat{b}^\dagger_\mathbf{k} \right].
\end{equation}
The spatial eigenfunctions $\mathcal{H}_{\mathbf{k}}(\mathbf{x})$ are constructed to satisfy the orthonormality condition:
\begin{align}\label{calH}
    \int d^n x \sqrt{f} \mathcal{H}^*_{\mathbf{k}}(\mathbf{x})\mathcal{H}_{\mathbf{k}^\prime}(\mathbf{x})=\delta^n(\mathbf{k}-\mathbf{k}^\prime)\equiv
    \begin{cases}
        \delta_{k k'}\prod_{j=1}^{n-1}\delta_{m_j m_j^\prime}, & \quad {\text{for discrete }} k, \\
        \delta(k-k')\prod_{j=1}^{n-1}\delta_{m_j m_j^\prime}, & \quad {\text{for continuous }} k,
    \end{cases}
\end{align}
where the integration is performed over the spatial volume equipped with the measure $\sqrt{f}$. Following Ref.~\cite{Parker2009}, we define the scalar inner product as
\begin{equation}\label{innerPdef}
    (F_1, F_2) \equiv i \int_{\mathbf{x}} a^n(t) \left( F_1^* \partial_t F_2 - F_2 \partial_t F_1^* \right).
\end{equation}
Based on this definition, we impose the orthonormality conditions on the mode functions:
\begin{equation}\label{orthonormal}
\begin{aligned}
    (u_{k}\mathcal{H}_{\mathbf{k}}, u_{k^\prime}\mathcal{H}_{\mathbf{k}^\prime}) &= \delta^n(\mathbf{k}-\mathbf{k}^\prime), \\
    (u_{k}\mathcal{H}_{\mathbf{k}}, u^*_{k^\prime}\mathcal{H}^*_{\mathbf{k}^\prime}) &= 0.
\end{aligned}
\end{equation}
The orthonormality relations in Eq.(\ref{calH}) and (\ref{orthonormal}) lead to the Wronskian normalization condition for the temporal mode functions: \(u^*_{k}\dot{u}_{k}-\dot{u}^*_{k}u_{k}=1/(ia^n(t)) \). We note that this condition alone is insufficient to determine the set of mode functions $u_k$ uniquely. The necessary additional condition will be discussed following Eq.~(\ref{cacr}). 
Using the inner product defined in Eq.(\ref{innerPdef}), the creation and annihilation operators \( \{\hat{b}^\dagger_\mathbf{k}, \hat{b}_\mathbf{k}\} \) are obtained via projection: 
\begin{equation}
    \hat{b}_\mathbf{k}=\left(u_{k}(t)\mathcal{H}_{\mathbf{k}}(\mathbf{x}),\hat{\phi}(t,\mathbf{x})\right),\quad 
    \hat{b}^\dagger_\mathbf{k}=-\left(u^*_{k}(t)\mathcal{H}^*_{\mathbf{k}}(\mathbf{x}),\hat{\phi}(t,\mathbf{x})\right). 
\end{equation}
For the Lagrangian density given in Eq.~(\ref{L}), the conjugate momentum of the field \(\hat{\phi}\) is defined as $\hat{\pi}=\frac{\delta\mathcal{L}}{\delta(\partial_t\hat{\phi})}=\hbar\sqrt{f}a^n(t)\partial_t\hat{\phi}$. To quantize the field, we impose the standard equal-time canonical commutation relations: 
\begin{equation}\label{ccr}
\begin{aligned}[c]
    &\left[\hat{\phi}(t,\mathbf{x}),\hat{\pi}(t,\mathbf{x}^\prime)\right]=i\hbar\delta^n(\mathbf{x}-\mathbf{x}^\prime); \\
    &\left[\hat{\phi}(t,\mathbf{x}),\hat{\phi}(t,\mathbf{x}^\prime)\right]=\left[\hat{\pi}(t,\mathbf{x}),\hat{\pi}(t,\mathbf{x}^\prime)\right]=0, 
\end{aligned}
\end{equation}
which leads to the commutation relations of creation and annihilation operators 
\begin{equation}\label{cacr}
\begin{aligned}[c]
    [\hat{b}_\mathbf{k},\hat{b}^\dagger_{\mathbf{k}^\prime}]&=
    -\left[\left(u_{k}\mathcal{H}_{\mathbf{k}}(\mathbf{x}),\hat{\phi}(t,\mathbf{x})\right),\left(u^*_{k'}\mathcal{H}^*_{\mathbf{k}'}(\mathbf{x}'),\hat{\phi}(t,\mathbf{x}')\right)\right]\\
    &=-\int_{\mathbf{x}}\int_{\mathbf{x}'}a^{2n}(t)\mathcal{H}^*_{\mathbf{k}}(\mathbf{x})\mathcal{H}_{\mathbf{k}'}(\mathbf{x}')
    \left(u^*_{k}\dot{u}_{k'}[\hat{\phi}(t,\mathbf{x}'),\partial_t\hat{\phi}(t,\mathbf{x})]-\dot{u}^*_{k}u_{k'}[\hat{\phi}(t,\mathbf{x}),\partial_t\hat{\phi}(t,\mathbf{x}')]\right)\\
    &=\delta^n(\mathbf{k}-\mathbf{k}^\prime); \\
    [\hat{b}_\mathbf{k},\hat{b}_{\mathbf{k}^\prime}]&=[\hat{b}^\dagger_{\mathbf{k}},\hat{b}^\dagger_{\mathbf{k}^\prime}]=0. 
\end{aligned}
\end{equation}
These expressions represent the standard commutation relations for bosonic creation and annihilation operators in the FLRW spacetime. However, it is important to note that the concept of particles in curved spacetime is not always well-defined \cite{Wald1995}. For comoving observers measuring proper time $t$, the set \( \{\hat{b}^\dagger_\mathbf{k}, \hat{b}_\mathbf{k}\} \) corresponds to creation and annihilation operators of well-defined particles only if the associated mode functions \( u_{k}(t) \) are selected as positive-frequency modes.


We now proceed to solve Eq.~(\ref{Hk}) to obtain the explicit form of $\mathcal{H}_{\mathbf{k}}(\mathbf{x})$. By applying the method of separation of variables, the Laplace-Beltrami operator for the metric $f_{\alpha\beta}$ decomposes into radial and angular components: $\nabla^2=\nabla_r^2+r^{-2}\nabla_\theta^2$. Here, $\nabla_r^2$ and $\nabla_\theta^2$ represent the Laplace-Beltrami operators corresponding to the radial coordinate $r$ and the $(n-1)$-sphere angular coordinates $\theta_{1,\dots,n-1}$, respectively. 
The eigenvalue equation for the angular component $\nabla_\theta^2$ is given by
\begin{equation}
    \nabla^2_\theta Y_{m_1,\cdots,m_{n-1}}=-m_{n-1}(m_{n-1}+n-2)Y_{m_1,\cdots,m_{n-1}},
\end{equation}
where $Y_{m_1,\cdots,m_{n-1}}(\theta_1,\cdots,\theta_{n-1})$ denotes the normalized $(n-1)$-dimensional spherical harmonics \cite{Higuchi1987}. The eigenvalue equation for the radial part $\nabla_r^2$ takes the form:
\begin{equation}
    \frac{\sqrt{1-\kappa r^2}}{r^{n-1}}\frac{d}{dr}\left(r^{n-1}\sqrt{1-\kappa r^2}\frac{d\mathcal{K}_{k,m_{n-1}}(r)}{dr}\right)
    =\left(\frac{m_{n-1}(m_{n-1}+n-2)}{r^2}-k^2\right)\mathcal{K}_{k,m_{n-1}}(r).
\end{equation}
The resulting radial eigenfunctions $\mathcal{K}_{k,m_{n-1}}(r)$ and eigenvalues $k^2$ are:
\begin{align}
    \mathcal{K}_{k,m_{n-1}}(r)=&\begin{cases}
        r^{1/2-\mathcal{C}}P^{\sigma}_{\ell_+ +\mathcal{C}-1/2}(\sqrt{1-|\kappa|r^2});&\quad \kappa>0\\
        r^{1/2-\mathcal{C}}J_{\sigma}(\ell_0 r);&\quad \kappa=0\\
        r^{1/2-\mathcal{C}}P^{\sigma}_{i\ell_- -1/2}(\sqrt{1+|\kappa|r^2});&\quad \kappa<0 
    \end{cases}\\
    k^2=&\begin{cases}
        \ell_+(\ell_+ +2\mathcal{C})|\kappa|;&\quad \kappa>0\\
        \ell_0^2;&\quad \kappa=0\\
        (\ell_-^2 +\mathcal{C}^2)|\kappa|,&\quad \kappa<0 
    \end{cases}\label{k}
\end{align}
where we have defined $\mathcal{C}\equiv(n-1)/2$ and $\sigma=m_{n-1}+\mathcal{C}-1/2$, with $\ell_+\in \mathbb{N}$ and $\ell_{0,-}\in [0,+\infty)$. In these expressions, $P^{m}_{\ell}$ are the associated Legendre polynomials, $J_m$ are the Bessel functions of the first kind, and $P^{m}_{i\ell-1/2}$ are the conical functions. 
The full orthonormal eigenfunctions are defined as $\mathcal{H}_{\mathbf{k}}(\mathbf{x}) = h_{k,m_{n-1}}\mathcal{K}_{k,m_{n-1}}(r)Y_{m_1,\cdots,m_{n-1}}(\theta_1,\cdots,\theta_{n-1})$, where $h_{k,m_{n-1}}$ are normalization coefficients. Since the radial functions $\mathcal{K}_{k,m_{n-1}}(r)$ are real and the spherical harmonics satisfy the relation $Y^*_{m_1,m_2,\cdots,m_{n-1}}=(-1)^{m_1}Y_{-m_1,m_2,\cdots,m_{n-1}}$, the complex conjugate of $\mathcal{H}_{\mathbf{k}}(\mathbf{x})$ is given by
\begin{equation}\label{H_star}
    \mathcal{H}^*_{\mathbf{k}}(\mathbf{x})=(-1)^{m_1}\mathcal{H}_{\bar{\bf k}}(\mathbf{x}),
\end{equation}
where $\bar{\mathbf k}$ denotes the index set $(k,-m_1,m_2,\cdots,m_{n-1})$.

\subsubsection{\rm 3. Particle production for the linearly expanding universe}
We proceed to analyze particle production by comparing the vacuum states corresponding to the pre- and post-expansion. Before the expansion, the field operator \( \hat{\phi} \) is expanded in terms of the creation and annihilation operators \( \{\hat{b}^\dagger_{\mathbf{k},\mathrm{i}}, \hat{b}_{\mathbf{k},\mathrm{i}}\} \) and the mode functions \( u_{k,\mathrm{i}}(t) \). After the expansion, the field is expressed using the set \( \{\hat{b}^\dagger_{\mathbf{k},\mathrm{f}}, \hat{b}_{\mathbf{k},\mathrm{f}}\} \) and mode functions \( u_{k,\mathrm{f}}(t) \): 
\begin{equation}
\begin{aligned}
    \hat{\phi}(t,\mathbf{x})
    &= \int dk \sum_{m_1}\cdots\sum_{m_{n-1}}
    \left[ u_{k,\mathrm{i}}(t)\mathcal{H}_{\mathbf{k}}(\mathbf{x})\hat{b}_{\mathbf{k},\mathrm{i}}
    + u_{k,\mathrm{i}}^*(t)\mathcal{H}^*_{\mathbf{k}}(\mathbf{x})\hat{b}_{\mathbf{k},\mathrm{i}}^\dagger \right], \\[3pt]
    \hat{\phi}(t,\mathbf{x})
    &= \int dk \sum_{m_1}\cdots\sum_{m_{n-1}}
    \left[ u_{k,\mathrm{f}}(t)\mathcal{H}_{\mathbf{k}}(\mathbf{x})\hat{b}_{\mathbf{k},\mathrm{f}}
    + u_{k,\mathrm{f}}^*(t)\mathcal{H}^*_{\mathbf{k}}(\mathbf{x})\hat{b}_{\mathbf{k},\mathrm{f}}^\dagger \right].
\end{aligned}
\end{equation}
Both sets of mode functions, \( u_{k,\mathrm{i}}(t) \) and \( u_{k,\mathrm{f}}(t) \), satisfy the equation of motion given in Eq.~(\ref{eom}). We define the positive-frequency modes by imposing the following asymptotic behaviors:
\( u_{k,\mathrm{i}}(t) \propto e^{-ikl t/t_{\rm i}} \quad \text{for } t < t_{\rm i}, \quad \text{and} \quad u_{k,\mathrm{f}}(t) \propto e^{-ikl t/t_{\rm f}} \quad \text{for } t > t_{\rm f}. \)
These boundary conditions identify \( u_{k,\mathrm{i}} \) as the modes defining the initial vacuum ($t < t_{\rm i}$) and \( u_{k,\mathrm{f}} \) as those defining the final vacuum ($t > t_{\rm f}$). Consequently, the particle number operators are defined as \(\hat{b}^\dagger_{\mathbf{k},\mathrm{i}}\hat{b}_{\mathbf{k},\mathrm{i}}\) for the initial state and \(\hat{b}^\dagger_{\mathbf{k},\mathrm{f}}\hat{b}_{\mathbf{k},\mathrm{f}}\) for the final state.

A Bogoliubov transformation connects these two sets of mode functions:
\begin{equation}
    u_{k,\mathrm{f}}(t)=\alpha_{k} u_{k,\mathrm{i}}(t)+\beta_{k} u_{k,\mathrm{i}}^*(t),
\end{equation}
where the normalization requires \(|\alpha_{k}|^{2}-|\beta_{k}|^{2}=1\). Substituting this transformation into the field expansion allows us to derive the corresponding transformation for the operators, utilizing the property in Eq.~(\ref{H_star}):
\begin{equation}\label{bogoliu}
    \hat{b}_{\mathbf{k},\mathrm{f}}=\alpha^*_{k}\hat{b}_{\mathbf{k},\mathrm{i}}-\beta^*_{k}(-1)^{m_1}\hat{b}^\dagger_{\bar{\mathbf k},\mathrm{i}},
\end{equation}
We assume the initial state contains no particles before the expansion, meaning the field occupies the vacuum state associated with $\hat{b}_{\mathbf{k},\mathrm{i}}$. Using the relation in Eq.~(\ref{bogoliu}), the expectation value of the particle number operator $\hat{b}^\dagger_{\mathbf{k},\mathrm{f}}\hat{b}_{\mathbf{k},\mathrm{f}}$ after the expansion is derived as:
\begin{equation}\label{N}
    N_k=\langle\Omega|\hat{b}^\dagger_{\mathbf{k},\mathrm{f}}\hat{b}_{\mathbf{k},\mathrm{f}}|\Omega\rangle=|\beta_{k}|^2,
\end{equation}
where $|\Omega\rangle$ denotes the vacuum state of the initial operators $\hat{b}_{\mathbf{k},\mathrm{i}}$. The resulting nonzero value of $N_k$ signifies the production of particles induced by the expansion.

For the linearly expanding FLRW spacetime, known as the coasting universe, the scale factor is defined as
\begin{align}
    a(t)=
    \begin{cases}
        t_{\rm i}/l, &\quad t < t_{\rm i},\\[3pt]
        t/l, &\quad t_{\rm i} \leqslant t \leqslant t_{\rm f},\\[3pt]
        t_{\rm f}/l, &\quad t_{\rm f} < t,
    \end{cases}
\end{align}
where \( 1/l \) characterizes the expansion rate. The initial and final times are defined as \( t_{\rm i/f} = l a_{\rm i/f} \), where \( a_{\rm i/f} \) denotes the scale factors before and after the expansion, respectively.
In the interval \( t_{\rm i} \leqslant t \leqslant t_{\rm f} \), the equation of motion for the mode function \( u_{k}(t) \) is given by
\begin{equation}\label{costingEoM}
    \ddot{u}_{k}(t)
    + \frac{n}{t}\dot{u}_{k}(t)
    + \frac{(kl)^2}{t^2}u_{k}(t) = 0.
\end{equation}
This equation is invariant under time rescaling \( t \to \lambda t \). It admits analytic solutions for arbitrary \( n \), with the two linearly independent solutions taking the form:
\begin{equation}
    f_{\pm}(t) = t^{-\mathcal{C} \pm \sqrt{\mathcal{C}^2 - (kl)^2}}.
\end{equation}
The behavior of these solutions depends on the value of \( kl \): for \( kl < \mathcal{C} \), they exhibit polynomial time dependence, whereas for \( kl > \mathcal{C} \), they become log-periodic functions of \( t \). This dynamical bifurcation is underpinned by the hidden SU(1,1) symmetry of the system. Specifically, we introduce the algebraic generators \(\hat{K}_{0,1,2}\) defined as:
\begin{equation}
\begin{aligned}
    \hat K_{0(1)}&=\partial_t^2+\frac{n}{t}\partial_t+\frac{k^2l^2}{t^2}\mp \frac{t^2}{16}, \\
    \hat K_2&=-\frac{i}{2}\left(t\partial_t+\frac{n+1}{2}\right).
\end{aligned}
\end{equation}
These operators form a representation of the SU(1,1) algebra, reflecting a hidden symmetry intrinsic to the time-scaling invariance of the equation of motion. This representation is characterized by the Bargmann index \(1/2\pm\sqrt{\mathcal{C}^2-(kl)^2}/2\) \cite{deAlfaro1976}. Similar to the \((2+1)\)-dimensional case presented in the main text, \(\hat{K}_2\) serves as the generator for time-scaling, while \(\hat{K}_{0(1)}\) governs the dynamics. In the following analysis, we demonstrate that this index signifies a bifurcation into two distinct dynamical behaviors separated by the critical value \(kl=\mathcal{C}\), consistent with the momentum-resolved particle generation being strictly governed by the dimensionless parameter \(kl\).

To determine the coefficient $\beta_{k}$, we solve the equation of motion given in Eq.~(\ref{costingEoM}). The explicit solution for the mode function $u_{k,\mathrm{i}}(t)$ is given by:
\begin{align}
    u_{k,\mathrm{i}}(t)&=
    \begin{cases}
        \displaystyle \frac{(l/t_{\rm i})^{\mathcal{C}}}{\sqrt{2k}} e^{-i kl t/t_{\rm i}}, &\; t<t_{\rm i}\\[8pt]
        \displaystyle A_1 t^{-\mathcal{C}-\sqrt{\mathcal{C}^2-(kl)^2}}+A_2 t^{-\mathcal{C}+\sqrt{\mathcal{C}^2-(kl)^2}}, &\; t_{\rm i}\leqslant t\leqslant t_{\rm f}\\[8pt]
        \displaystyle A_3 \frac{(l/t_{\rm i})^{\mathcal{C}}}{\sqrt{2k}}e^{-i kl t/t_{\rm f}}+A_4 \frac{(l/t_{\rm i})^{\mathcal{C}}}{\sqrt{2k}}e^{i kl t/t_{\rm f}}, &\; t_{\rm f}<t
    \end{cases}
\end{align}
while the post-expansion mode function $u_{k,\mathrm{f}}(t)$ is defined for $t > t_{\rm f}$ as:
\begin{equation}
    u_{k,\mathrm{f}}(t)=\frac{(l/t_{\rm i})^{\mathcal{C}}}{\sqrt{2k}} e^{-{i kl t/t_{\rm f}}}.
\end{equation}
We impose continuity conditions on the mode function and its time derivative at the transition times $t_{\rm i}$ and $t_{\rm f}$:
\begin{equation}
\begin{aligned}
    \left.u_{k}\right|_{t\rightarrow {t_{\rm i}}^+}&=\left.u_{k}\right|_{t\rightarrow {t_{\rm i}}^-},\quad
    \left.\dot{u}_{k}\right|_{t\rightarrow {t_{\rm i}}^+}=\left.\dot{u}_{k}\right|_{t\rightarrow {t_{\rm i}}^-}, \\
    \left.u_{k}\right|_{t\rightarrow {t_{\rm f}}^+}&=\left.u_{k}\right|_{t\rightarrow {t_{\rm f}}^-},\quad
    \left.\dot{u}_{k}\right|_{t\rightarrow {t_{\rm f}}^+}=\left.\dot{u}_{k}\right|_{t\rightarrow {t_{\rm f}}^-}.
\end{aligned}
\end{equation}
Solving these boundary matching equations yields the coefficients $A_{3}$ and $A_{4}$:
\begin{equation}
\begin{aligned}
    A_3&=\cosh\left(\sqrt{\mathcal{C}^2-(kl)^2}\ln(t_{\rm f}/t_{\rm i})\right)-i\frac{\sinh\left(\sqrt{\mathcal{C}^2-(kl)^2}\ln(t_{\rm f}/t_{\rm i})\right)}{\sqrt{\mathcal{C}^2/(kl)^2-1}},\\
    A_4&=e^{-2ikl}\frac{\sinh\left(\sqrt{\mathcal{C}^2-(kl)^2}\ln(t_{\rm f}/t_{\rm i})\right)}{\sqrt{1-(kl)^2/\mathcal{C}^2}}.
\end{aligned}
\end{equation}
Comparing the late-time behavior of $u_{k,\mathrm{i}}(t)$ with the Bogoliubov transformation $u_{k,\mathrm{i}}(t)=\alpha^*_{k} u_{k,\mathrm{f}}(t)-\beta_{k} u_{k,\mathrm{f}}^*(t)$, we identify the Bogoliubov coefficients as $\alpha_{k}=A^*_3$ and $\beta_{k}=-A_4$. Consequently, the particle number expectation value $N_k = |\beta_k|^2$ from Eq.~(\ref{N}) is calculated as:
\begin{equation}\label{Nk}
\begin{aligned}
    N_k=
    \begin{cases}
        \displaystyle\frac{\sin^2\left(\sqrt{(kl)^2-\mathcal{C}^2}\ln(t_{\rm f}/t_{\rm i})\right)}{(kl)^2/\mathcal{C}^2-1};&kl>\mathcal{C}\\
        \displaystyle\frac{\sinh^2\left(\sqrt{\mathcal{C}^2-(kl)^2}\ln(t_{\rm f}/t_{\rm i})\right)}{1-(kl)^2/\mathcal{C}^2},&kl<\mathcal{C}
    \end{cases}
\end{aligned}
\end{equation}
which reduces to Eq.(4) in the main text when $n=2$.
\begin{figure}[h]
    \includegraphics[width=0.45\textwidth]{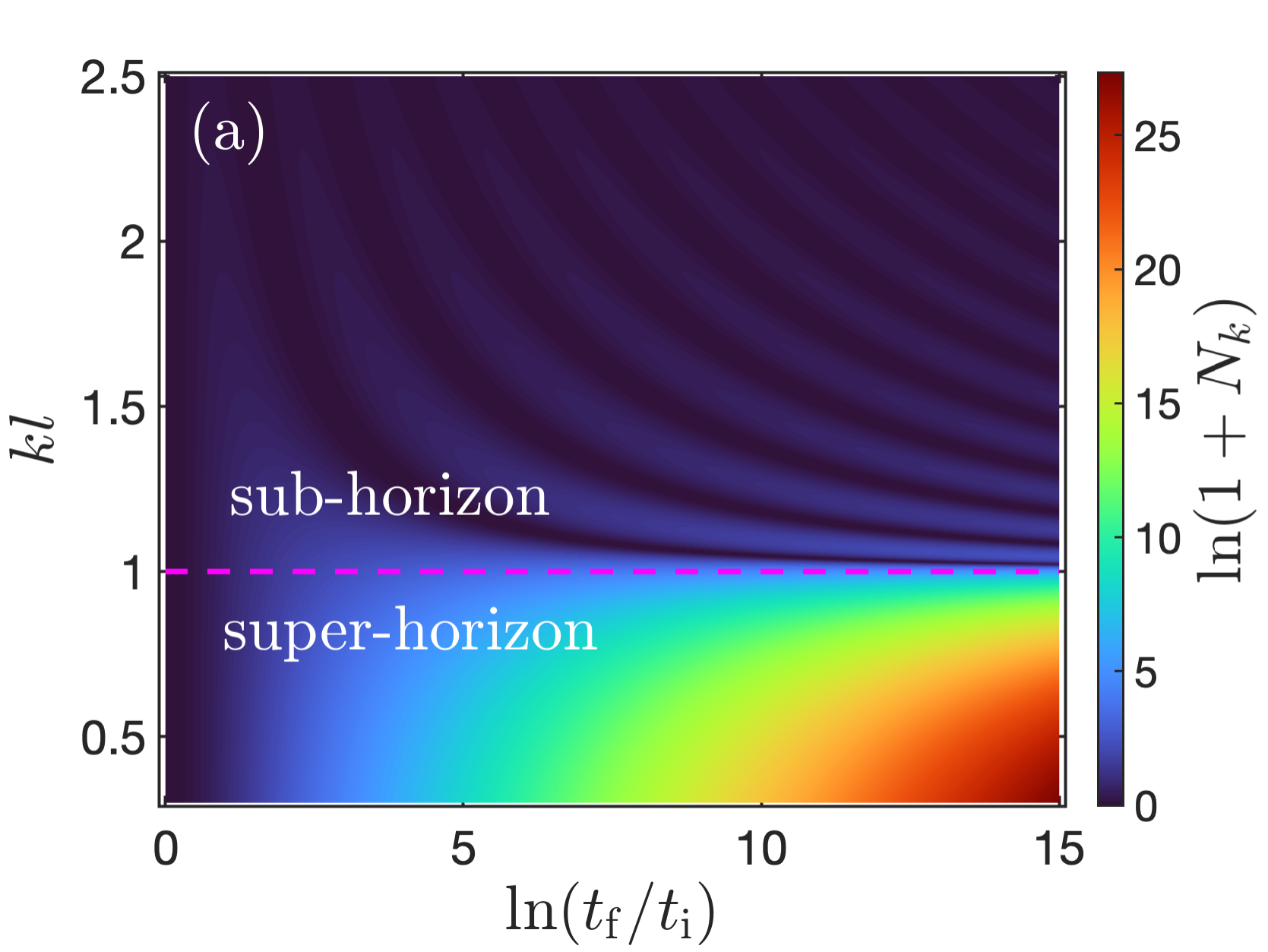}
    \includegraphics[width=0.45\textwidth]{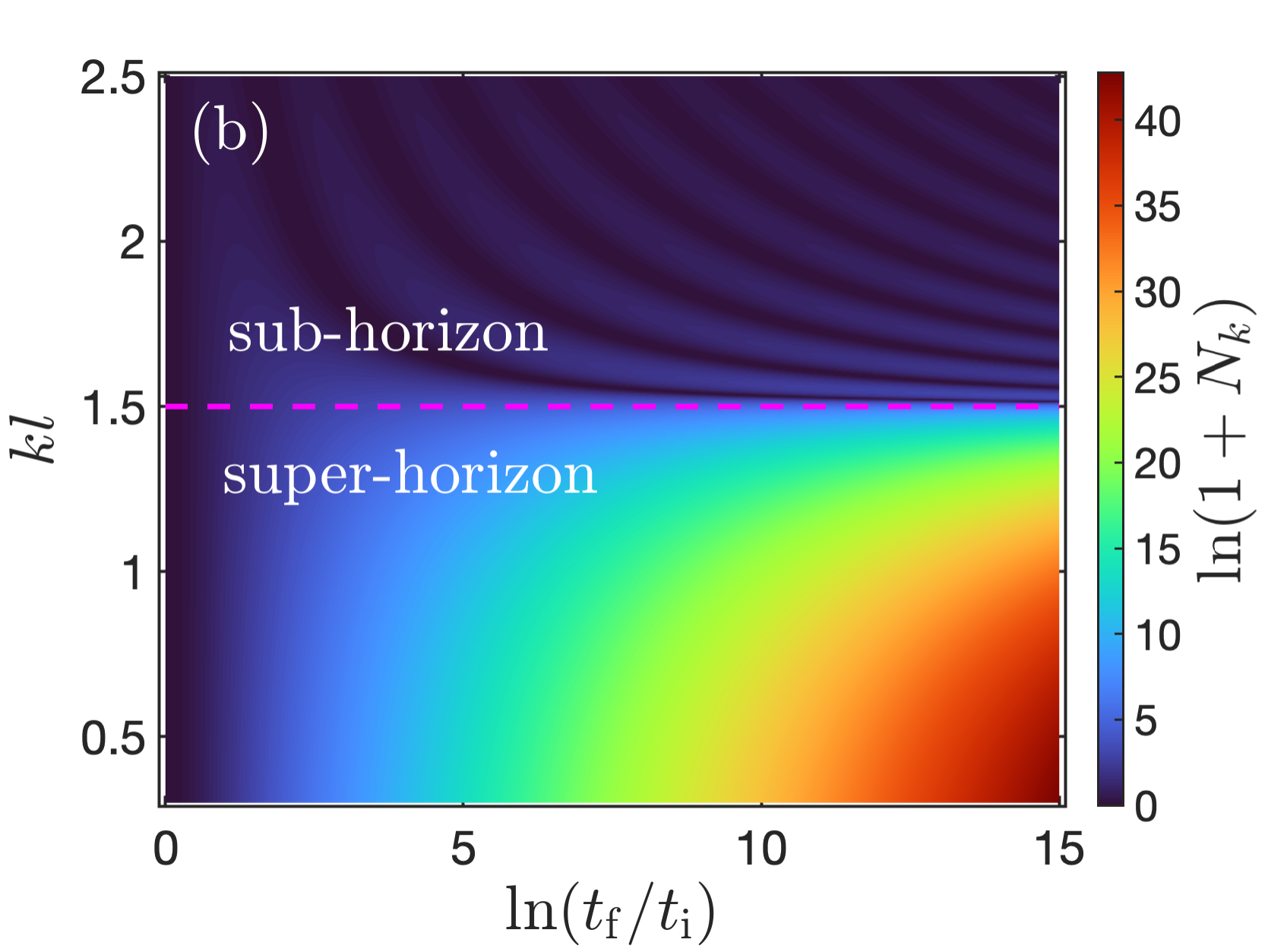}
    \caption{Post-expansion phonon production in the $(n+1)$-dimensional analog coasting universe as a function of $kl$ and $\ln(t_{\rm f}/t_{\rm i})$. The color intensity represents $\ln(N_{k}+1)$. We present two representative cases, (a) $n=3$ and (b) $n=4$, to demonstrate the universality of the results. The emergence of two distinct regimes is robust across dimensions: In the sub-horizon regime ($kl>{\cal C}$), $N_{k}$ depends log-periodically on $t_{\rm f}/t_{\rm i}$, whereas in the super-horizon regime ($kl<{\cal C}$), it exhibits a power-law dependence. The boundary between these regimes is marked by the purple dashed line at the critical value $kl={\cal C}$, where ${\cal C}\equiv(n-1)/2$.}
    \label{figureS1}
\end{figure}

Figure~\ref{figureS1} illustrates the post-expansion phonon production for an $(n+1)$-dimensional analog coasting universe, plotted against $kl$ and the expansion duration $\ln(t_{\rm f}/t_{\rm i})$. The color map represents the magnitude of $\ln(N_{k}+1)$. By comparing cases (a) $n=3$ and (b) $n=4$, we demonstrate that the dynamical behavior is universal regardless of dimension. Specifically, the dynamics bifurcate into two distinct regimes separated by a critical value ${\cal C}=(n-1)/2$ (indicated by the purple dashed line): a sub-horizon regime ($kl>{\cal C}$) where $N_{k}$ is a log-periodic function of $t_{\rm f}/t_{\rm i}$, and a super-horizon regime ($kl<{\cal C}$) characterized by a power-law growth.

Furthermore, the spatial curvature $\kappa$ determines the admissible range of wavenumbers $k$, as shown in Eq.~(\ref{k}). For non-negative curvature ($\kappa\geqslant 0$), the spectrum is unbounded with $k\geqslant 0$. In contrast, for negative curvature ($\kappa<0$), a lower bound emerges as $k\geqslant \mathcal{C}\sqrt{|\kappa|}$. A notable example is the Milne universe~\cite{Milne1934, Yamamoto1995, BenoitLevy2012, Baumann2022}, a special case of the linearly expanding FLRW universe with $\kappa=-1/l^{2}$; strictly following this bound implies that all modes satisfy $kl\geqslant \mathcal{C}$, thereby confining the system entirely to the sub-horizon regime.

\subsection{III. Momentum-integrated particle density of sub- and super-horizon modes for $(n+1)$ dimensional FLRW spacetime}
Using flat space (\(\kappa=0\)) as an example, we now proceed to calculate the momentum-integrated particle densities for the sub- and super-horizon modes. In flat space, these densities are defined as 
\begin{align}
    n_{\rm super} &= \int_{0}^{\mathcal{C}/l} N_{k} S_{n} k^{n-1} dk,\label{nsuperdef} \\
    n_{\rm sub} &= \int_{\mathcal{C}/l}^{\Lambda} N_{k} S_{n} k^{n-1} dk,\label{nsubdef}
\end{align}
where \( \Lambda \) is the ultraviolet cutoff, and 
\( S_{n} = 2\pi^{n/2}/\Gamma(n/2) \) denotes the surface area of the unit \(n\)-sphere. For the sake of the following discussion, we define \( r \equiv \ln(t_{\rm f}/t_{\rm i}) \). 

\subsubsection{\rm 1. Super-horizon}
We now determine the explicit form of the momentum-integrated particle density for the super-horizon modes. Substituting the particle number derived in Eq.(\ref{Nk}) into the integral definition in Eq.(\ref{nsuperdef}) yields:
\begin{equation}
    n_{\rm super}=\frac{S_{n}\mathcal{C}^n}{l^n}\frac{(\mathcal{C}r)^2}{n}\;_2F_3\left(1,1;\frac{3}{2},2,\frac{n}{2}+1;(\mathcal{C}r)^2\right),
\end{equation}
where $\;_2F_3$ denotes the generalized hypergeometric function \cite{Volkmer2023}. To analyze the behavior of this density in the regime of long-time expansion, we consider the asymptotic expansion of the hypergeometric function in the limit \( x \to \infty \):
\begin{equation}
\begin{aligned}
    \;_2F_3\left(1,1;\frac{3}{2},2,\frac{n}{2}+1;x^2\right)=&\frac{n \left(-2\ln(x)+\psi^{(0)}\left(\frac{1}{2}\right)+\psi ^{(0)}\left(\frac{n}{2}\right)-i\pi \right)}{4x^2} +O\left(x^{-4}\right)\\
    &+e^{-\frac{n}{2}i\pi} \frac{\Gamma\left(\frac{n}{2}+1\right)}{4} \frac{e^{-2x}}{x^{n/2+2}} \left(1+\frac{n(n-6)}{16x} +O\left(x^{-2}\right)\right)\\
    &+\frac{\Gamma\left(\frac{n}{2}+1\right)}{4} \frac{e^{2x}}{x^{n/2+2}} \left(1-\frac{n(n-6)}{16x} +O\left(x^{-2}\right)\right),
\end{aligned}
\end{equation}
where $\psi^{(0)}(x)$ represents the digamma function. In the large-\( r \) limit (where \(x = \mathcal{C}r\)), the expression is dominated by the term containing \( e^{2x} \), which corresponds to the power-law growth characteristic of the super-horizon regime. Neglecting the sub-leading terms, the asymptotic expression for the particle density simplifies to:
\begin{equation}\label{sup}
    n_{\rm super}\approx\frac{(\pi\mathcal{C})^{n/2}}{4l^n}\frac{(t_{\rm f}/t_{\rm i})^{n-1}}{\ln^{n/2}(t_{\rm f}/t_{\rm i})}.
\end{equation}
The asymptotic behavior of $n_{\rm super}$ across different dimensions is illustrated in Fig.~\ref{figureS2}(b). 

\subsubsection{\rm 2. Sub-horizon}
For dimensions \( n > 2 \), an exact closed-form expression for \( n_{\rm sub} \) in terms of elementary functions is unavailable. However, we can derive an approximate analytical expression by analyzing the asymptotic behavior of the particle number. Recall from Eq.~(\ref{Nk}) that \( N_k \approx \mathcal{C}^2(kl)^{-2} \) for large momenta (\( k \gg \mathcal{C}/l \)) due to the rapid oscillation of the sine function depending on \( \ln(t_{\rm f}/t_{\rm i}) \). Conversely, for small momenta (\( k \sim \mathcal{C}/l \)), the behavior is \( N_k \approx \mathcal{C}^2\ln^2(t_{\rm f}/t_{\rm i}) \). Consequently, the integral is dominated by contributions from two distinct regions: the ultraviolet limit where \( k \sim \Lambda \) (typically large \( \Lambda l \)) and the infrared limit where \( k \sim \mathcal{C}/l \) (characterized by large \( \ln(t_{\rm f}/t_{\rm i}) \)).

To evaluate the integral, we introduce the cutoff parameter \( s \equiv \sqrt{(\Lambda l)^2 - \mathcal{C}^2} \) and perform a change of variables to \( q = \sqrt{(kl)^2 - \mathcal{C}^2} \). With these substitutions, the integral can be approximated as:
\begin{equation}
\begin{aligned}
    n_{\rm sub}&=\frac{S_{n}\mathcal{C}^n}{l^n}
    \left[\int_{0}^{\mathcal{C}}\frac{\sin^2(qr)}{q}\left[1 + \left(\frac{q}{\mathcal{C}}\right)^2 \right]^{(n-2)/2}dq+  \int_{\mathcal{C}}^{s}\frac{\sin^2(qr)}{q}\left[1 + \left(\frac{q}{\mathcal{C}}\right)^2 \right]^{(n-2)/2}dq
    \right]\\
    &\approx\frac{S_{n}\mathcal{C}^n}{l^n}\left[
    \int_{0}^{\mathcal{C}}\frac{\sin^2(qr)}{q}dq+ \int_{\mathcal{C}}^{s}\frac{\sin^2(qr)}{q}\left(\frac{q}{\mathcal{C}}\right)^{n-2}dq\right].
\end{aligned}
\end{equation}
These integrals yield analytic solutions in terms of special functions:
\begin{equation}
\begin{aligned}
    \int_{0}^{\mathcal{C}}\frac{\sin^2(qr)}{q}dq &= \frac{1}{2}\!\left[\ln(2\mathcal{C}r) + \gamma - \mathrm{Ci}(2\mathcal{C}r)\right], \\
    \int_{\mathcal{C}}^{s}\frac{\sin^2(qr)}{q}\left(\frac{q}{\mathcal{C}}\right)^{n-2}dq &= \frac{1}{2 (n-2)}+\frac{\left({s}/{\mathcal{C}}\right)^{n-2}}{2 (n-2)}+\left.\frac{\left({i}/{2}\right)^n \Gamma (n-2,-2 i x)+\left(-{i}/{2}\right)^n \Gamma (n-2,2 i x)}{(\mathcal{C}r)^{n-2}}\right|^{\mathcal{C}r}_{x=sr},
\end{aligned}
\end{equation}
where \( \mathrm{Ci}(x) \) is the cosine integral, \( \gamma \approx 0.577216 \) is the Euler-Mascheroni constant, and \( \Gamma(a, x) \) is the incomplete gamma function. Their asymptotic expansions for large arguments are given by:
\begin{align}
    \mathrm{Ci}(x) &\sim \sum_{p=0}^{\infty} \frac{p!}{2} \left(\frac{e^{ix}}{(ix)^{p+1}} + \frac{e^{-ix}}{(-ix)^{p+1}}\right),\\
    \Gamma (n-2,x) &= (n-3)!e^{-x}\sum_{p=0}^{n-3}\frac{x^p}{p!}.
\end{align}
In the limit of large \( r \equiv \ln(t_{\rm f}/t_{\rm i}) \), terms independent of \( \ln(r) \) become subleading. The leading-order behavior is found to be:
\begin{equation}
    n_{\rm sub}\approx \frac{S_{n}\mathcal{C}^n}{2l^n}
    \ln\!\left[\ln(t_{\rm f}/t_{\rm i})\right].
\end{equation}
This result confirms the universality of the logarithmic growth for sub-horizon modes across all dimensions \( n \geq 2 \). Including the next-to-leading order (constant) term for \( n > 2 \), the expression becomes:
\begin{equation}\label{sub}
    n_{\rm sub}\approx \frac{S_{n}\mathcal{C}^n}{2l^n}\left[\ln\!\left[\ln(t_{\rm f}/t_{\rm i})\right]
    +\frac{\left({\Lambda l}/{\mathcal{C}}\right)^{n-2}}{n-2}\right].
\end{equation}
The second term effectively accounts for the contribution from the high-momentum region (\( k \sim \Lambda \)), while the logarithmic term arises from the low-momentum region (\( k \sim \mathcal{C}/l \)). Physical consistency requires that the second term not be neglected when \(\ln(r) < (\Lambda l/\mathcal{C})^{n-2}/(n-2)\).

\subsubsection{\rm 3. Numerical results}
\begin{figure}[h]
    \includegraphics[width=0.45\textwidth]{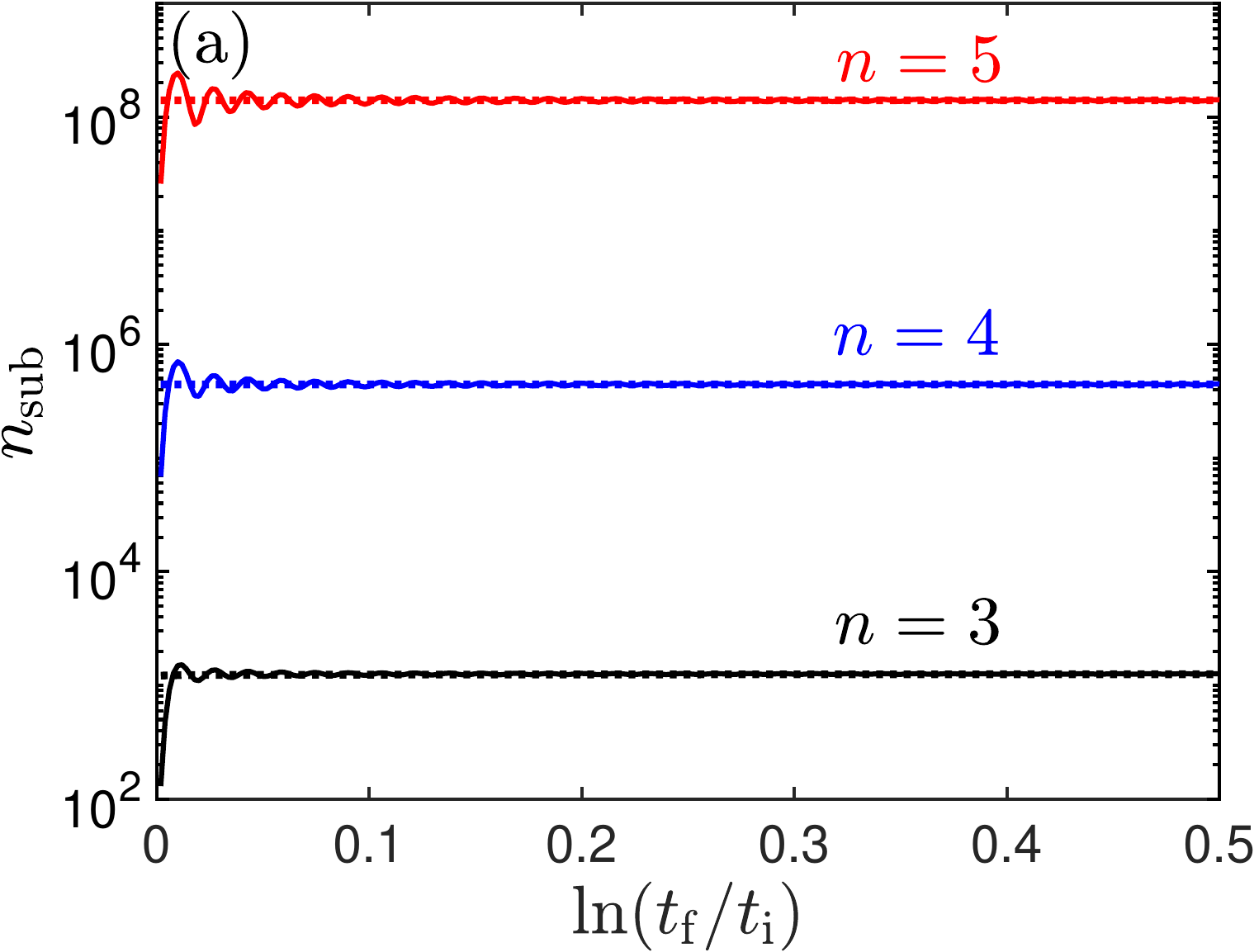}
    \includegraphics[width=0.45\textwidth]{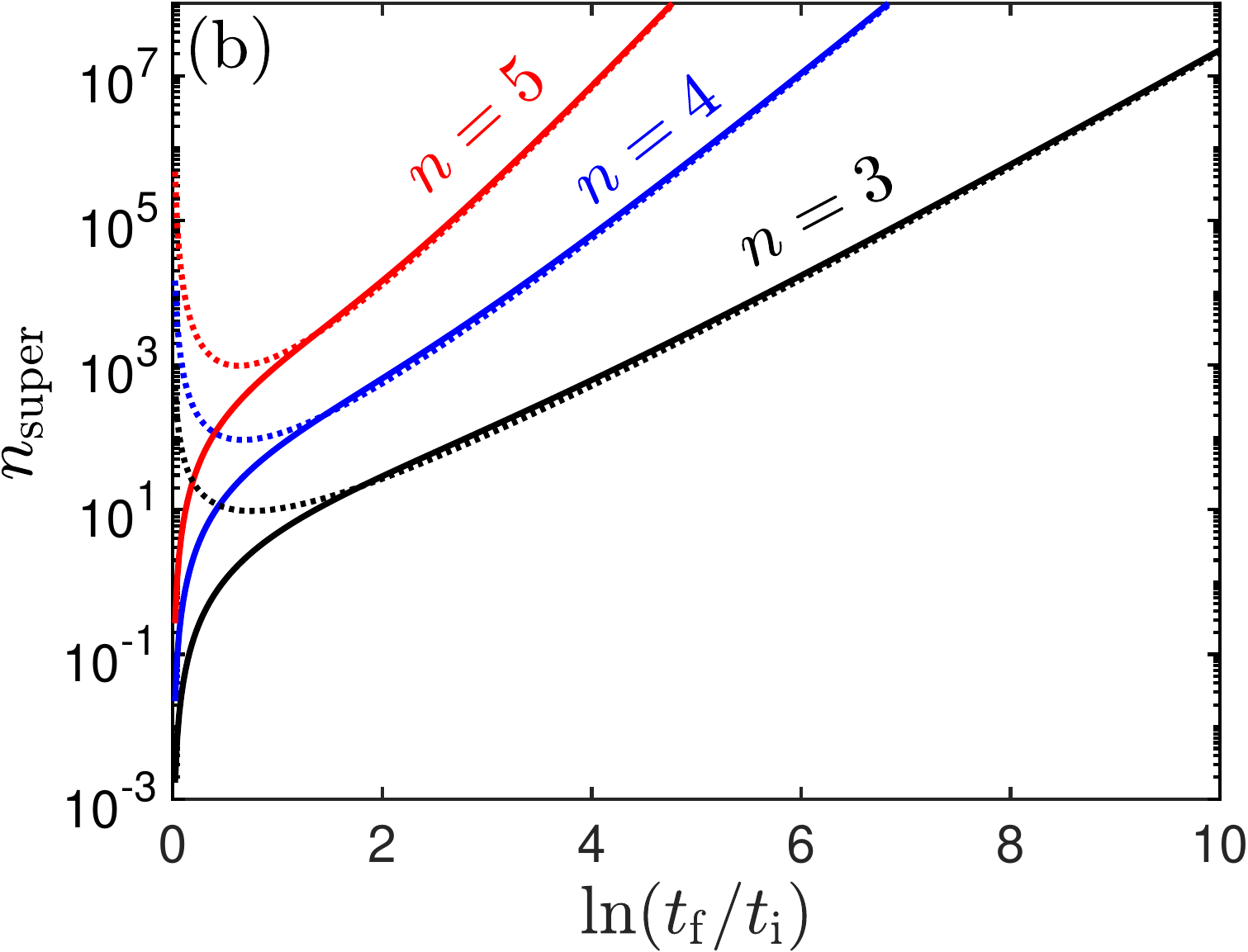}
    \caption{Phonon number density in (a) the sub- and (b) the super-horizon regime for the $(n+1)$ dimensional analog coasting universe. The solid curves represent the exact phonon number densities in the super-horizon regime $n_{\rm super}$, and in the sub-horizon regime $n_{\rm sub}$. The dotted curves are the asymptotic limit of $n_{\rm super}$ and $n_{\rm sub}$, respectively. The ultraviolet cutoff $\Lambda=200/l$. We take $l$ as the unit of length.}
    \label{figureS2}
\end{figure}
To validate the universality of our results, we compute the phonon number densities and compare them to their asymptotic expressions in \((n+1)\)-dimensional FLRW spacetimes, distinguishing between the sub-horizon and super-horizon regimes. The time evolution in these regimes is characterized by the logarithmic expansion factor \(\ln(t_{\rm f}/t_{\rm i})\). 
\begin{itemize}
    \item \textbf{Sub-horizon regime }($n_{\text{sub}}$): As illustrated in Fig.~\ref{figureS2}(a), the phonon number density in the sub-horizon regime is depicted by solid curves (exact numerical results) and dotted curves (asymptotic behavior). The exact density displays transient oscillations at early times but rapidly saturates, converging perfectly to the constant asymptotic limit. This behavior confirms that short-wavelength modes are largely insensitive to the expansion of the universe.

    \item \textbf{Super-horizon regime }($n_{\text{super}}$): In Fig.~\ref{figureS2}(b), we display the phonon number density in the super-horizon regime for various spatial dimensions ($n=3, 4, 5$). The density grows monotonically with the expansion factor $\ln(t_{\rm f}/t_{\rm i})$, driven by the amplification of long-wavelength modes. Similar to the sub-horizon case, the solid curves (exact results) show excellent agreement with the dotted curves (asymptotic limits), confirming the theoretical scaling behavior in the presence of dominant super-horizon modes.
\end{itemize}
The excellent agreement between the exact numerical integration of $N_k$ and the asymptotic expressions derived in Eqs.~(\ref{sup}) and (\ref{sub}) for dimensions $n=3,4,5$ confirms the validity of our analytical approximations across different dimensions.

%